\documentclass[aps,floatfix,10pt,twocolumn,preprintnumbers,amsmath,amssymb,superscriptaddress,nofootinbib,longbibliography,prb]{revtex4-2}

\usepackage{graphicx}

\usepackage{bm}
\usepackage{dsfont}
\usepackage[usenames,dvipsnames]{xcolor}
\usepackage{pstricks}
\usepackage[tight]{subfigure}
\usepackage{verbatim}
\usepackage{units}
\usepackage{multirow}
\usepackage{enumitem}
\usepackage{mathrsfs}
\usepackage{leftidx}
\usepackage{xspace}
\usepackage{braket}
\usepackage{bbm}
\usepackage{physics}
\usepackage{mathtools}
\usepackage{amsfonts,amssymb,amsmath}
\usepackage{tabularx}
\usepackage[normalem]{ulem}
\usepackage{bbold}
\usepackage{capt-of}
\usepackage{hyperref}
\hypersetup{colorlinks=true,linktoc=all,linkcolor=blue,breaklinks=true,citecolor=blue,urlcolor=blue}

\usepackage[english]{babel}
\addto\captionsenglish{}

\usepackage{hyperref}

\AtBeginDocument{%
\newwrite\bibnotes
\def\bibnotesext{Notes.bib}
\immediate\openout\bibnotes=\jobname\bibnotesext
\immediate\write\bibnotes{@CONTROL{REVTEX42Control}}
\immediate\write\bibnotes{@CONTROL{%
        apsrev42Control,author="08",editor="1",pages="1",title="0",year="1"}}
\if@filesw
\immediate\write\@auxout{\string\citation{apsrev42Control}}%
\fi
}%

\newcommand{\inv}{\text{i}}

\renewcommand{\L}{\text{L}}
\renewcommand{\P}{\text{P}}
\newcommand{\C}{\text{C}}
\newcommand{\dT}{\delta T}
\newcommand{\R}{\text{R}}
\newcommand{\dS}{\ensuremath{\delta S}}
\newcommand{\bx}{\mathbf{x}}
\newcommand{\bR}{\mathbf{R}}
\newcommand{\bA}{\mathbf{A}}
\newcommand{\bB}{\mathbf{B}}
\newcommand{\bM}{\bar{\mathbf{M}}}
\newcommand{\cO}{\mathcal{O}}
\newcommand{\cS}{\mathcal{S}}
\newcommand{\cT}{\mathcal{T}}

\newcommand*{\tW}{\tilde{W}}
\newcommand*{\etaC}{\eta_\text{Carnot}}
\newcommand{\nr}{\bm{\nabla}_\bR}
\newcommand{\pr}[1][1]{\partial_{R_{#1}}}

\newcommand{\Refs}[1]{Refs.~\cite{#1}}
\providecommand{\Ref}[1]{Ref.~\cite{#1}}
\renewcommand{\Ref}[1]{Ref.~\cite{#1}}
\newcommand{\App}[1]{Appendix~\ref{#1}}
\newcommand{\Eq}[1]{Eq.~(\ref{#1})}
\newcommand{\Eqs}[1]{Eqs.~(\ref{#1})}
\newcommand{\eq}[1]{(\ref{#1})}
\newcommand{\Fig}[1]{Fig.~\ref{#1}}

\newcommand{\Figs}[1]{Figs.~\ref{#1}}
\newcommand{\Sec}[1]{Sec.~\ref{#1}}
\renewcommand{\ket}[1]{|#1\rangle}
\renewcommand{\bra}[1]{\langle#1|}

\renewcommand{\Ket}[1]{\bm{|}#1\bm{)}}
\renewcommand{\Bra}[1]{\bm{(}#1\bm{|}}
\renewcommand{\Braket}[2]{\bm{(}#1\bm{|}#2\bm{)}}
\newcommand{\KetBra}[2]{\bm{|}#1\bm{)}\bm{(}#2\bm{|}}
\newcommand{\zin}{z^{\inv}}
\newcommand{\zia}{z^{\inv}_\alpha}
\newcommand{\za}{z_\alpha}
\newcommand{\one}{\mathds{1}}
\newcommand{\fpOp}{(-\one)^N}
\newcommand{\cD}{\mathcal{D}}
\newcommand{\pH}{\mathcal{P}_{\mathcal{H}}}
\newcommand{\cP}{\mathcal{P}}
\newcommand*{\melt}[3]{\Bra{#1}#2\Ket{#3}}
\newcommand{\comal}{\alpha}
\newcommand{\comL}{\L}
\newcommand{\comR}{\R}
\newcommand{\TC}{\text{TC}}
\newcommand{\NTC}{\text{NTC}}
\newcommand{\CN}{\C}
\newcommand{\PN}{\P,N}
\newcommand{\Pp}{\P,p}
\newcommand{\gamc}{\gamma_{\text{c}}}
\newcommand{\gamcp}{\gamma_{\text{c}/\text{p}}}
\newcommand{\gamcpa}{\gamma_{\text{c}/\text{p}\comal}}
\newcommand{\gamca}{\gamma_{\text{c}\comal}}
\newcommand{\gamcR}{\gamma_{\text{c}\comR}}
\newcommand{\gamcL}{\gamma_{\text{c}\comL}}
\newcommand{\gampR}{\gamma_{\text{p}\comR}}
\newcommand{\gampL}{\gamma_{\text{p}\comL}}
\newcommand{\gamcia}{\gamma^{\inv}_{\text{c}\comal}}
\newcommand{\gamcap}{\gamma_{\text{c}\comal'}}
\newcommand{\gamcapp}{\gamma_{\text{c}\comal''}}
\newcommand{\gamp}{\gamma_{\text{p}}}
\newcommand{\gampa}{\gamma_{\text{p}\comal}}

\newcommand{\nz}{N_z}
\newcommand{\nza}{N_{z\comal}}
\newcommand{\nzap}{N_{z\comal'}}
\newcommand{\nzapp}{N_{z\comal''}}
\newcommand{\nzi}{N^{\inv}_z}
\newcommand{\nzia}{N^{\inv}_{z\comal}}
\newcommand{\nziap}{N^{\inv}_{z\comal'}}

\newcommand{\nzR}{N_{z\comR}}
\newcommand{\nzL}{N_{z\comL}}
\newcommand{\pzR}{p_{z\comR}}

\newcommand{\nziR}{N^{\inv}_{z\comR}}
\newcommand{\nziL}{N^{\inv}_{z\comL}}
\newcommand{\mua}{\mu_\alpha}
\newcommand{\muap}{\mu_{\alpha'}}
\newcommand{\muL}{\mu_{\text{L}}}
\newcommand{\muR}{\mu_{\text{R}}}
\newcommand{\Ta}{T_\alpha}

\newcommand{\Tapp}{T_{\alpha''}}
\newcommand{\TL}{T_{\text{L}}}
\newcommand{\TR}{T_{\text{R}}}
\newcommand{\Gamap}{\Gamma_{\alpha'}}
\newcommand{\Gama}{\Gamma_\alpha}
\newcommand{\GamL}{\Gamma_{\text{L}}}
\newcommand{\GamR}{\Gamma_{\text{R}}}
\newcommand{\Vb}{V_{\text{b}}}
\newcommand{\Wa}{W^{\alpha}}
\newcommand{\tWa}{\tilde{W}^{\alpha}}
\newcommand{\Wap}{W^{\alpha'}}

\newcommand{\WR}{W^{\R}}
\newcommand{\IOa}{I_{\cO\comal}}
\newcommand{\INa}{I_{N\comal}}
\newcommand{\IHa}{I_{H\comal}}
\newcommand{\IQa}{I_{Q\comal}}
\newcommand{\IOL}{I_{\cO\comL}}
\newcommand{\INL}{I_{N\comL}}

\newcommand{\IOR}{I_{\cO\comR}}
\newcommand{\INR}{I_{N\comR}}
\newcommand{\IHR}{I_{H\comR}}
\newcommand{\AOa}{\mathbf{A}_{\cO\comal}}
\newcommand{\ANa}{\mathbf{A}_{N\comal}}
\newcommand{\AHa}{\mathbf{A}_{H\comal}}
\newcommand{\AHR}{\mathbf{A}_{H\comR}}
\newcommand{\AOaC}{\mathbf{A}^{\C}_{\cO\comal}}

\newcommand{\AHaC}{\mathbf{A}^{\C}_{H\comal}}
\newcommand{\AOaP}{\mathbf{A}^{\P}_{\cO\comal}}

\newcommand{\AHaP}{\mathbf{A}^{\P}_{H\comal}}
\newcommand{\aCNOa}{a_{\cO\comal}^{\CN}}
\newcommand{\aPNOa}{a_{\cO\comal}^{\PN}}
\newcommand{\aPpOa}{a_{\cO\comal}^{\Pp}}
\newcommand{\aCNNa}{a_{N\comal}^{\CN}}

\newcommand{\aCNHa}{a_{H\comal}^{\CN}}
\newcommand{\aCTCHa}{a_{H\comal}^{\CN,\TC}} 
\newcommand{\aCNTCHa}{a_{H\comal}^{\CN,\NTC}} 
\newcommand{\aCNHR}{a_{H\comR}^{\CN}}
\newcommand{\aPNHa}{a_{H\comal}^{\PN}}
\newcommand{\aPNHR}{a_{H\comR}^{\PN}}
\newcommand{\aPpHa}{a_{H\comal}^{\Pp}}
\newcommand{\aPpHR}{a_{H\comR}^{\Pp}}
\newcommand{\taCNHa}{\tilde{a}_{H\comal}^{\CN}}
\newcommand{\aOa}{a_{\cO\comal}}
\newcommand{\BOa}{\mathbf{B}_{\cO\comal}}
\newcommand{\BNa}{\mathbf{B}_{N\comal}}
\newcommand{\BHa}{\mathbf{B}_{H\comal}}
\newcommand{\BHR}{\mathbf{B}_{H\comR}}
\newcommand{\BNR}{\mathbf{B}_{N\comR}}
\newcommand{\BHRC}{\mathbf{B}^{\C}_{H\comR}}
\newcommand{\BHRP}{\mathbf{B}^{\P}_{H\comR}}
\newcommand{\BOaC}{\mathbf{B}^{\C}_{\cO\comal}}

\newcommand{\BHaC}{\mathbf{B}^{\C}_{H\comal}}
\newcommand{\BOaP}{\mathbf{B}^{\P}_{\cO\comal}}

\newcommand{\BHaP}{\mathbf{B}^{\P}_{H\comal}}
\newcommand{\BCNTCHa}{B_{H\comal}^{\CN,\NTC}} 
\newcommand{\bxC}{\bx^{\text{C}}}
\newcommand{\bxP}{\bx^{\text{P}}}
\newcommand{\teq}{\text{eq}}
\newcommand{\cEa}{\mathcal{E}^{\alpha}}

\newcommand{\cEeq}{\mathcal{E}^{\teq}}
\newcommand{\ciEeq}{\bar{\mathcal{E}}^{\teq}}
\newcommand{\gamceq}{\gamma_{\text{c},\teq}}
\newcommand{\nzeq}{N_{\teq}}
\newcommand{\nzieq}{N^{\inv}_{\teq}}
\newcommand{\delnzsqeq}{\delta N^2_{\teq}}
\newcommand{\delnzisqeq}{(\delta N^{\inv}_{\teq})^2}
\newcommand{\lamnzcbeq}{\lambda\!N^3_{\teq}}
\newcommand{\lamnzicbeq}{(\lambda\!N^{\inv}_{\teq})^3}

\newcommand{\zeq}{z_{\teq}}
\newcommand{\zieq}{z^{\inv}_{\teq}}
\newcommand{\pzeq}{p_{\teq}}
\newcommand{\Weps}{\mathcal{W}_{\epsilon}}
\newcommand{\WLam}{\mathcal{W}_{\Lambda}}
\newcommand{\Work}{\mathcal{W}}
\newcommand{\QR}{\mathcal{Q}_{\R}}
\newcommand{\beps}{\bar{\epsilon}}
\newcommand{\bLam}{\bar{\Lambda}}
\newcommand{\bVb}{\bar{V}_{\text{b}}}
\newcommand{\bU}{\bar{U}}
\newcommand{\bTL}{\bar{T}_{\text{L}}}
\newcommand{\bTR}{\bar{T}_{\text{R}}}
\newcommand{\bmua}{\bar{\mu}_{\alpha}}
\newcommand{\bmuL}{\bar{\mu}_{\text{L}}}

\newcommand{\TN}{T}
\newcommand{\sudag}{\boldsymbol{\dagger}}

\begin{document}

\title{Geometric energy transport and refrigeration with driven quantum dots}

\author{Juliette Monsel}
\email{monsel@chalmers.se}
\affiliation{Department of Microtechnology and Nanoscience (MC2), Chalmers University of Technology, S-412 96 G\"oteborg, Sweden}

\author{Jens Schulenborg}
\affiliation{Center for Quantum Devices, Niels Bohr Institute, University of Copenhagen, DK-2100 Copenhagen}

\author{Thibault Baquet}
\affiliation{Department of Microtechnology and Nanoscience (MC2), Chalmers University of Technology, S-412 96 G\"oteborg, Sweden}

\author{Janine Splettstoesser}
\affiliation{Department of Microtechnology and Nanoscience (MC2), Chalmers University of Technology, S-412 96 G\"oteborg, Sweden}

\date{\today}

\begin{abstract}
    We study geometric energy transport in a slowly driven single-level quantum dot weakly coupled to electronic contacts and with strong onsite interaction, which can be either repulsive or attractive. Exploiting a recently discovered fermionic duality for the evolution operator of the master equation, we provide compact and insightful analytic expressions of energy pumping curvatures for any pair of driving parameters. This enables us to systematically identify and explain the pumping mechanisms for different driving schemes, thereby also comparing energy and charge pumping. We determine the concrete impact of many-body interactions and show how particle-hole symmetry and fermionic duality manifest, both individually and in combination, as system-parameter symmetries of the energy pumping curvatures. Building on this transport analysis, we study the driven dot acting as a heat pump or refrigerator, where we find that the sign of the onsite interaction plays a crucial role in the  performance of these thermal machines.
\end{abstract}

\maketitle

\section{Introduction}

Energy transport in driven mesoscopic systems is of high interest from the perspective of two very different research fields. First, while steady-state transport spectroscopy is a well established experimental tool to characterize quantum devices, opportunities arising from the combination of AC driving~\cite{Schawlow1982Jul,Haight1995Jan,Splettstoesser2006Aug,Xu2007Aug,Hoffmann2009Apr,Reckermann2010Jun,Rohwer2011Mar,Wang2013Oct,Fuks2015May,Wiedenmann2016Jan,Riwar2016Jun,Koski2018Jul,Hays2018Jul,Luu2018Mar,Dartiailh2021Jan,Krojer2022Jan, Vigneau2022Feb} \emph{and} heat or energy transport~\cite{leSueur2010Jul,Altimiras2010Jan,Altimiras2010Nov,Halbertal2016Nov,Rossello2015Mar,Marguerite2019Nov,Krahenmann2019Sep,Fletcher2019Nov,Bauer2021Nov} in spectroscopy are promising but have been less explored. Of specific interest in this context is whether these two ``knobs" can be leveraged in adiabatic energy pumping~\cite{Haupt2013Nov, Kolodrubetz2018Apr,Hasegawa2020May,Simons2020Dec}, which is expected to give new insights,  especially when factoring in the geometric nature of pumping due to slow driving~\cite{Calvo2012Dec,Haupt2013Nov,Pluecker2017Apr,Pluecker2017Nov,Hasegawa2020May}. 
Second, periodic driving to pump controlled energy flows in mesoscopic conductors is at the heart of realizing cyclic thermal machines at the nanoscale. A broad analysis of such cyclic small-scale engines has been put forward in very different types of devices, see e.g. \Refs{Alicki1979May,Feldmann2003Jul,Allahverdyan2008Apr,Kosloff2014Apr,Kosloff2017Mar,Myers2022Jan,Hino2021Feb} and references therein. 
Driven quantum dots are one of the most basic setups in which a cyclic operation for, e.g., heat engines~\cite{Juergens2013Jun,Dare2016Jan} and motors~\cite{Bustos-Marun2013Aug,Calvo2017Oct,Bruch2018May,Ludovico2014Apr,Bustos-Marun2019Aug,Bhandari2020Oct,Ribetto2021Apr,Bhandari2021Jul} can be implemented.
Geometric charge pumping by slowly driving such dots has been analyzed in great detail in theory~\cite{Aleiner1998Aug,Aono2004Sep,Splettstoesser2005Dec, Splettstoesser2006Aug,Schiller2008Jan,Moskalets2008Jul,Riwar2010Nov,Battista2011Mar,Taguchi2016,Hasegawa2017Jan,Hasegawa2018Mar} and experiment~\cite{Pothier1992Jan,Leek2005Dec,Roche2013Mar,Pekola2013Oct}. However, energy pumping has only been studied addressing specific aspects~\cite{Arrachea2007Jun,Ren2010Apr,Haupt2013Nov,Ludovico2016Feb, Hasegawa2020May,Riwar2021Jan}, whereas a detailed, full-fledged analytical study is still missing, even for the simplest case of a single-level quantum dot.

\begin{figure}[ht!]
    \centering
    \includegraphics{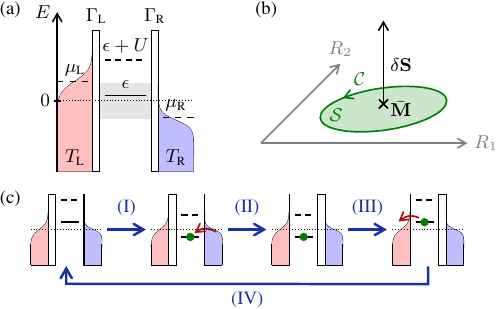}
    \caption{\label{fig_model}
        (a) Energy landscape of a spin-degenerate, single-level quantum dot with level position $\epsilon$ and local two-particle interaction strength $U$, tunnel coupled to two reservoirs $\alpha = \L,\R$ at temperature $\Ta$ and chemical potential $\mua$, where $\muL-\muR=\Vb$, with coupling strengths $\Gamma_{\text{L}/\text{R}} = \Gamma(1 \pm \Lambda)/2$, where $\Gamma$ is a fixed typical tunneling frequency and $\Lambda$ is the left-right asymmetry. The sketch shows the usual situation of repulsive Coulomb interaction $U > 0$, but we also consider the attractive case $U < 0$. (b) Representation of the closed driving path $\cal C$ in parameter space of a pair of generic driving parameters $R_1,R_2$, encircling the surface with vector $\delta\mathbf{S}$ and area $\cal S = |\delta\mathbf{S}|$ around a working point defined by time-averaged parameters $\bM = (\beps,\,\bU,\, \bVb,\, \bLam,\, \bTL)$. The pumping procedure is detailed in \Sec{sec_pumpingintro}. 
        (c) Example of a driving cycle as considered in \Sec{sec_refrigerator}, implementing a refrigerator with $\TL > \TR$ and $\muL = \muR$ and driving parameters $\epsilon$ and $\Lambda$, the latter indicated by vertical lines of different thickness.
    }
\end{figure}

This paper therefore systematically analyzes geometric energy pumping through a single-level quantum dot with possibly strong onsite interaction---which can be either repulsive or attractive---and weak tunnel coupling to electronic reservoirs. We address the characteristics of energy pumping for transport spectroscopy as well as cyclically operated quantum dot heat engines. Appealing to both the geometric character of the driving and to general symmetries of open fermionic systems, we derive analytical expressions directly linking the intuitively understandable, well-known DC linear response properties of the dot~\cite{Staring1993Apr,Dzurak1993Sep} to the pumped charge and energy. This enables an in-depth pumping analysis revealing the pumped energy to be significantly more insightful into the resonant and off-resonant dot dynamics than the charge, especially for attractive interaction. Moreover, we directly relate the efficiency of the dot operated as a driven refrigerator or heat pump to the dot's steady-state thermovoltage. This makes particularly transparent how local electron pairing due to attractive onsite interaction would influence this operation.

To realize geometric charge and energy pumping, we assume the system to be slowly driven in time by the periodic modulation of an arbitrary pair of parameters, see \Fig{fig_model}. This includes any finite, and possibly driven interaction strength, bias voltage~\cite{Reckermann2010Jun,Calvo2012Dec,Placke2018Aug} and temperature differences between the leads, with the latter two being particularly relevant when operating the system as an engine, a refrigerator or a heat pump. Namely, time-dependent driving can result in charge and energy flow against the current direction imposed by biases. Beyond thermoelectrics, geometric pumping currents which superpose bias-induced stationary currents have also been shown to be useful as a spectroscopic tool~\cite{Reckermann2010Jun,Calvo2012Dec}.

We also include the less usual \textit{attractive} interaction into our analysis. In locally confined systems, an electron pair can experience such attractive interaction if, by bringing the pair closer together, \emph{additional electrons} repelling off the pair are enabled to redistribute in a way that overcompensates the added Coulomb energy between the pair~\cite{Placke2018Aug}. This effect was shown experimentally for quantum dots in \Refs{Cheng2015May,Hamo2016Jul,Cheng2016Dec,Prawiroatmodjo2017Aug}, thus providing a simple, tunable platform to test the effect of possibly strong electron-pairing on various transport phenomena. In particular, the energy current through such a dot \emph{directly} probes the interaction energy ---in contrast to the charge current~\cite{Schulenborg2016Feb,Vanherck2017Mar}--- and attractive interaction has been predicted to exhibit unusual features already in stationary thermoelectric transport~\cite{Schulenborg2020Jun}. It is thus not only fundamentally interesting to extend the previous work on geometric charge pumping with attractive interaction~\cite{Placke2018Aug} to energy pumping. In fact, it is also expected and confirmed by our work that interaction-induced pairing can have a considerable effect on the operation and performance of a dot as a driven thermal machine.  

To efficiently cover the many different system parameter regimes, we make use of a recently discovered dissipative symmetry of the master equation of fermionic systems~\cite{Schulenborg2016Feb}, which is referred to as fermionic duality. This duality relation has proven useful to describe heat- and energy transport not only in time-dependently driven~\cite{Vanherck2017Mar, Schulenborg2018Dec,Ortmanns2022} but also in stationary systems~\cite{Schulenborg2017Dec,Schulenborg2020Jun,Bruch2021Sep}. Importantly, fermionic duality is valid for arbitrary local interaction strengths, as well as for any bias voltage or temperature difference between the leads; it hence yields general, yet compact and easy to interpret analytical results for the complete geometric formulation of the pumped energy, which were until now lacking. We thereby systematically identify ---both for strong repulsive or attractive onsite interaction--- different pumping mechanisms~\cite{Placke2018Aug} and pinpoint the nature of the contributions of different modes of the dissipative dot dynamics. This in particular exposes the parameter regimes in which the pumped energy differs nontrivially from the pumped charge and highlights the impact of many-body interactions. Moreover, our formalism clearly identifies symmetries between the pumped energy at different working points and with different interaction signs, due to particle-hole symmetry and due to fermionic duality itself. 
Equipped with these general physical insights into energy pumping, we then address the slowly driven quantum dot specifically as a refrigerator or heat pump. 
Our analytical approach relates the performance of this quantum-dot machine directly the to dot's steady-state Seebeck coefficient, its equilibrium charge fluctuations, and its typical RC time~\cite{Splettstoesser2010Apr}. This provides intuitive, yet quantitative predictions for the output power and efficiency, and elucidates how the interplay between onsite interaction, its sign, and the driving frequency affects these quantities.

The paper is organized as follows. We introduce the theoretical approach in \Sec{sec_theory}. Section \ref{sec_analytical_results} then sets up the full analytical formulation of the pumping transport equations relying on the fermionic duality and discusses general properties of (energy) pumping based on these analytical results. Specific features of energy pumping for different driving schemes are discussed in detail in \Sec{sec_results_repulsive} for a quantum dot with repulsive onsite interaction and in  \Sec{sec_results_attractive} for a quantum dot with attractive onsite interaction. Finally, in \Sec{sec_refrigerator}, we analyze the driven quantum dot as a refrigerator or heat pump, where heat is pumped out of the cold and into the hot contact, and identify the performance characteristics for repulsive and attractive quantum dots. Readers more inclined towards this last part may also skip the discussion in \Sec{sec_results}, as \Sec{sec_refrigerator} focuses on a different system parameter regime. Further details and derivations are presented in extensive appendices.

\section{Theoretical approach}\label{sec_theory}

We start by introducing the model for the quantum dot and the time-dependent driving leading to pumping. This also includes an overview of the master equation approach our theoretical analysis is based on.

\subsection{Quantum-dot model}\label{sec_dotmodel}

We consider a single-level quantum dot, with a spin-degenerate energy level $\epsilon$, tunnel coupled to an environment as represented in \Fig{fig_model}(a). The total Hamiltonian $\hat{H}_{\text{tot}} = \hat{H} + \sum_\alpha\hat{H}_\alpha + \hat{H}_{\text{tun}}$ describes the isolated quantum dot $(\hat{H})$, two macroscopic electronic contacts $(\hat{H}_{\alpha=\text{L,R}})$, and their tunnel coupling to the dots $(\hat{H}_{\text{tun}})$. The dot Hamiltonian reads
\begin{equation}
    \hat{H}=\sum_{\sigma=\uparrow,\downarrow}\epsilon\hat{N}_\sigma+U\hat{N}_\uparrow\hat{N}_\downarrow.
\end{equation}
Here, the occupation number operator for the different spin directions is given by $\hat{N}_\sigma=\hat{d}^\dagger_\sigma\hat{d}^{}_\sigma$, with the dot annihilation operator $\hat{d}_\sigma$ for spin $\sigma$ and $\hat{N}=\sum_\sigma \hat{N}_\sigma$. 
We account for---possibly strong---onsite interaction between electrons, which is characterized by the interaction strength $U$. Importantly, in the present paper, we analyze both the standard situation of repulsive Coulomb interaction, where $U>0$, but also the case where the onsite interaction is attractive, $U<0$. Such quantum dots with attractive interaction have recently been studied theoretically~\cite{Koch2006Feb,Koch2007May,Costi2010Jun,Andergassen2011Dec,Taraphder1991May} and have been realized in experiment~\cite{Cheng2015May,Hamo2016Jul,Cheng2016Dec,Prawiroatmodjo2017Aug}.
The possible states of the dot are the empty state $\ket{0}$, the single occupied state with either spin up $(\ket{\!\uparrow})$ or spin down $(\ket{\!\downarrow})$, or the state of double occupation $\ket{2}$.

The two macroscopic electronic contacts $\alpha=\text{L,R}$ are assumed to be simple spin-degenerate metallic leads. The corresponding Hamiltonian of the lead $\alpha$ is given by
\begin{equation}
    \hat{H}_\alpha=\sum_{k,\sigma}\varepsilon_{\alpha,k}^{} \hat{c}_{\alpha,k,\sigma}^\dagger\hat{c}_{\alpha,k,\sigma}^{}\label{H_lead}
\end{equation}
with the annihilation operator $\hat{c}_{\alpha,k,\sigma}$ acting on states with orbital quantum number $k$. The lead occupation number operator is given by $\hat{N}_\alpha=\sum_{k,\sigma} \hat{c}_{\alpha,k,\sigma}^\dagger\hat{c}_{\alpha,k,\sigma}^{}$. The Fermi functions $f^{\pm}_\alpha(E)=\left(1+\exp[\pm(E-\mua)/k_\text{B}\Ta]\right)^{-1}$ with the electrochemical potential $\mua$ and the temperature $\Ta$ quantify the corresponding, average particle$(+)$/hole$(-)$ occupation per mode with energy $E$.

The state of the quantum dot can change in time due to tunneling to or from the leads. The corresponding tunneling Hamiltonian is assumed to be spin-independent,
\begin{equation}
    \hat{H}_\text{tun}=\sum_{\alpha,k,\sigma}V_{\alpha,k}^{}\hat{c}_{\alpha,k,\sigma}^\dagger\hat{d}_\sigma^{}+\text{h.c.}\label{eq_hamiltonian_tun}
\end{equation}
Given the simple metallic contacts described above, the tunneling Hamiltonian $\hat{H}_\text{tun}$ conserves spin and charge. On an energy scale given by the internal dot splittings, the density of states in the contacts and coupling strength $V_{\alpha,k}$ typically vary little around the chemical potential. We therefore simplify the typical tunneling rate $\Gamma_{\alpha}=\pi\rho_\alpha|V_{\alpha}|^2$ also as energy-independent, i.e. with $V_{\alpha,k}\equiv V_{\alpha}$ (wideband limit).

The focus of our studies is on the---experimentally relevant---weak-coupling regime, $\Gama \ll k_\text{B}\Ta$. Since we are particularly interested in the role of electron-electron interaction for energy pumping, we furthermore mostly study (but are not limited to) the case $|U|\gg k_\text{B}\Ta$, especially for the detailed analysis in \Sec{sec_results}. Also note that we here only consider devices in which the energy flow via coupling between electrons and bosonic degrees of freedom (phonons, photons, etc.) is negligible. Finally, we henceforth set $\hbar = k_\text{B} = |e| = 1$.

\subsection{Adiabatic charge and energy pumping}\label{sec_pumpingintro}

This paper deals with adiabatic charge and energy pumping, i.e., charge and energy transport across the quantum dot due to the slow periodic modulation of system parameters. We hence analyze the time-resolved particle- and energy currents into one contact $\alpha$,
\begin{gather}
    \INa(t) = \partial_t\langle\hat{N}_\alpha\rangle \quad,\quad \IHa(t) = \partial_t\langle\hat{H}_\alpha\rangle,
\end{gather}
averaged over one period $2\pi/\Omega$ of the driving, $\Delta N_\alpha=\int_0^{2\pi/\Omega}dt \INa(t)$ and $\Delta H_\alpha=\int_0^{2\pi/\Omega}dt \IHa(t)$. We denote the expectation value with respect to the total-system state as  $\langle\bullet\rangle$.
The adiabatically pumped transport quantities\footnote{This study can straightforwardly be extended to heat currents, $\IQa(t) = \IHa(t) - \mua(t) \INa(t)$, see also \App{app_macro_timedep}.} $\Delta N_\alpha$ and $\Delta H_\alpha$ are geometric quantities~\cite{Altshuler1999Mar,Mottonen2008Apr,Yuge2012Dec,Thingna2014Sep,Pluecker2017Apr,Pluecker2017Nov} and require the modulation of at least a pair of driving parameters encircling a finite surface in parameter space, as indicated in \Fig{fig_model}(b). 

A pure pumping current is obtained when modulating the dot and coupling parameters $\epsilon(t),\,U(t),\,\Gama(t)$. This can in practice be achieved by externally driven local gate voltages. In the theoretical description, their time-dependence enters directly into the Hamiltonian parameters introduced in the previous \Sec{sec_dotmodel}. 
With regards to the dot-contact coupling, it was previously found for our setup~\cite{Placke2018Aug} that the variation of the combined coupling strength, $\Gamma(t) = \GamL(t)+\GamR(t)$ does not lead to any pumping. We hence fix $\Gamma(t) = \Gamma$, and consider the left-right asymmetry $\Lambda(t)=(\GamL(t)-\GamR(t))/(\GamL(t)+\GamR(t))$ as the only coupling-related pumping parameter.

Beyond pure pumping, we also consider a more generic situation with finite, possibly time-dependent voltage biases and temperature differences. The adiabatic pumping currents are then the first-order in driving frequency contributions in addition to the zeroth-order currents, the latter stemming from steady-state biases or rectification effects.
Driven electrochemical potentials $\mua(t)$ are routinely realized in ac transport experiments. However, the time-dependent modulation of temperatures $\Ta(t)$ of electronic contacts requires well-controlled heating and cooling~\cite{Thierschmann2015Oct,Josefsson2018Oct}. The theoretical treatment of these macroscopic parameters can here be performed by replacing the parameters in the Fermi functions by time-dependent parameters. A more detailed justification of this procedure is given in \App{app_macro_timedep}, see also Refs.~\cite{Luttinger1964Sep,Eich2014Sep,Tatara2015May}. For practical reasons, we choose to modulate the bias voltage $V_\text{b}(t)=\muL(t)-\muR(t)$ symmetrically with respect to both leads, thereby fixing the average potential $\mu_0=(\muL+\muR)/2 \equiv 0$ as reference energy. Temperature driving is instead only performed on the left reservoir $\TL(t)$; the right contact remains at a fixed temperature $\TR\equiv \TN$, which we typically use as energy unit.

Adiabatic pumping due to any of these driving parameters is geometric and does not depend on the detailed time dependence of the driving, but only on the area of the enclosed surface\footnote{At higher orders in frequency, already a single driven dot parameter is sufficient since the required asymmetry is provided by the slight lagging behind of the particle current compared to the driving~\cite{Cavaliere2009Sep}}. Since the precise boundary shape of this surface is not essential for our discussion, we assume an analytically convenient, sinusoidal form
\begin{equation}
    R_i(t)=\bar{R}_i+\delta R_i\sin(\Omega t+\varphi_i)\label{Ri}
\end{equation}
with frequency $\Omega$ for the periodically driven parameters $\epsilon(t)/\TN,\,U(t)/\TN,\,V_\text{b}(t)/\TN,\,\Lambda(t),\,\TL(t)/\TN$. The working point defined by the constant contribution $\bar{R}_i$, and the amplitude $\delta R_i$ are set independently for all different modulated parameters. The phases $\varphi_i$ need to differ between the different $R_i$ to achieve a finite encircled surface in parameter space and hence a possibly finite pumped quantity. The adiabaticity condition reads~\cite{Reckermann2010Jun}
\begin{align}
    \Omega\delta R_i \ll\Gamma.
\end{align}
For better readability, we will later collect the set of any two driving parameters for a specific protocol in a three-component vector $\mathbf{R}=(R_1(t),\,R_2(t),\, 0)^\text{T}$. We call any pair of two parameters a \textit{driving scheme}.

\subsection{Master equation and fermionic duality}\label{sec_master_equation}
To describe the dynamics of the driven quantum-dot system and to calculate the sought-for transported charge and energy, we resort to a master equation approach~\cite{Splettstoesser2006Aug,Cavaliere2009Sep,Reckermann2010Jun}. This is valid in the here relevant regime of weak coupling, $\Gamma\ll \min_\alpha\left\{\Ta\right\}$, and for up to moderately slow driving, $0<\Omega \delta R_i/\Gamma\lesssim 1$. Performing an expansion in $\Omega/\Gamma$, the time-evolution of the reduced quantum-dot density matrix $\rho(t)$ reads
\begin{equation}
    0 = W\Ket{\rho^{(0)}(t)} \quad,\quad \partial_t\Ket{\rho^{(\ell-1)}(t)} =  W\Ket{\rho^{(\ell)}(t)} .\label{eq_master}
\end{equation}
This introduces the \emph{instantaneously time-dependent} kernel superoperator $W$, which describes transitions due to tunneling events between dot and reservoirs. The kernel acts on the reduced density operator $\Ket{\rho(t)} = \hat{\rho}(t)$. We choose a notation with rounded kets $\Ket{x}$ for operators $\hat{x}$ in Liouville space, and with rounded bras for the corresponding covectors $\Bra{x}\bullet=\Tr\{\hat{x}^\dagger\bullet\}$ with respect to the Hilbert-Schmidt scalar product. The driving frequency expansion is denoted with the superscript $(\ell)$, representing the $\ell$-th order in $\Omega/\Gamma$.
The solution for the density operator $\Ket{\rho(t)}=\sum_{\ell=0}^{\infty}\Ket{\rho^{(\ell)}(t)}$ then follows from the master equation~\eq{eq_master}, including the instantaneous stationary state $\Ket{\rho^{(0)}(t)} \equiv\Ket{z}$ with respect to the modulated system parameters at some time $t$, as well as the corrections $\Ket{\rho^{(\ell)}(t)}$ in $\ell$-th order in $\Omega/\Gamma$. 

Note that due to charge- and spin-conserving tunneling, the occupation probabilities $P_i(t) = \bra{i}\rho(t)\ket{i}$, namely the diagonal elements of the reduced density matrix in the basis of dot energy eigenstates, decouple from the coherences, namely the off-diagonal elements $\bra{i}\rho(t)\ket{j}$. Since the pumped charge and energy ---as the here relevant observables--- are also diagonal in the energy eigenbasis, the coherent dynamics is completely discarded, and we only determine the time-dependent probabilities to obtain the relevant transport quantities.
The required matrix representation of the kernel $W$ in the remaining probability subspace consists of transition rates between the dot energy eigenstates of zero $[\Ket{0} = \ket{0}\bra{0}]$, single~\footnote{The spin degree of freedom does not play any role in this work and we therefore do not separately treat the different spin state occupations in the singly-occupied mixed state.} $[\Ket{1} = \frac{1}{2}\sum_{i=\uparrow,\downarrow}\ket{i}\bra{i}]$, and double $[\Ket{2} = \ket{2}\bra{2}]$ occupation. The rates can be calculated using the lead decomposition $W=\sum_\alpha \Wa$ and Fermi's golden rule for each lead-resolved kernel $\Wa$ separately. They are given by\footnote{The basis is trace-normed, but not \emph{orthonormal}, since $\Braket{1}{1} = 1/2$. Therefore, $W = \sum_{ij}\frac{\Bra{i}W\Ket{j}}{\Braket{i}{i}\Braket{j}{j}} \Ket{i}\Bra{j} $, see Suppl. of \cite{Schulenborg2016Feb}.}
\begin{eqnarray}
    \Wa_{10} = 2\Gama f^+_\alpha(\epsilon) & \ \ \ \ \ \ \  & \Wa_{21} = \Gama f^+_\alpha(\epsilon+U)\notag\\
    \Wa_{01} = \Gama f^-_\alpha(\epsilon) & & \Wa_{12} =2 \Gama f^-_\alpha(\epsilon+U). \label{Wa_elts}
\end{eqnarray}
The diagonal elements of the matrices representing $\Wa$ and, hence, $W$ are obtained from these rates by total probability conservation, dictating $\Bra{\one}\Wa = \Bra{\one}W = 0$ with identity operator $\hat\one$. The instantaneous time-dependence of $W$ comes from the driving parameters [\Eq{Ri}] entering the expressions of the tunneling rates and Fermi functions in \Eq{Wa_elts}.

While the solution for the occupation probabilities as well as for the transport quantities can in principle be straightforwardly carried out, such a straightforward approach typically yields rather long, inaccessible expressions that offer little clues towards further, parameter-regime specific simplifications and approximations. An improved understanding based on more compact and insightful analytical results---which are until now not available---can, however, be obtained from a recently discovered fermionic duality relation. This has proven to be particularly useful in understanding dot systems driven out of equilibrium~\cite{Schulenborg2016Feb,Vanherck2017Mar,Schulenborg2018Dec,Ortmanns2022}, and observables influenced by strong many-body effects, such as the energy being affected by a large Coulomb interaction \mbox{$|U|/\TN \gg 1$}~\cite{Schulenborg2017Dec,Schulenborg2020Jun}.

Specifically, this dissipative symmetry for the kernel $W$ of the master equation---the fermionic duality---connects the hermitian conjugate of $W$ to the kernel of a \textit{dual} model, in which all energies are inverted, see \App{app_duality}.
This allows to straightforwardly write the kernel in its eigenbasis~\cite{Schulenborg2017Dec,Schulenborg2018Dec}
\begin{align}
    W=-\gamp\KetBra{p}{p'}-\gamc\KetBra{c}{c'},\label{eq_kernel}
\end{align}
which can be interpreted in a meaningful way. The eigenvalues of the  kernel are the negative of the relaxation rates, which one identifies as the charge relaxation rate $\gamc$, the parity rate $\gamp$ and the additional eigenvalue 0, which corresponds to the stationary state. The fermionic duality cross-relates the relaxation rates to each other. In particular, the parity rate $\gamp=2\Gamma$ connects to the zero-eigenvalue of the kernel through the duality-based decay rate relation between any two eigenmodes $x,y$,
\begin{equation}
    \gamma_y=2\Gamma-\gamma^\inv_x. \label{eq_rate_duality}    
\end{equation}
The superscript ``$\inv$''  indicates the dual model with \emph{inverted} energies
\begin{equation}
    \epsilon^\inv=-\epsilon\quad,\quad U^\inv=-U\quad,\quad\mu^\inv_\alpha=-\mua.
\end{equation} 
Temperature and coupling rate $\Gamma$ are not inverted by the dual transform.
The charge relaxation rate $\gamc=f^+_\epsilon + f^-_U$, where we use the compact notation for combinations of Fermi functions $f^\pm_\epsilon=\sum_\alpha\Gama f^\pm_\alpha(\epsilon)$ and $f^\pm_U=\sum_\alpha\Gama f^\pm_\alpha(\epsilon+U)$, is self-dual following the relation of Eq.~(\ref{eq_rate_duality}).

In a similar way, the right and left eigenvectors of the kernel are interconnected 
through the duality-based cross-relation
\begin{equation}
    \Bra{y'}=\left[\left(-\hat\one\right)^{\hat{N}}\Ket{x^\inv}\right]^\dagger=\left[\mathrm{e}^{i\pi\hat{N}}\Ket{x^\inv}\right]^\dagger.\label{eq_vector_duality}
\end{equation} 
This relation leads to the compact form of charge and parity decay as well as the stationary state $\Ket{z}$, given by 
\begin{gather}
    \Ket{z} = \frac{1}{\Gamma\gamc}\left[f_\epsilon^-f_U^-\Ket{0}+f^+_\epsilon f^-_U\Ket{1}+f^+_\epsilon f^+_U\Ket{2}\right]\label{eq_eigensystem_right}\\
    \Ket{p} = \Ket{(-\one)^{N}} \quad,\quad \Ket{c} = \frac{1}{2}(-\hat\one)^{\hat{N}}\left[\Ket{N}-\nzi\Ket{\one}\right],\notag
\end{gather}
and the corresponding left eigenvectors 
\begin{gather}
    \Bra{z'} = \Bra{\one}\quad,\quad\Bra{p'} = \Bra{\zin(-\one)^{N}}\notag\\
    \Bra{c'} = \Bra{N}-\nz\Bra{\one}\label{eq_eigensystem_left}.
\end{gather}
Equations~\eq{eq_eigensystem_right} and \eq{eq_eigensystem_left} contain the dot occupations
\begin{equation}
    \nz = \Braket{N}{z} \quad,\quad \nzi = \Braket{N}{\zin}\label{eq_occupations}
\end{equation}
with respect to the stationary state $\Ket{z}$ and with respect to the dual stationary state $\Ket{\zin}$, that is, the stationary state in the dual model with inverted energies, see \App{app_duality} for explicit analytical expressions.

Our following analysis of energy pumping benefits from this approach using the duality-based eigendecomposition of the kernel $W$ in \Eqs{eq_kernel}-\eq{eq_occupations}. The basic starting point is to write both the zeroth-order solution $\Ket{\rho^{(0)}}$ as well as the first-order correction in frequency $\Ket{\rho^{(1)}}$ in terms of the instantaneous eigenmodes and the corresponding decay rates:
\begin{equation}
    \Ket{\rho^{(0)}} = \Ket{z}\quad,\quad \Ket{\rho^{(1)}} = \frac{1}{\tW}\partial_t\Ket{z},\label{eq_first_correction}
\end{equation}
with the pseudo-inverse of the kernel
\begin{equation}
    \frac{1}{\tW}=-\frac{1}{\gamp}\KetBra{p}{p'}-\frac{1}{\gamc}\KetBra{c}{c'}. 
\end{equation}
This pseudo-inverse was constructed exploiting that it only ever acts on vectors $\Ket{x}$ orthogonal to the zero-eigenvalue subspace, i.e., with $\Braket{\one}{x} = 0$. Finally, for energy pumping in a refrigerator scheme as discussed in \Sec{sec_refrigerator}, we also require second-order corrections in frequency,
\begin{equation}
    \Ket{\rho^{(2)}} = \frac{1}{\tW}\partial_t\left[\frac{1}{\tW}\partial_t\Ket{z}\right].
\end{equation}
This correction constitutes a limitation for thermodynamically efficient pumping~\cite{Juergens2013Jun}.

\section{Transport equations}\label{sec_analytical_results}

\subsection{Transport of a generic observable due to time-dependent driving}\label{sec_Transport}

Based on the theoretical approach introduced in \Sec{sec_theory}, we now calculate the stationary current as well as the first-order in $\Omega/\Gamma$ correction leading to pumping, for an arbitrary observable. One can write the $\ell$-th order contribution to the current into contact $\alpha$ as
\begin{equation}
    \IOa^{(\ell)}=\Bra{\cO}\Wa\Ket{\rho^{(\ell)}},    \label{eq_obs_current}
\end{equation}
that is, in terms of a local dot observable $\hat{\cO}$, as long as this observable is conserved~\cite{Saptsov2014Jul}. This means that the flow of the expectation value of the observable $\hat{\cO}$ out of the local dot system equals the sum over the currents into all contacts. It is naturally fulfilled for the dot charge as local observable and the resulting charge currents into the contacts. For the energy current, Eq.~\eq{eq_obs_current} is only valid if no energy is stored on the tunneling barriers, that is, when $\langle \hat{H}_{\text{tun}}\rangle(t)$ is constant. The latter is true for the here considered, weakly coupled dot~\cite{Schulenborg2018} as long as non-electronic energy flow due to, e.g., dissipation to phonons is negligible. By contrast, in strong-coupling situations, the time-dependent storage of energy on the barriers can play an important role~\cite{Wu2008Nov,Ludovico2014Apr,Haughian2018Feb}.

The ingredients of \Eq{eq_obs_current} are known for stationary charge and energy currents ($\ell=0$) and have been analyzed exploiting the fermionic duality in detail~\cite{Schulenborg2017Dec}. Also, time-dependent charge- and energy currents through a quantum dot have been analyzed after a rapid switch in the parameters~\cite{Schulenborg2016Feb,Vanherck2017Mar,Schulenborg2018Dec}. By contrast, a detailed analysis of the pumping currents exploiting the fermionic duality has been missing.

Collecting the time-dependent parameters in the driving scheme $\bR = (R_1(t),R_2(t),0)^T$ as prescribed at the end of \Sec{sec_pumpingintro}, we can write the pumping current as~\cite{Calvo2012Dec}
\begin{equation}\label{current}
    \IOa^{(1)}  =  \Bra{\cO}\Wa\frac{1}{\tW}\nr\Ket{z}\cdot\partial_t\bR\,,
\end{equation}
with the gradient defined as $\nr = (\pr[1],\pr[2],0)^T$. The resulting transported observable per driving period equals a \textit{geometric phase}, as pointed out by Ning, Haken and Landsberg~\cite{Ning1992Apr,Landsberg1993Jan}, see also \Refs{Sinitsyn2009Apr,Pluecker2017Apr,Pluecker2017Nov}:
\begin{subequations}
    \begin{align}
        \Delta\cO_\alpha^{(1)} & = \int_0^{2\pi/\Omega}\dd t \, \IOa^{(1)}(t) =  \oint_\mathcal{C} \dd\bR\cdot  \AOa(\bR)\label{geometric phase A}\\
        & =  \iint_\cS \dd \textbf{S}\cdot \BOa(\bR)\ ,\label{geometric phase B}
    \end{align}
\end{subequations}
with the geometric connection
\begin{equation}
    \AOa(\bR)=\Bra{\cO}\Wa\frac{1}{\tW}\nr\Ket{z}\ ,\label{eq_obs_connection}
\end{equation}
and the pseudo-magnetic field, also called the pumping curvature,\footnote{By using more general differential forms than $\nr\times\bullet$, one can generalize Stokes' theorem (see for example Ref.~\cite{Arnold1978}, chapter 7) to connect \Eq{geometric phase A} with \Eq{B_O,alpha} for any discrete dimensional vector $\bR$, corresponding to driving arbitrarily many parameters simultaneously. Our treatment, however, focuses on the minimal and experimentally relevant situation of two-parameter driving, for which the driving cycle describes a path $\cal C$ enclosing the surface $\cS$ of a flat, two-dimensional plane in parameter space as shown in \Fig{fig_model}(b).}
\begin{eqnarray}
    \BOa(\bR) & = & \nr\times \AOa(\bR).\label{B_O,alpha}
\end{eqnarray}

The pumping curvatures $\BHa(\bR)$ and $\BNa(\bR)$ for the energy $(\hat{\cO}=\hat{H})$ and charge $(\hat{\cO}=\hat{N})$ will be analyzed in detail in the following \Sec{sec_results} for different sets of pumping parameters, using the insights from the fermionic duality relation. However, before addressing any specific transport observable, we note that the time-dependent driving of parameters can excite the quantum system away from the stationary state in two different ways. Namely,
\begin{align}
    \frac{1}{\tW}\nr\Ket{z} = \bxC \Ket{c} + \bxP\Ket{p}\label{eq_mode_decomp}
\end{align}
with a charge-type excitation and a parity-type excitation. Exploiting the duality-based kernel decomposition, the amplitudes of these excitations can be written as
\begin{subequations}\label{eq_x}
    \begin{align}
        \bxC &= -\frac{1}{\gamc}\Bra{N}\nr\Ket{z} = -\frac{1}{\gamc} \nr \nz,\label{x_c}\\
        \bxP &= -\frac{1}{\gamp}\Bra{\zin\fpOp}\nr\Ket{z} \label{x_p_overlap}\\
        &= -\frac{1}{4 \gamp}\nr p_z - \frac{1}{2\gamp}\left(\nzi - 1\right)\nr \nz,\label{x_p}
    \end{align}
\end{subequations}
with the stationary quantum dot parity $p_z = \Braket{\fpOp}{z}$, see \App{app_duality} for explicit analytical expressions. The function $\bxC$ for the charge-like excitation is expectedly proportional to the inverse of the charge relaxation rate $\gamc$, and the parity-like term $\bxP$ proportional to the inverse of the parity rate $\gamp$.
However, when expressing $\bxP$ in terms of observables as in \eq{x_p}, we find gradients of both the stationary parity $p_z$ and occupation number $\nz$, with the latter also entering via the dual occupation $\nzi$.
This observable decomposition in \eq{x_p}, as well as the vector-overlap form in  \Eq{x_p_overlap}, will offer valuable insights for the interpretation of energy pumping in the remainder of this paper.

Further exploiting the mode decomposition \eq{eq_mode_decomp}, we can split the geometric connection of a specific observable $\hat{\cO}$, transported into a specific contact $\alpha$,  $\AOa(\bR)$, into contributions
$\AOa(\bR) = \AOaC(\bR) + \AOaP(\bR)$ coming, respectively, from the charge- and parity-like system excitations due to the driving,
\begin{gather}
    \AOa^{\C/\P}(\bR) = \melt{\cO}{\Wa}{c/p} \bx^{\C/\P}.\label{decomposition_A_step}
\end{gather}
The crucial point is now that the eigenmode decomposition given in \Eqs{eq_kernel}-\eq{eq_occupations} for the \emph{full} kernel $W$ only derives from probability conservation, from the local equilibrium state in each individual lead $\alpha$, and from fermionic duality~\cite{Schulenborg2017Dec,Schulenborg2018Dec}. The decomposition can hence be carried out in exactly the same way for the lead-resolved kernel $\Wa$. This yields a lead-resolved stationary state $\Ket{\za}$, dual state $\Ket{\zia}$, and their associated lead-resolved dot occupation numbers 
\begin{equation}
    \nza = \Braket{N}{\za}\quad,\quad \nzia = \Braket{N}{\zia}.
\end{equation}
These quantities together determine the full set of lead-resolved eigenvectors of $\Wa$ as well as the corresponding charge rate $\gamca = \Gama [f^+_\alpha(\epsilon) + f^-_\alpha(\epsilon + U)]$ and parity rate $\gampa = 2\Gama$. In particular, we again obtain the parity mode $\Ket{p_\alpha} = \Ket{p} = \Ket{\fpOp}$ as a parameter-independent right eigenvector. When inserted together with \Eq{eq_x} into \Eq{decomposition_A_step}, this eigenvector yields
\begin{subequations}\label{decomposition_A}
    \begin{align}
        \AOaC(\bR) &= \aCNOa \nr \nz,\\
        \AOaP(\bR) &= \aPpOa \nr p_z + \aPNOa \nr \nz,
    \end{align}
\end{subequations}
where
\begin{subequations}\label{coeffs_decomposition_A}
    \begin{gather}
        \aCNOa = \frac{\gamca}{\gamc} \Braket{\cO}{c_\alpha} + \frac{\gampa}{\gamc}\Braket{p'_\alpha}{c} \Braket{\cO}{p},\label{coeffs_decomposition_A_C}\\
        \aPpOa =  \frac{\gampa}{4 \gamp}\Braket{\cO}{p}\;,\, \aPNOa = \frac{\gampa}{2\gamp}\left(\nzi - 1\right)\Braket{\cO}{p}.\label{coeffs_decomposition_A_P}
    \end{gather}
\end{subequations}
This set of equations readily shows the requirements that a certain observable must fulfill in order to make the different excited modes (charge-like and parity-like) visible. The second term in \Eq{coeffs_decomposition_A_C} and the two parity-like terms from \Eq{coeffs_decomposition_A_P} in particular only contribute if the observable $\hat{\cO}$ is sensitive to many-body effects. Geometric \emph{charge} pumping $(\hat{\cO} = \hat{N})$ in the here studied quantum dot system is therefore not sensitive to the parity mode excitation, as $\Braket{N}{p} = \Braket{N}{\fpOp} = 0$ leaves only the first term in \Eq{coeffs_decomposition_A_C} to contribute. The consequence is that the charge pumping current is directly proportional to the inverse of the charge relaxation rate~\cite{Splettstoesser2008May,Reckermann2010Jun}. Moreover, since a finite bias voltage or temperature difference generally causes the lead-resolved eigenvectors of $\Wa$ to differ from those of the full kernel, $\Bra{p_\alpha'}\neq\Bra{p'}$ and $\Ket{c_\alpha}\neq\Ket{c}$, the ratio between charge-like and parity-like contributions generally also depends on the contact in which the transported observable is detected.

To obtain compact and insightful expressions for the pumped transport variables (such as the pumped charge and energy), we split the pumping curvature in a similar way, $\BOa(\bR) = \BOaC(\bR) + \BOaP(\bR)$, with $\BOa^{\C/\P} (\bR) = \nr \times  \AOa^{\C/\P} (\bR) $.
The pumping curvature then generally reads as
\begin{align}
    \BOaC(\bR) &=  \nr \aCNOa \times\nr \nz,\label{decomposition_B}\\
    \BOaP(\bR) &= \nr \aPpOa \times\nr p_z  + \nr \aPNOa \times\nr \nz.\notag
\end{align}
Equation \eq{decomposition_B} clarifies that a finite pumping current not only needs a finite parameter gradient of the stationary dot occupation and/or parity but in particular a gradient component orthogonal to $\nr\aOa$~\cite{Calvo2012Dec,Placke2018Aug}.

The analysis of charge and energy pumping put forward in this paper mainly relies on the functional form of \Eqs{decomposition_A} to \eq{decomposition_B} in order to understand the different pumping mechanisms and pumping schemes.

\subsection{Energy pumping}\label{sec_analytical_energy}

To support our detailed pumping analysis for various parameter drivings and external conditions in \Sec{sec_results}, we employ the fermionic-duality-based approach to highlight general analytical features relevant for energy pumping.

\subsubsection{Geometric connection}\label{sec_geometric_energy}

We begin with analytical results for the geometric connection of energy pumping and their general physical implications, in particular in comparison to charge pumping. Equations \eq{decomposition_A} and \eq{coeffs_decomposition_A} yield
\begin{subequations}\label{aH}
    \begin{align}
        \aCNHa &=\aCTCHa + \aCNTCHa,\label{aH_CN}\\
        \!\aPpHa &= \frac{U\Gama}{4\Gamma},\label{aH_Pp}\\
        \!\aPNHa &= \frac{U\Gama}{2\Gamma}(\nzi - 1),\label{aH_PN}
    \end{align}
\end{subequations}
where we have denoted
\begin{subequations}
    \begin{align}
        \aCTCHa &= \cEa \frac{\gamca}{\gamc},\label{aH_CN TC} \\
        \aCNTCHa &= \frac{U}{2}\frac{\gampa}{\gamc}(\nzia - \nzi)\label{aH_CN NTC}.
    \end{align}
\end{subequations}
The term $\aCTCHa$ includes the characteristic energy 
\begin{equation}
    \cEa = \epsilon + \frac{U}{2}(2 - \nzia),\label{eq_tight_coupling_energy}
\end{equation} 
as a function of the above introduced, lead-resolved dual occupation number $\nzia$.

To interpret \Eq{aH}, we first note that charge pumping only senses the charge-like excitation $(\ANa \sim \bxC)$:
\begin{align}\label{aN}
    \ANa(\bR) = \aCNNa\nr N_z \quad,\quad \aCNNa &= \gamca/\gamc.
\end{align}
Based on the expression of $\aCNNa$, we identify $\aCTCHa$ as the \textit{tight-coupling} contribution to energy pumping. This term is ``tightly coupled'' to the charge current through multiplication by the characteristic energy $\cEa$, which in fact equals the (stationary) Seebeck coefficient of the quantum dot~\cite{Schulenborg2017Dec}. The other term in the charge-like contribution, $\aCNTCHa$, contributes only if $\nzia \neq \nzi$, i.e., when a stationary temperature difference or bias voltage is applied across the dot. This and all other contributions to energy pumping given in Eqs.~(\ref{aH}) are uniquely due to the onsite interaction $U$, and hence vanish in a non-interacting quantum dot.

To gain additional insight into the properties of the \emph{parity contribution} $\AHaP(\bR)$ to the geometric connection, let us return to its original definition in \Eq{decomposition_A_step};
\begin{align}
    \AHaP(\bR) &= \melt{H}{\Wa}{p}\bxP = -U\gampa\bxP\notag\\
    &= \frac{U\gampa}{\gamp}\Bra{\zin\fpOp}\nr\Ket{z}.\label{eq_A_p_overlap}
\end{align}
We first note that this contribution fulfills a symmetry under the dual transform, namely under the sign inversion of all energies. Concretely, using the product rule of $\nr$ together with the eigenvector orthogonality $\Braket{\zin\fpOp}{z} = \Braket{z\fpOp}{\zin} = 0$, hermiticity of the states $z = z^\dagger$ and $\zin = (\zin)^\dagger$, and parity superselection dictating $[z,\fpOp] = [\zin,\fpOp] = 0$, we find $\bxP$ to be antisymmetric with respect to the transformation to the dual model:\footnote{This assumes an energy inversion \emph{prior} to taking the gradient $\nr$, the latter not commuting with energy inversion.}
\begin{align}
    \bx^{\P,\inv} &= -\frac{1}{\gamp}\Bra{z\fpOp}\nr\Ket{\zin}\notag\\
    &= +\frac{1}{\gamp}\Bra{\zin\fpOp}\nr\Ket{z} = -\bxP.\label{x_p_anti}
\end{align}
This and the sign inversion of $U$ under the dual transform means that $\AHaP(\bR) = -U\gampa\bxP$ is self-dual, i.e., identical when transforming to the dual model. In \Sec{sec_results_attractive}, we further translate this to a symmetry of the parity-like energy pumping curvature $\BOaP(\bR)$ for repulsive vs. attractive onsite interaction of equal strength. 

Secondly, we can derive from the general form of Eq.~(\ref{eq_A_p_overlap}) and the behavior of $\Ket{z}$ and $\Ket{\zin}$, that $\AHaP(\bR)$ is non-zero only for specific parameter regions, meaning that the geometric connection for energy pumping is in many cases dominated by the charge-mode contribution $\AHaC(\bR)$.
Let us consider a parameter regime, where both dot transition energies $\epsilon$ and $\epsilon + U$ lie outside the bias window, $|\epsilon|,|\epsilon + U| > |\Vb|/2$. The dot state is then either in the empty or the doubly occupied state, but never in the single occupied state. The dual state $\Ket{\zin}$ describes, by its inverted nature, a nearly opposite occupation compared to $\Ket{z}$, i.e., an empty dual state $\nzi = 0$ for a fully occupied stationary dot state $\nz = 2$ and vice versa.
A similar situation occurs for strong interaction: $|U|>|\Vb|$ and $|U| \gg \Ta$. Assuming a repulsive onsite interaction $U > 0$ on the dot, the fictitious dual model with inverted energies (characterizing $\Ket{\zin}$) has a stationary state that is characterized by an attractive interaction, $U^\text{i}=-U$, and is hence dominated by electron pairing~\cite{Schulenborg2017Dec,Schulenborg2018Dec}. Therefore the dual model exhibits a single, sharp two-particle transition at $\epsilon + U/2 = 0$. This means that the dual steady-state probability for single occupation is strongly suppressed.\footnote{An equivalent statement can be made for a quantum dot model with attractive interaction $U < 0$. In this case, it is the stationary state of the actual dot model $\Ket{z}$ that is always either given by the empty or the double-occupied state while the probability of single occupation is strongly suppressed.}
For both of these situations, the overlap in \Eq{eq_A_p_overlap} can then, independently of the sign of the interaction $U$, be well approximated by
\begin{equation}
    \AHaP(\bR) \rightarrow \frac{U\gampa}{\gamp}\left[P^{\inv}_{0}\nr P_{0} + P^{\inv}_{2}\nr P_{2}\right],\label{eq_p_overlap}
\end{equation}
where $P_{0} = \Braket{0}{z},P_{2} = \Braket{2}{z}$ are the probabilities for the dot to be empty $(0)$ or doubly occupied $(2)$ in the stationary state $\Ket{z}$, and $P^{\inv}_{0} = \Braket{0}{\zin}, P^{\inv}_{2} = \Braket{2}{\zin}$ are the corresponding duals in the state $\Ket{\zin}$. The crucial point is now that for $|\Vb| < |U|$ and $|U| \gg \Ta$, or if both $\epsilon$ and $\epsilon + U$ are outside the bias window, $P^{\inv}_{0}$ and $P^{\inv}_{2}$ are only sizable for parameters in which $P_{0}$ and $P_{2}$ are stably suppressed, meaning $P^{\inv}_{0}\nr P_{0},P^{\inv}_{2}\nr P_{2} \rightarrow 0$. Equation \eq{eq_p_overlap} thereby implies that the parity contribution $\AHaP(\bR)$ to pumping of energy and, in fact, the parity contribution $\AOaP(\bR)$ to pumping of any observable becomes negligible in all the above mentioned parameter regimes.

The key physical insight derived from \Eq{eq_p_overlap} is that unless \emph{both} dot resonances $\epsilon$, $\epsilon + U$ are close to or within a bias window $|\Vb| > |U|$, the time-dependent energy current due to the slow driving, as represented by the geometric connection $\AHa(\bR)$, is well approximated by its charge-mode contribution,
\begin{equation}
    \AHa(\bR) \approx \aCNHa\nr \nz =\AHaC(\bR).\label{eq_pure_charge_geometric}
\end{equation}
In other words, the corrections at the first order in $\Omega/\Gamma$ to the time-dependent energy and charge current become proportional, $\IHa^{(1)}(t) \approx C(t)\INa^{(1)}(t)$. The factor $C(t) = \aCNHa$ only depends on the momentary system parameter configuration, but not on the time-derivative of that configuration as determined externally via the driving frequency $\Omega$.
The energy pumping curvature in this regime is accordingly governed by the charge component, $\BHa(\bR) \rightarrow \BHaC(\bR)$. Since $\aCNHa$ consists only of the  Seebeck coefficient $\cEa$ and terms proportional to dual occupation numbers $\nzi$, $\nzia$, our detailed energy pumping analysis in \Sec{sec_results} can, in many parts, simply refer to the intuitive and well-known physics governing these quantities.

Let us point out that it is rather surprising at first glance that the pumped energy through a system with  comparably large interaction strength $|U|$ is due to the charge-mode excitation only, in extended parameter regions. Indeed, in contrast to the charge current, the energy current is \emph{directly sensitive} to many-body effects via the Coulomb energy that can be transferred through the quantum dot even in the absence of net charge current. The suppression of parity-like terms can nevertheless be understood when distinguishing close-to-steady-state from excited-state dot dynamics. As clarified in \Ref{Schulenborg2018Dec}, a maximally excited---or unstable---quantum-dot state with respect to the environment is in fact closely approximated by the dual, inverted state $\Ket{\zin}$ if $|U| \gg |\Vb|,\Ta$ or if $|\epsilon|$, $|\epsilon + U| \gg |\Vb|/2$, $\Ta$. For example, a doubly occupied dot would be maximally unstable if the stationary dot state is close to an empty state. The overlap \eq{eq_A_p_overlap} thus expresses that (as long as the notion of a maximally unstable state is meaningful) the parity mode only enters for excitations close to maximal instability. Such maximally unstable states can be created, e.g., with fast level switches as studied in Refs.~\cite{Schulenborg2016Feb,Vanherck2017Mar,Schulenborg2018Dec}. The key difference of adiabatic pumping with respect to the fast-switching case is that slow driving alone cannot induce such a strong excitation away from the steady state. The suppressed parity-like contribution to geometric energy pumping simply reflects this fact.

The analysis changes however for sufficiently large bias voltage, $|\Vb| > |U|$, if $\epsilon$ and $\epsilon + U$ lie either inside or at an edge of the bias window. In this case, the overlap between stationary and inverted stationary state is in general finite.  
However, a pumped energy current deviating from a pure charge-mode contribution for $|U| \gg T$ requires not only large biases $|\Vb| \geq |U|$, but also at least $\epsilon$ or $\epsilon + U$ to be close to a resonance with one lead potential, $\muL$ or $\muR$, since otherwise the gradient $\nr\Ket{z}$ vanishes.\footnote{If both transition energies are well inside the bias window, the small coupling asymmetry $\Lambda \approx 0$ considered in \Sec{sec_results} implies that all possible dot states become equally stable or unstable in the stationary limit of this strongly bias driven configuration, $\Ket{z} \rightarrow \Ket{\one}/4$. Then, $\Ket{\zin} \rightarrow \Ket{z}$, but since $\Ket{z} \rightarrow \Ket{\one}/4$ results in a vanishing gradient $\nr\Ket{z} \rightarrow 0 $, the overlap \eq{eq_A_p_overlap} still approximately vanishes.} If this resonance condition is fulfilled any driving affecting it can then in principle also excite parity-like terms, as further illustrated in \Sec{sec_results_repulsive}. In this case, the above pointed out proportionality $\IHa^{(1)}(t) \approx C(t)\INa^{(1)}(t)$ between the first-order corrections of charge and energy current breaks down. This physically means either that $\IHa^{(1)}(t)$ can be finite while $\INa^{(1)}(t) = 0$, or that $C(t)$ would not anymore be determined by the system properties alone, but also by the driving speed set by the frequency $\Omega$ and amplitudes $\delta R_i$ in \Eq{Ri}.

\subsubsection{Pumping mechanisms}\label{sec_mechanisms}

\begin{figure*}
    \includegraphics[width=\linewidth]{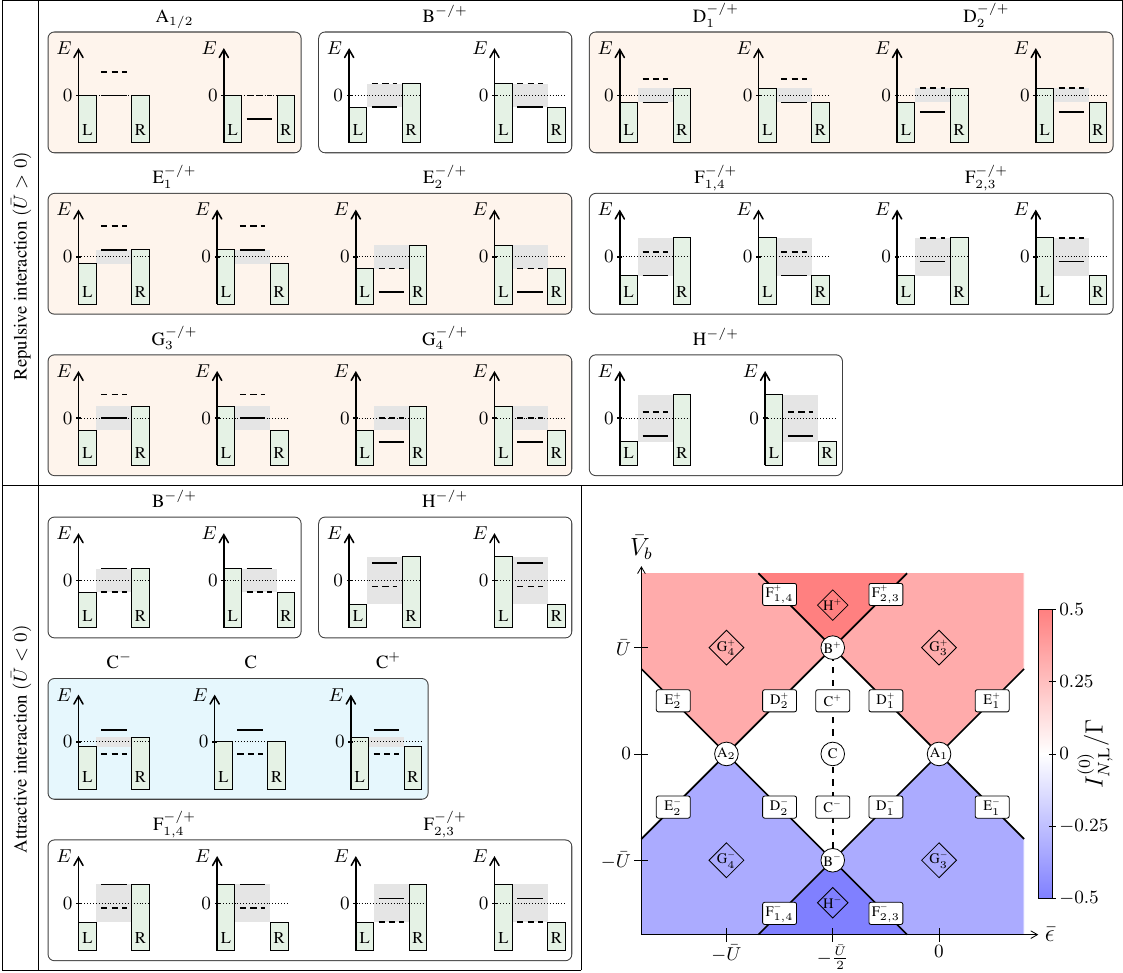}
    \captionof{figure}{
        \label{fig_pumping mechanisms}
        Electrochemical potential configurations corresponding to the discussed pumping mechanisms for a repulsive interaction (top rows) and an attractive one (bottom rows). The solid black line indicates the transition energy $\epsilon$ while the dashed one corresponds to $\epsilon + U$. The mechanisms specific to the repulsive case are highlighted in light orange while the one specific to the attractive case is highlighted in light blue. The mechanisms are also indicated on a map of the parameter space $(\beps, \bVb)$ in the bottom-right corner, with circles for points at double resonances, the rectangles for lines of single resonances and diamonds for surfaces, namely regions in the absence of resonance. The colormap indicates the stationary current from the left reservoir into the quantum dot,  $\INL^{(0)}$. For simplicity, the sketch is done for zero-temperature reservoirs.}
    \vskip0.5cm
    \setlength{\tabcolsep}{4pt}
    \renewcommand{\arraystretch}{1.1}
    \begin{tabularx}{\linewidth}{cX}
        \hline\hline
        Mechanism & Description\\
        \hline
        A & Double resonance at zero bias voltage\\
        B & Double resonance at $|\bVb| = |\bU|$\\
        C & Particle-hole symmetric point, namely $\beps=-\bU/2$\\
        D & Single resonance on the side of the Coulomb diamond (bias window in between the two transition energies)\\
        E & Single resonance while the bias window is not in between the transition energies\\
        F & Single resonance while the other transition energy lies within the bias window\\
        G & Single transition energy in the bias window\\
        H & Both transition energies in the bias window\\
        \hline\hline
    \end{tabularx}
    \captionof{table}{\label{tab_mechanisms}
        Description of the different pumping mechanisms.
    }
\end{figure*}

To facilitate the detailed discussion of the features of energy pumping in \Sec{sec_results}, it is helpful to label interesting areas of the parameter space. We call a \textit{pumping mechanism}~\cite{Placke2018Aug} a quantum dot configuration\footnote{For notational simplicity, we will here and in later occasions drop the units of parameters occurring in $\bR$ and $\bM$, as introduced in \Sec{sec_pumpingintro}.} $\bM = (\beps,\,\bU,\, \bVb,\, \bLam,\, \bTL)$ to which a driving scheme is applied, such that it leads to charge or energy transport. In the following, we focus only on the impact of the three first parameters, $\beps,\,\bU,$ and $\bVb$, while $\bLam = 0$ and $\bTL = \TN$.  We classify the different mechanisms leading to energy pumping for both a repulsive interaction ($\bU >0$) and an attractive one ($\bU < 0$). We sketch the electrochemical potential configurations corresponding to these mechanisms and we indicate their locations in the $(\beps, \bVb)$-space in \Fig{fig_pumping mechanisms}. Extending the notations used in Ref.~\cite{Placke2018Aug} for charge pumping, we have named the mechanisms X$^\pm_s$. Here, X is a letter between A and H whose meaning is given in Table \ref{tab_mechanisms}. The superscript $\pm$ corresponds to the sign of the bias voltage $\bVb$. Finally, the subscript $s$ takes the value $s=1$ if the transition energy $\epsilon$ is resonant with the Fermi level of one of the reservoirs, and $s=2$ if the transition energy  $\epsilon + U$ is resonant. The subscripts $s=3$ and $s=4$ indicate that a single transition energy, $\epsilon$ for $s=3$ and $\epsilon + U$ for $s=4$, lies within the bias window. 

As will be discussed in more detail in \Sec{sec_results}, the mechanisms A, D, E and G are specific to the repulsive case while C is present only in the attractive one. B, F and H are common to both cases. Note that B and F are the two mechanisms involving parity-like pumping contributions.

\subsubsection{Symmetry of pumping curvatures}\label{sec_symmetry}

The pumping curvatures plotted in \Figs{fig_pumping_curvature_repulsive_R} as well as \ref{fig_pumping curvature attractive R}, and discussed in full detail in \Sec{sec_results}, exhibit several interesting symmetries under particular changes of the working point parameters $\beps$, $\bVb$, $\bU$. This subsection provides the key analytical results underlying these symmetries, based on a rigorous derivation in \App{app_symmetry_ph}.

Depending on the driving scheme, the pumping curvature in \Figs{fig_pumping_curvature_repulsive_R} and \ref{fig_pumping curvature attractive R} is either symmetric or anti-symmetric under the parameter transform $\cT:(\epsilon, U, \Vb, \Ta, \Lambda) \mapsto (-\epsilon - U, +U, -\Vb, +\Ta, +\Lambda)$. This parameter transform can be shown to correspond to a particle-hole transform. Since the quantum-dot system we consider here is particle-hole symmetric, also the pumped observables $\Delta N_\alpha$ and $\Delta H_\alpha$, which are obtained from a surface integral over the respective curvature, are symmetric under a particle-hole transform. 

We start by addressing driving schemes with time-independent interaction $U$ and driving of any two of the parameters $(\epsilon, \Vb, \Ta, \Lambda)$. The transform $\cT$ leaves the area enclosed by the cycle invariant, but may affect the cycle orientation. This goes along with sign changes of the curvature that are found to be [\App{app_symmetry_ph}], 
\begin{align}
    \cT\BHa(\bR) = \sigma(R_1)\sigma(R_2)\BHa(\bR),\label{B_sym}
\end{align}
with the introduced signs
\begin{equation}\label{def_sigma}
    \sigma(R_i) = \left\{\begin{array}{ll}
        -1 &\text{if }R_i\in \{\epsilon, \Vb\}\\
        +1 &\text{if }R_i \in \{\TL, \Lambda\}
    \end{array}\right. .
\end{equation}

By contrast, when driving $U(t)$, the particle-hole transformed driving cycle always involves an effectively time-dependent $\epsilon$, $\cT \epsilon = -\epsilon - U(t)$, regardless of which second driving parameter $R_2$ is chosen next to $U$. In this case, $\cT$ bends the driving surface out of the $(U,R_2)$-plane at fixed $\epsilon$, generally affecting both cycle orientation and enclosed area. Remarkably, though, we show explicitly in \App{app_symmetry_ph} that the additional $\epsilon$-driving can still be straightforwardly accounted for by modifying \Eq{B_sym} to\footnote{For notational simplicity, we will here and in later occasions drop the third component of $\mathbf{R}$, which is always zero.}
\begin{align}\label{B_eps_B_U}
    \cT\BHa(\{U, R_2\})         &= \sigma(R_2)\left[\BHa(\{U, R_2\})\right.\notag\\
    &\phantom{=\sigma(R_2)\left[\right.}\left.- \BHa(\{\epsilon, R_2\})\right]\ .
\end{align} 
In the special case in which we drive $R_1 = \epsilon$ and $R_2 = U$, this results in
\begin{equation}\label{B_eps_U_sym}
    \cT\BHa(\{\epsilon, U\}) = -\BHa(\{\epsilon, U\}).
\end{equation}
The above argument also straightforwardly extends to charge pumping. The only difference is that unlike for the energy observable $\hat{H}$, the transform $\cT$ has no effect on the dot occupation operator $\hat{N}$ itself. As a result, the charge pumping curvature relations analogous to \Eqs{B_sym}, \eq{B_eps_U_sym} and \eq{B_eps_B_U} all have an additional minus sign on their respective right hand sides, see \App{app_symmetry_ph}.

Finally, it is interesting to note that while we have physically motivated \Eqs{B_sym}, \eq{B_eps_U_sym} and \eq{B_eps_B_U} for the full curvatures, we also find these relations to hold separately for the charge and parity components, $\BHa^{\P/\C}(\bR)$. This is because up to an overall sign, $\cT$ also acts as a particle-hole transform on each left and right eigenvector of the kernels $W, \Wa$ individually.

\section{Characteristics of energy pumping}\label{sec_results}

We now proceed with a detailed analysis of energy pumping based on the analytical results obtained using fermionic duality in the previous sections. We show plots of the energy pumping curvature $\BHR(\bR)$ of the right contact for all different driving schemes in \Figs{fig_pumping_curvature_repulsive_R} and \ref{fig_pumping curvature attractive R}. Figures \ref{fig_charge_pumping_curvature_repulsive_R} and \ref{fig_charge_pumping_curvature_attractive_R} in \App{app_charge} also display the corresponding charge pumping curvatures $\BNR(\bR)$ for relevant comparisons between charge and energy pumping. These pumping curvatures are calculated by setting $\alpha = \R$ and  $\hat\cO = \hat N, \hat H$ in \Eq{decomposition_B}, namely
\begin{equation}
    \BNR(\bR) = \BNR^\C(\bR) =\nr\frac{\gamcR}{\gamc} \times \nr \nz, \label{BNR}
\end{equation}
and
\begin{align}
    \BHRC(\bR) &=  \nr \aCNHR \times\nr \nz,\label{BHR}\\
    \BHRP(\bR) &= \nr \aPpHR \times\nr p_z  + \nr \aPNHR \times\nr \nz\notag,
\end{align}
where the coefficients $a$ are given by \Eqs{aH}.

Furthermore, for sufficiently small driving amplitudes, the geometric phase can be approximated by the bilinear response $\Delta \cO_\alpha^{(1)} \approx \BOa({\bR})\cdot \mathbf{\dS}$, where the vector $\mathbf{\dS}$ is orthogonal to the surface $\cS$ and has the encircled area of $\cS$ as its norm [\Fig{fig_model}(b)]. Therefore, in this limit, the amount of pumped energy or charge per driving cycle is directly proportional to the pumping curvature plotted in the figures.

\subsection{Quantum dot with repulsive onsite interaction}
\label{sec_results_repulsive}

The energy pumped through the quantum dot per driving cycle shows a very rich behavior as a function of the driving parameters, as well as of the constant working point parameters. In \Fig{fig_pumping_curvature_repulsive_R}, we show the pumping curvature for the energy pumped through the quantum dot with \textit{repulsive} onsite interaction for all possible choices of pairs of driving parameters (indicated by $\mathbf{R}$ on top of each panel). All curvatures are plotted as functions of the working point dot level position $\beps$ and bias voltage $\bVb$. This choice of representation is motivated by the fact that it allows all features to be compared to the well-known Coulomb diamonds characterizing the steady-state transport through interacting quantum dots.
We fix the average temperature to be equal in both contacts $\bTL = \bTR \equiv \TN$ and much smaller than the average Coulomb interaction $\bU=30\TN$, a regime where Coulomb interaction effects are dominant and clearly visible. For simplicity, the working point coupling asymmetry is furthermore chosen to be zero, $\bLam=0$.

\subsubsection{Specific features of energy pumping and comparison to charge pumping}\label{sec_features_repulsive}

As discussed on a formal level in \Sec{sec_symmetry}, we observe that the energy pumping curvature presented in all panels is particle-hole symmetric or antisymmetric depending on the driving scheme, except if $U$ is one of the driving parameters. Furthermore, in comparison with charge pumping, see \App{app_charge}, we identify a number of new pumping mechanisms, see also \Sec{sec_mechanisms}, with features specific to energy pumping. Generally, we find features at \emph{double-resonances}, in particular at those crossing points of the Coulomb diamonds, which we classify as mechanisms B, at \emph{single-resonances}, namely the lines confining the Coulomb diamonds, which are due to mechanisms D, E, and F, as well as \emph{non-resonant} features, appearing in the surfaces outside the Coulomb diamonds, classified as mechanisms G and H.

We start by discussing the features that occur when the coupling asymmetry is kept constant $\Lambda=\bar{\Lambda}$, beginning specifically at zero bias voltage, that is with mechanism A. This mechanism, at a double resonance with an equilibrium environment, is one of the mechanisms in charge pumping with the largest magnitude [see \Refs{Reckermann2010Jun,Calvo2012Dec} and \App{app_charge}]. By contrast, the pumped energy precisely at these points is vanishingly small. One rather finds that any feature occurring in the vicinity of $\bVb/\TN=0$ is not due to a unique double-resonance mechanism, but rather occurs as part of the lines associated with mechanisms D and E, which require a single resonance, only, as further discussed below.
The reason for this is the mostly tightly-coupled energy flow
, see \Sec{sec_geometric_energy} and in particular \Eq{aH_CN},
\begin{equation}
    \AHa(\bR) \approx (\cEa \gamca/\gamc)\nr \nz, \label{AH tight-coupling}
\end{equation}
which yields the characteristic energy $\cEa$ related to the Seebeck coefficient [\Eq{eq_tight_coupling_energy}] as the dominant contribution to the energy current in a near-equilibrium environment. Since this energy coefficient is suppressed close to the resonances, so is the energy effectively carried by the pumped charge.

The situation is reversed for mechanism B, corresponding to a double resonance with a large bias $|\Vb| \approx |U|$, where not only the charge mode but also the parity mode starts to contribute to energy transport, see \Sec{sec_geometric_energy}. At this double resonance, driving any parameter other than $\Lambda$  leads to pumping mostly by changing the balance between transport at one resonance with energy $\epsilon$ vs. the other one with energy $\epsilon + U$. This has a negligible effect on the net charge transport as long as the coupling is symmetric, but it does affect the transported energy stemming from two resonances, with equally important charge-like and parity-like contributions, see also \Sec{sec_parity_repulsive}.

\begin{figure*}
    \includegraphics{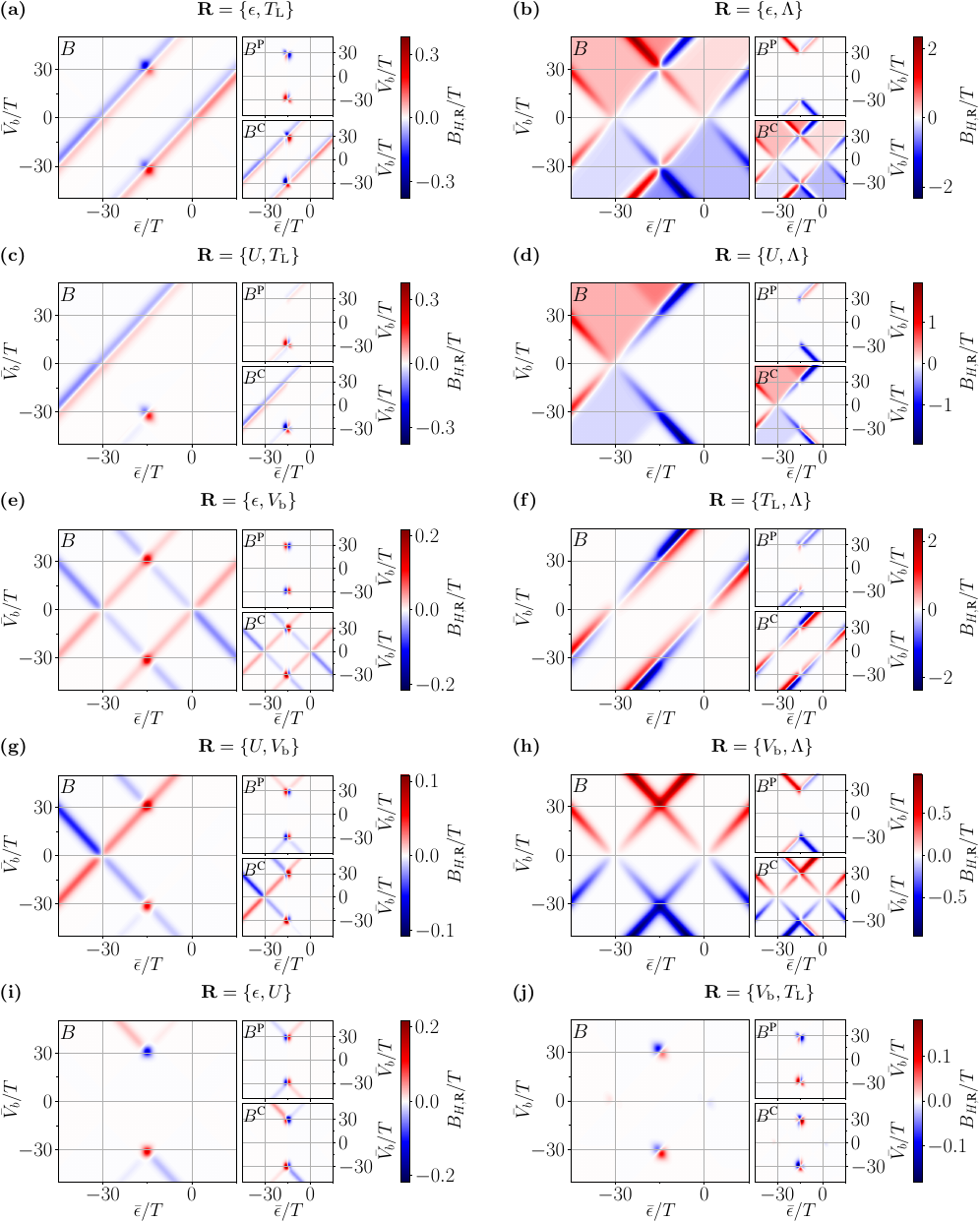}
    \caption{\label{fig_pumping_curvature_repulsive_R} 
        Pumping curvatures $\BHR$ of the repulsive system ($\bU > 0$) in the $(\beps, \bVb)$-space for all possible two-parameter driving schemes $\bR = \{R_1, R_2\}$. We have plotted the relevant component of $\BHR$, which is the one perpendicular to the driving plane defined by $(R_1, R_2)$, in units of $\TN$. In each subfigure, the largest panel corresponds to the total pumping curvature, the top-right one to the parity contribution $\BHRP$ and the bottom-right one to the charge contribution $\BHRC$.
        The driving is done around the working point $\bM = (\beps, \bU, \bVb, \bTL, \bLam)$ with $\bLam = 0$, $\bTL = \bTR = \TN$ and $\bU = 30 \TN$. 
        We have taken $\mu_{\L/\R} = \pm \Vb/2$, with the reference chemical potential ${\mu}_0$ set to zero. 
    }
\end{figure*}

As a next step, we analyze the most common single-resonance features in the pumped energy, namely the lines occurring due to mechanisms D, E, F. While strongly suppressed in charge pumping for any driving scheme apart from those involving $\Lambda$, energy pumping always exhibits at least some of those features except when only driving macroscopic variables $\TL$ and $\Vb$. To understand this, we note that for all these mechanisms with constant $\Lambda$, the rate of particle transport between the dot and the \emph{non-resonant} lead $\alpha'$ is fixed in strength and direction, i.e.  $\pr[i] \Wap \approx 0$, since the transport resonance lies either significantly below or above the respective chemical potential $\muap$ on the scale of the temperature. This results in the driving of only one \textit{effective} parameter associated with the near-resonant lead $\alpha$ around a steady-state value at the working point~\cite{Reckermann2010Jun}. The charge current therefore averages out over one cycle, see Appendix \ref{app_non_resonant} for an explicit derivation.

Conversely, the net pumped \emph{energy} for the single-resonance mechanisms D, E, F can still be finite when work done on the particles, via driven local dot energies $\epsilon$ and/or $U$, affects the transported energy time-dependently, see also \App{app_work}. This formally corresponds to the \emph{direct} $(\epsilon,U)$ dependence of the prefactors $a$ in the geometric connection for energy pumping [\Eqs{aH}], i.e., \emph{not} the one entering implicitly via the occupation numbers, which depend on time only at resonance. A nonzero cycle-averaged pumped energy moreover requires the energy during a particle transfer \emph{from} the resonant lead to differ from the energy for the corresponding reversed process \emph{to} the lead at the same rate. This asymmetry is established by the second driving parameter next to $\epsilon$ or $U$. We also show pumping curvatures for a non-interacting dot in \App{app_non_interacting}, providing even simpler examples of how work done on the electron in the dot in resonance with a single lead results in finite energy pumping in the absence of charge pumping. By contrast, since driving the lead parameters alone does not do any work \emph{on the transferred particles}, we accordingly find that $\BHa(\{\Vb,\TL\})$ vanishes for the single-resonance mechanisms. This is evident from \Fig{fig_pumping_curvature_repulsive_R}(j), and formally derived in \App{app_non_resonant}.

A particularly striking case in which work done on the dot electrons causes nonzero energy pumping curvatures, $\BHa(\bR) \neq 0$, in the absence of cycle-averaged charge pumping is the driving scheme involving only dot parameters, $\mathbf{R}=\{\epsilon,U\}$ [\Fig{fig_pumping_curvature_repulsive_R}(i)]. Here, mechanisms B and F yield finite energy pumping without charge pumping for the reasons explained above. Charge pumping, on the other hand, can in any case only occur for mechanisms A and B, at double resonances. With mechanism B already ruled out, we find that mechanism A also cannot contribute to the curvature $\BNa(\{\epsilon,U\})$ because the dot is coupled to a constant equilibrium environment, $\Vb = \TL - \TN = 0$, during the entire driving cycle. The two leads $\alpha = \text{L,R}$ individually act equivalently to the sum of leads, and the transport situation is fully symmetric. Hence, the cycle-averaged particle current vanishes, see \App{app_work}.

The comparison between charge and energy pumping changes when the couplings $\Gama$ are driven via their relative asymmetry $\Lambda$, as apparent in the explicit formulas given in Appendix \ref{app_non_resonant}, in particular \Eqs{eq_geometric_alt2} and \eq{eq_geometric_alt3}. First, the pumped charge for mechanisms D, E, F at a single resonance is generally finite, see \Fig{fig_charge_pumping_curvature_repulsive_R} in \App{app_charge}. The non-vanishing cycle average stems from the fact that the tunneling asymmetry $\Lambda(t)$  enters the time-dependent charge current nonlinearly, regardless of the resonance with the lead $\alpha$. This is because $\Lambda(t)$ influences the tunneling current both directly, via the tunneling rates themselves, and indirectly by affecting the relative influence of each lead on the dot state.

A second characteristic of $\Lambda$-driving which clearly distinguishes energy pumping from charge pumping is the surface features due to mechanisms G and H, appearing for $\epsilon$ or $U$ as second driving parameter [\Figs{fig_pumping_curvature_repulsive_R}(b,d)]. These plateaus correspond to a completely \emph{non-resonant} situation outside the Coulomb diamond, in which the kernels of both leads $\alpha = \text{L,R}$ and their derived quantities (apart from $\nzia$) are independent of any parameter except the couplings, i.e. $\pr[i] \Wa /\Gamma\approx \pr[i] W /\Gamma \approx \pr[i] \gamca  /\Gamma\approx\pr[i]\nza \approx 0$ for $R_i \neq \Lambda$. Since the charge pumping curvature [\Eq{BNR}] 
depends on the parameters exclusively through these quantities, driving $\Lambda$ and any other parameter $R_i \neq \Lambda$ effectively equals single-parameter geometric driving with $\Lambda$ only, which vanishes by definition. By contrast, a non-zero pumped energy is still possible because it directly depends on $\epsilon$ and $U$ via the tight-coupling energy $\cEa$ given in \Eq{eq_tight_coupling_energy}. The physical explanation again lies in the work done on the dot electrons. Namely, while a modulation of $\Lambda$ does not affect how much charge is transported across the dot in one cycle, a simultaneous modulation of $\Lambda$ and $\epsilon$ or $U$ still means that the energy taken out of one lead by a particle hopping to the dot is not the same as the energy which the same particle carries when hopping to the other lead.

Finally, let us end this subsection with a remarkable similarity between charge and energy pumping which occurs for temperature driving: compared to the case of constant temperature, all single- and double resonance features exhibit additional sign changes within the feature itself [\Figs{fig_charge_pumping_curvature_repulsive_R}(a,c,f,j) and \Figs{fig_pumping_curvature_repulsive_R}(a,c,f,j)], that is when the transition energy, $\beps$ or $\beps + \bU$, crosses the resonance with the left lead ($\bar{\mu}_\L = \bVb/2$). Indeed, a temperature change always affects the lead-electron occupation asymmetrically around the chemical potential: driving the temperature to, e.g., a slightly higher value, the chance of finding electrons at some energy $E - \muL > 0$ above the chemical potential rises, whereas the probability for electrons at $-(E - \muL) > 0$ below this potential lowers by the same amount. Consequently, the temperature-driving effect on electron transport to or from the lead inverts when crossing the resonance, and as such, it affects both the pumped charge and energy.

\subsubsection{Parity-like contributions to energy pumping}
\label{sec_parity_repulsive}
Our general analysis in \Sec{sec_geometric_energy} has already revealed, on a formal level, that for the here relevant case of strong repulsive interactions $U \gg \Ta$, parity-like excitations only enter at large bias $|\Vb| \geq U$ and with at least one of the two dot transition energies $\epsilon$ and $\epsilon + U$ being near-resonant with one of the lead potentials. The two smaller side panels of every subfigure of \Fig{fig_pumping_curvature_repulsive_R}, which show the charge-like $\BHRC(\bR)$ and parity-like $\BHRP(\bR)$ contributions to the pumping curvature separately, explicitly confirm this behavior. Namely, we observe parity-like contributions at \emph{point-like} features around double resonances at $\bVb = |\bU|$ (mechanism B) with $U$ and $\Lambda$ constant, or at \emph{line-like} features when $U$ or $\Lambda$ are among the driving parameters (mechanism F). In the following, we further elucidate these features.

As explained above, a constant $\Lambda$ with only a single resonant lead implies that any finite cycle-averaged energy transport due to the time-dependent driving must involve work done on the dot electrons. For constant $U$, this means that $\epsilon$ must be a driving parameter, and as such, this driving modifies both transition energies $\epsilon$ and $\epsilon + U$ equally. The energy transferred in addition to the steady-state energy current is hence tightly coupled to the charge transported back and forth between the dot and resonant lead, i.e., each particle on average carries the same amount of pumped energy. Consequently, the parity component to the pumping curvature disappears for mechanism F with constant $U$, $\Lambda$ [\Figs{fig_pumping_curvature_repulsive_R}(a,e,j)]. 

For mechanism B with the dot in resonance with \emph{both} leads, any two driving parameters generally affect the balance between electron transfer at energy $\epsilon$ and energy $\epsilon + U$.
The parity-like contribution to this energy pumping is then determined by how much pumped energy is transferred near a state of maximal instability (this is a well-defined concept in the vicinity of the special points $|\bVb| = |\bU|$); the dot state in fact rapidly switches between empty and doubly occupied when crossing the double resonance. In other words, the large bias voltage provides a finite probability even for slow driving to excite transport near maximal instability.

A driven $U$ differs from the other parameters in that it selectively affects only transport with transition energy $\epsilon + U$. Even in the single-resonance case for mechanism F, the driving-induced, time-dependent deviation from the steady-state energy current thereby generates non-tightly coupled contributions, i.e., not every transported charge carries the same energy, allowing the parity mode to contribute [see \Eqs{aH_Pp} and \eq{aH_PN}]. A driven coupling asymmetry $\Lambda$ instead always \emph{equally} affects the rates of transition via both energies $\epsilon$ and $\epsilon + U$, but as such still gives rise to parity-like time-dependent corrections to the energy current. Indeed, as stated before, a \emph{cycle-averaged} contribution from these parity-like corrections only arises if they are accompanied by a resonant effect that modulates the balance between transport at the two transition energies $\epsilon$ and $\epsilon + U$. When driving only $U$ and $\Lambda$, this demands a lead potential resonant with the transition energy $\epsilon + U$, explaining why the parity contribution due to mechanism F in \Fig{fig_pumping_curvature_repulsive_R}(d) disappears for $\beps < -\bU/2$. Driving the temperature $\TL$ of the left lead and $U$ or $\Lambda$ analogously requires a resonance with the left lead to give finite parity-like pumped energy per cycle, so that $\BHaP(\bR)$ vanishes for $\bVb > \bU \gg \TL$ and $\beps < -U/2$ as well as for $\bVb < -\bU$ and $\beps > -\bU/2$, see \Figs{fig_pumping_curvature_repulsive_R}(c,f).

\subsection{Quantum dot with attractive onsite interaction}
\label{sec_results_attractive}
\begin{figure*}
    \includegraphics{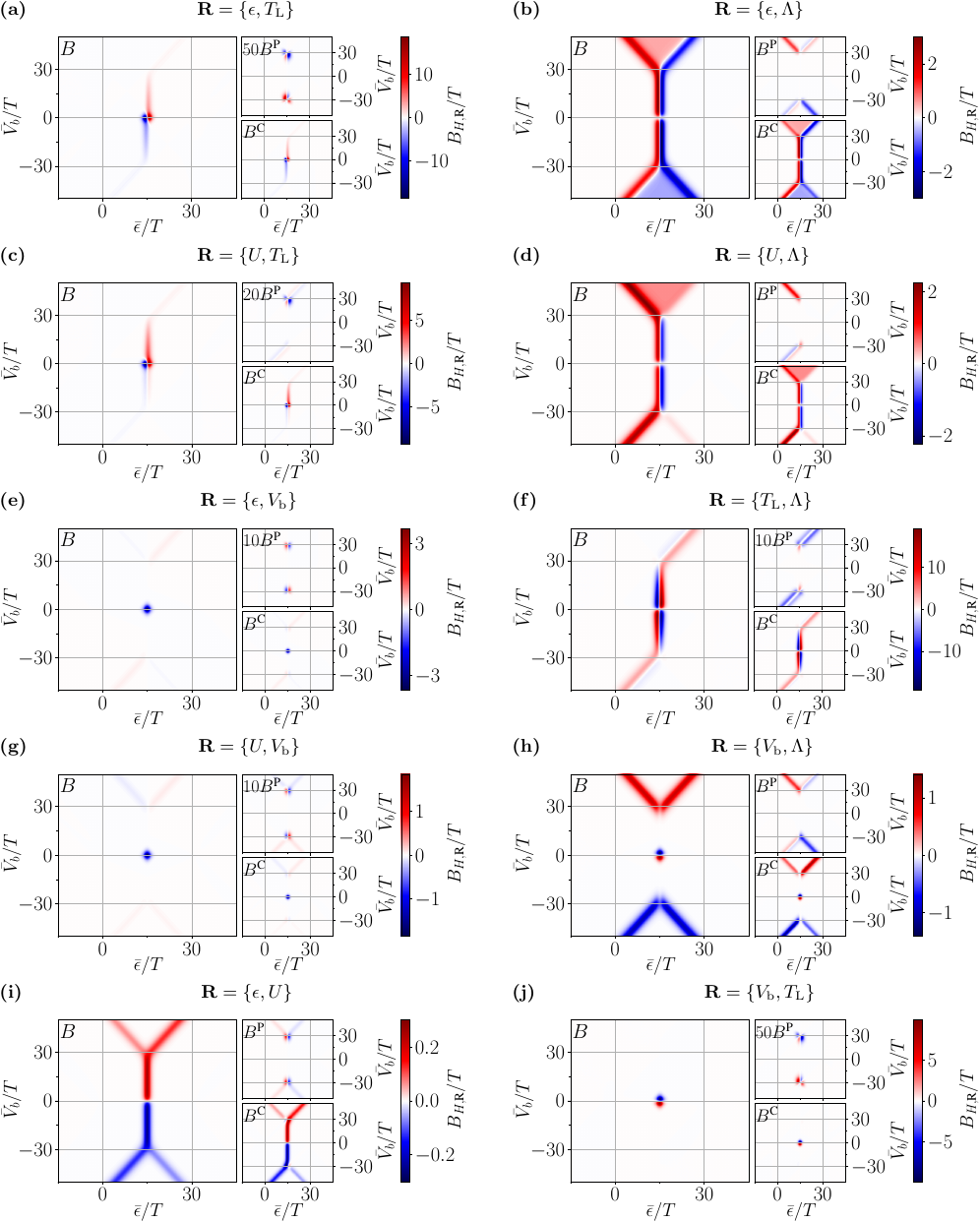}
    \caption{\label{fig_pumping curvature attractive R}
        Pumping curvature $\BHR$ of the attractive system ($\bU < 0$) in the $(\beps, \bVb)$-space for all possible two-parameter driving schemes $\bR = \{R_1, R_2\}$. We have plotted the component of $\BHR$ perpendicular to the driving plane defined by $(R_1, R_2)$, in units of $\TN$. In each subfigure, the largest panel corresponds to the total pumping curvature, the top-right one to the parity contribution $\BHRP$ and the bottom-right one to the charge contribution $\BHRC$. $\BHRP$ has been multiplied by a factor indicated in each panel to make its features visible using the same color scale as in the other panels. The parameters are the same as in \Fig{fig_pumping_curvature_repulsive_R}, except that $\bU = -30 \TN$.
    }
\end{figure*}

We now analyze the same driving schemes as above for a quantum dot with an \emph{attractive} onsite interaction, $U<0$, and we identify the major differences and similarities with the more standard case of a repulsive  onsite interaction.  Results for the energy pumped through a quantum dot with $U<0$ are shown in \Fig{fig_pumping curvature attractive R} for the different driving schemes. We consider similar working points as previously, namely $\bTL = T$, $\bLam = 0$, $\bU = -30 T$, and plot the curvatures in the $(\beps, \bVb)$-space.

\subsubsection{Specific feature of attractive onsite interaction: the two-particle resonance}\label{sec_attractive_C}
A unique feature of attractive onsite interaction is the pairing induced two-particle resonance at the particle-hole symmetric point $\epsilon = -U/2$~\cite{Schulenborg2020Jun}. At this point, when $\Lambda = 0$ and $\TL= \TN$, the dot switches from a stable empty state (for $\epsilon/\TN \gg -U/(2\TN)$) to a stable double occupation (for $\epsilon/\TN \ll -U/(2\TN)$), regardless of bias voltage (up to $\Vb\approx|U|$). Interestingly, this transition in the dot occupation $\nz$ still occurs around $\epsilon/\TN \approx -U/(2\TN)$ and depends little on $\Vb$, even for small but finite coupling asymmetries and temperature differences. 

As justified in \Sec{sec_geometric_energy}, only the charge mode contributes to the energy pumping for $|\Vb| < |U|$, namely $\BHa(\bR) \approx \nr \aCNHa \times \nr\nz$. Therefore, as visible in \Fig{fig_pumping curvature attractive R}, the energy pumping curvatures exhibits features only along this two-particle resonance line for these biases, corresponding to the mechanism C defined in \Sec{sec_mechanisms}. However, the charge pumping, that has already been addressed in detail in \Ref{Placke2018Aug} for $U < 0$, only happens for $|\Vb|/\TN \approx 0$ and $\epsilon/\TN \approx -U/(2\TN)$, see \Fig{fig_charge_pumping_curvature_attractive_R}, in Appendix \ref{app_charge}.   

In this section, we compare charge and energy pumping close to the two-particle resonance, $\epsilon/\TN \approx -U/(2\TN)$, and we start by analyzing the pumping curvatures close to zero bias voltage, $\Vb/\TN \ll 1$. First and foremost, energy pumping cannot be understood by simply relating it to charge pumping. As a matter of fact, unlike for a repulsive onsite interaction, the contribution $\nr \aCTCHa \times \nr \nz$, tightly-coupled to charge pumping [\Eq{aH_CN TC}], \emph{does not} dominate the energy pumping curvature. On the contrary, the features of the energy pumping curvature mostly arise from the non-tightly-coupled contribution $\nr \aCNTCHa \times \nr \nz$ [\Eq{aH_CN NTC}], even at $|\Vb|/\TN \approx 0$. This contribution is, however, less straightforward to interpret for $U < 0$ than for $U >0$ since it contains the ratio of $(\nzia - \nzi)$ and $\gamc$, which is finite while both quantities individually are suppressed for mechanism C.  
To understand better energy pumping, we therefore conduct an in-depth analysis of $\BHa(\bR)$ for every driving scheme based on expressions derived from the explicit formulas in Appendix \ref{app_two_particle_resonance_energy}. In particular, we give analytical justifications of the features observed in \Fig{fig_pumping curvature attractive R}, found to be as follows.

Focusing on mechanism C$^\pm$, at finite bias voltage $|U| > |\Vb| > \TN$, we note that charge pumping is always suppressed, even though the system may be in a two-particle resonance with the combined two-lead system at $\beps = -\bU/2$. This is because even close to this resonance, the dot is nevertheless off-resonant with \emph{both lead potentials individually}, namely $(\epsilon + U/2)/\TN \ll |\mu_\alpha|/\TN$. The time-dependent charge flow is then at most affected by a single driving parameter related to the left-right balance of charge transfer, and single-parameter driving results in a vanishing cycle average, see Appendix \ref{app_two_particle_resonance_charge} for an analytical derivation.

Finite energy pumping for mechanism C$^\pm$ can, by contrast, still be achieved by doing work on the dot or by modulating the left-right lead coupling balance in a voltage-biased environment, see Appendix \ref{app_two_particle_resonance_energy}. A crucial difference from the $|\bVb|/T \approx 0$ case occurs for a driven bias voltage $\Vb$ [\Figs{fig_pumping curvature attractive R}(e,g,h,j)], which always features suppressed energy pumping. This originates from the fact that for $|\Vb| > \TN$ any further modulation of an already \emph{symmetrically} applied bias $|U| > |\Vb(t)| > \TN$ has no bearing on neither charge nor energy transport via the two-particle resonance, i.e. for $\epsilon/\TN \approx -U/(2\TN)$. Driving $\Vb$ and any second parameter is hence equivalent to single-parameter driving even for the energy current, thereby prohibiting adiabatic pumping altogether.

\subsubsection{Comparison to repulsive onsite interaction}\label{sec_attractive_comparison}

We now look more generally at the features of energy pumping for a quantum dot with an attractive onsite interaction and compare it to the repulsive onsite interaction from \Sec{sec_results_repulsive}. First, the energy pumping curvature exhibits the same particle-hole symmetry or antisymmetry as in the repulsive case in each panel of \Fig{fig_pumping curvature attractive R}, except if $U$ is one of the driving parameters [\Sec{sec_symmetry}]. Furthermore, the parity-like component $\BHaP(\bR)$ of the energy pumping curvature is very similar for both interaction signs, as revealed by comparing the top-right panels in \Fig{fig_pumping_curvature_repulsive_R} to the corresponding panels in \Fig{fig_pumping curvature attractive R}. In particular, $\BHaP(\bR)$ for \emph{constant} attractive interaction $\bU < 0$ equals $\BHaP(\bR)$ evaluated for repulsive interaction $-\bU > 0$ and a working point $\beps$ shifted by $\bU = -|\bU|$.
This stems from a combination of the self-duality of $\AHaP$ [\Sec{sec_geometric_energy}] and the particle-hole symmetry relations [\Sec{sec_symmetry}], as derived in details in \App{app_symmetry_dual}. The parity-like pumping contributions can thus be understood in the same way as discussed for the repulsive dot in \Sec{sec_parity_repulsive}, and are not further discussed here.

We continue our comparison by noting that many mechanisms leading to charge or energy pumping for $U > 0$ are absent for $U < 0$. Indeed, due to the pairing induced by attractive interaction, any particle flow to and from the dot is almost always exponentially suppressed if at least one of the two dot transition energies $\epsilon$, $\epsilon + U$ lies significantly outside the bias window on the scale of $\TN$. Accordingly, sizable contributions due to mechanisms A, D, E and G are absent both in energy and charge pumping, see  \Fig{fig_pumping curvature attractive R} and \Fig{fig_charge_pumping_curvature_attractive_R} 
respectively. On the contrary, mechanism C at the two-particle resonance $\beps = -U/2$ is specific to attractive onsite interaction, and the most relevant mechanism for both charge and energy pumping [\Sec{sec_attractive_C}]. 

Mechanism B also does not contribute with distinct features as for $U > 0$, but rather appears as a continuation of mechanism C$^\pm$, with vanishing charge pumping but generally finite energy pumping. This is consistent with the fact that at $\beps = -\bU/2$ and strong attraction, $-U \gg \TN$, an effective double resonance situation with both leads combined is not only fulfilled for $|\Vb| = |U|$ specifically, but for any bias voltage $|\Vb| \leq |U|$.  

We furthermore observe that mechanisms F and H contribute to energy pumping at large bias, $|\Vb| \geq |U|$, in both cases. Since mechanism H corresponds to both dot transition energies $\epsilon$ and $\epsilon + U$ being inside the bias window, the sign of the interaction strength $U$ loses its significance, and we accordingly observe the same plateaus in the pumped energy as for repulsive interaction, see \Figs{fig_pumping curvature attractive R}(b,d) and \Figs{fig_pumping_curvature_repulsive_R}(b,d). The charge pumping curvature is likewise suppressed for both $U > 0$ and $U < 0$. 
The situation for mechanism F at constant $U,\Lambda$ is, again, qualitatively very similar to the case of repulsive interaction, both for charge and energy pumping. The difference is that for a working point $\beps$ below/above the particle-hole symmetric point, it is not the energetically higher/lower dot transition energy as for $U > 0$, but instead the energetically lower/higher energy that is in resonance with one of the leads. While irrelevant for constant $U$, a time-dependent $U(t) < 0$ does yield a different pumped energy than for $U(t) > 0$ because of this distinction.
This can be seen by, e.g., comparing  $(\epsilon,U)$-driving in \Fig{fig_pumping curvature attractive R}(i) to \Fig{fig_pumping_curvature_repulsive_R}(i). Repulsive interaction here features a relatively larger magnitude of pumped energy for $\bVb > 0$, $\beps + \bU/2 < 0$ as well as $\bVb < 0, \beps + \bU/2 > 0$, whereas attractive interaction yields a larger pumping current with the sign of $\beps + \bU/2$ reversed. 

The driving scheme $\bR = \{\epsilon, U\}$ gives similar results for both signs of the interaction: charge pumping is suppressed [\Fig{fig_charge_pumping_curvature_attractive_R}(i)] due to the constant equilibrium environment of the dot while energy can still be pumped [\Fig{fig_pumping curvature attractive R}(i)]. The reason is that the $(\epsilon,U)$-driving does work on the particles while temporarily occupying the dot. However, unlike for $U > 0$, energy pumping also happens at biases $|\Vb| < |U|$ due to mechanism C, see previous \Sec{sec_attractive_C}.

Addressing the effect of driven \emph{coupling asymmetry}, the obvious difference between repulsive and attractive interaction is that while the pumped energy is qualitatively the same for both interaction signs [\Figs{fig_pumping curvature attractive R}(b,d,f,h) and \Figs{fig_pumping_curvature_repulsive_R}(b,d,f,h)], net cycle-averaged charge pumping [\Figs{fig_charge_pumping_curvature_attractive_R}(b,d,f,h) and \Figs{fig_charge_pumping_curvature_repulsive_R}(b,d,f,h)] is finite only for $U > 0$ but suppressed for $U < 0$. The reason for this is the same as for mechanism C$^\pm$: with both lead potentials away from the two-particle resonance, the driving has no additional effect on top of the bias-induced stationary charge flow when averaged over a cycle, but it can still affect the net energy of electrons passing through.

Finally, when driving the temperature, like for $U > 0$, there is a sign change in the feature at the single-resonance with the left lead, but only for mechanism F since D and E are suppressed [\Figs{fig_pumping curvature attractive R}(a,c,f)]. However, the effect is less visible due to the high values of the pumping curvature for mechanism C. As discussed previously, the intensity of the branches, above and below the resonance with the left lead, is reversed since the transition energies are also inverted.

\section{Refrigerator and  heat pump}\label{sec_refrigerator}
The regime of slow driving studied here is especially interesting for periodically driven thermal machines \cite{Bhandari2020Oct, Miller2021May}.
Therefore, in this section, we analyze the device as a cyclic thermal machine, intended to further cool a bath colder than its environment (refrigerator) or to further heat a bath warmer than its surrounding (heat pump). The working substance is the quantum dot itself, and the contacts represent the hot and cold bath, assuming equal chemical potential, $\Vb = 0$. We use the results from \Sec{sec_analytical_results} and \Sec{sec_results} specific to these parameters to identify interesting driving operation points and to gain insights into the mechanisms at play.

\subsection{Driving scheme and thermodynamic quantities}

As a paradigmatic (but not the only possible) way to achieve refrigeration or heat pumping, we consider the driving scheme $(\epsilon, \Lambda)$, 
\begin{equation}
    \epsilon(t) = \beps + \delta\epsilon \sin(\Omega t)\quad,\quad\Lambda(t) = \delta \Lambda \sin(\Omega t + \phi),\label{eq_driving_protocol}
\end{equation}
which moves the dot potential $\epsilon(t)$ while modulating the left-right tunnel coupling asymmetry $\Lambda(t)$ with a driving phase $\phi$ relative to $\epsilon(t)$. As an illustrative example, let us for a moment assume an amplitude $\delta\Lambda = 1$, tuning the dot all the way from only coupled to the cold bath, and decoupled from the hot bath, to the opposite configuration, only coupled to the hot bath. For an appropriately chosen $\phi$, this enables to pump electrons against a temperature difference, making the device operate as a refrigerator or heat pump as sketched in \Fig{fig_model}(c): Namely, in step (I), the dot is only coupled to the cold bath while its transition energy is lowered below the Fermi level to let an electron tunnel in. The dot is then decoupled from the cold bath and coupled to the hot bath in step (II). Step (III) increases the dot transition energy above the Fermi level so that the electron tunnels into the hot bath. The final step (IV) completes the cycle, reverting to a dot only coupled to the cold bath.

To be in the adiabatic-response regime, the driving frequency $\Omega$ as well as the amplitudes $\delta\epsilon, \delta\Lambda$ are chosen such that $\Omega\delta\epsilon/\TN, \Omega\delta\Lambda \ll \Gamma$. The targeted heat pump and refrigerator driving cycles operate at zero bias voltage $\Vb = 0$, but are meaningful if a small but finite temperature difference $\bTL = \TN +\dT$, $\bTR = \TN$ with $0 < \abs{\dT} \ll \TN$ exists, unlike $\abs{\dT} = 0$ considered in \Sec{sec_results}.
The device performance is quantified via the heat $\QR$ emitted or absorbed to/from the right reservoir, and via the work $\Work$ provided to the device by the external driving, both in one driving cycle. In this setting, the refrigerator (heat pump) operation mode corresponds to $\dT > 0$ and $\QR > 0$ ($\dT < 0 $ and $\QR < 0$). The average cooling power is given by $\QR \Omega/2\pi$, and the coefficient of performance is defined as
\begin{equation}
    \eta = \abs{\frac{\QR}{\Work}},\label{eq_coeff_of_performance}
\end{equation}
whenever $\QR$ has the desired sign, while we set $\eta = 0$ otherwise.
The corresponding Carnot efficiency reads
\begin{equation}
    \etaC =  \frac{\TN}{\abs{\dT}}.
\end{equation}

Since $\muR = 0$, the heat current coincides with the energy current $\IHR$ [\App{app_macro_timedep}], so that the heat $\QR$ can be expressed, up to the first-order correction, as $\QR = \QR^{(0)} + \QR^{(1)}$ with 
\begin{equation}
    \QR^{(\ell)} = \Delta H_\R^{(\ell)} = \int_0^{2\pi/\Omega} \dd t \, \IHR^{(\ell)}(t),\label{eq_heat_order}
\end{equation}
where $\IHR^{(\ell)}(t)$ is defined in \Eq{eq_obs_current} with $\hat{\mathcal{O}} = \hat{H}$. The zeroth order $\QR^{(0)}$ stems from the instantaneous stationary heat flow $\IHR^{(0)}(t)$ induced by the temperature difference between the contacts; it is hence proportional to $\Gamma/\Omega$, and always detrimental to the device operation, since heat always flows from the hot to the cold bath in the steady state. The first-order correction $\QR^{(1)}$, on the other hand, describes the geometric pumping contribution. It is independent of both total coupling strength $\Gamma$ and driving frequency $\Omega$, and an appropriate driving phase $\phi$ between $\epsilon(t)$ and $\Lambda(t)$ results in a finite, cycle-averaged heat pumped from the cold to the hot contact.

The work can be split into two contributions\footnote{If the bias voltage was finite, there would be an additional chemical work contribution.}, $\Work = \Weps + \WLam$ coming from the driving of each parameter. With $\WLam \rightarrow 0$ in the weak-coupling regime\footnote{Beyond the weak-coupling regime, higher orders in the tunnel coupling induce a renormalization of the dot's population \cite{Splettstoesser2006Aug} which can then be related to a finite work cost. See also Ref.~\cite{Esposito2015Feb} for an expression of the work in the strong coupling regime for a non-interacting fermionic system.}, the work is done exclusively by the $\epsilon$-driving, 
\begin{equation}
    \Weps = \int_0^{\frac{2\pi}{\Omega}} \dd t\, \Braket{\partial_t H}{\rho(t)} = \sum_{\ell > 0}\Weps^{(\ell)},\label{eq_work_expansion}
\end{equation}
where the factor $\partial_t \hat{H} = \dot{\epsilon}(t)\hat{N}$ suppresses the $\ell = 0$ term $\sim\Gamma/\Omega$ in the driving-frequency expansion in orders
\begin{equation}
    \Weps^{(\ell)}=\int_0^{\frac{2\pi}{\Omega}} \dd t\, \dot{\epsilon}(t)\Braket{N}{\rho^{(\ell - 1)}(t)}.\label{eq_work_expansion_orders}
\end{equation}
Unlike for $\QR$, we cannot truncate the series \eq{eq_work_expansion} already at the first-order correction $(\ell = 1)$, despite the slow driving. \App{app_first_order_dT} namely shows that $\Weps^{(1)}$ is of first order in $\dT$, whereas $\Weps^{(2)}$ contains a zeroth-order term in $\dT$, making $\Weps^{(2)}$ the dominant contribution to the coefficient of performance \eq{eq_coeff_of_performance} for $0 < |\dT|/\TN \ll 1$. 
In fact, it was recently shown on more general grounds~\cite{Eglinton2022Feb} that a first-order expansion in both temperature difference $\dT$ and driving frequency $\Omega$ is typically not sufficient to adequately estimate the performance of a thermal machine.

\subsection{Limit of small driving amplitudes}

With a full analysis of work $\Weps$, heat $\QR$ and performance $\eta$ of the quantum dot device following in \Sec{sec_performance}, we first gain further analytical insight by considering only small driving amplitudes, such that $\delta \epsilon/\TN, \delta \Lambda \ll 1$ allows to Taylor expand all quantities up to leading order in these amplitudes. The vanishing bias voltage $\Vb = 0$, the assumption of strong interaction $|U|/\TN \gg 1$, and the small temperature difference $|\delta T| \ll \TN$ furthermore enables us to use the linear-response results from \Ref{Schulenborg2017Dec} and the general steady-state linearization approach from \Ref{Schulenborg2018Dec} to expand all quantities up to linear order in $\delta T$.

Based on our findings from \Sec{sec_Transport}, we derive in \App{app_first_order_dT} that the heat contributions for $|U| \gg \TN$ are, up to leading order in $\delta T$, given by
\begin{align}
    \QR^{(0)}  &\approx\frac{2\pi}{\Omega}\left[\left(\cEeq \INR^{(0)} -\kappa\dT\right)\left(1 - \frac{(\delta\Lambda)^2}{2}\right)\right.\label{energy_pumping}\\
    &\phantom{\approx} \left. + \frac{\delta\epsilon^2}{4}\left(\cEeq\partial^2_\epsilon \INR^{(0)} + \partial_\epsilon \INR^{(0)} - (\partial^2_\epsilon\kappa)\dT\right)\right] + \mathcal{O}(\dT^2)\notag\\
    \QR^{(1)} &= \BHR(\bR)\cdot \delta \mathbf{S} \approx -\frac{\delnzsqeq\cEeq}{2}\dS + \mathcal{O}(\delta T).\notag
\end{align}
The heat thereby depends on the Seebeck energy $cEeq$, the Fourier heat~\cite{Schulenborg2017Dec}
\begin{equation}
    \left.\kappa\right|_{\Lambda = 0} = \frac{1}{4}\gamp\delnzsqeq\delnzisqeq\frac{U^2}{4\TN^2},\label{eq_fourier_heat}
\end{equation}
on the oriented driving surface $\dS = \iint_\cS \dd \Lambda (\dd \epsilon/\TN) = -\pi \sin(\phi) (\delta\epsilon/\TN)\delta\Lambda$, and on the stationary charge current, defined by the $\ell = 0$ component of \Eq{eq_obs_current}:
\begin{align}
    \INR^{(0)} &\approx -\frac{1}{4}\gamceq\delnzsqeq\cEeq\frac{\delta T}{\TN^2} + \mathcal{O}(\delta T^2)\label{I_N0}.
\end{align}
All of the above quantities depend on the equilibrium charge rate $\gamceq$ ---which determines the inverse RC time of the dot \cite{Splettstoesser2010Apr}--- and  on the Seebeck energy $\cEeq$,
\begin{align}
    \gamceq &= \left.\gamc\right|_{\delta T,\Vb = 0} = \frac{\Gamma}{\Gama}\left.\gamca\right|_{\delta T,\Vb = 0}\notag\\
    \cEeq &= \left.\cEa\right|_{\delta T,\Vb = 0} = \epsilon + \frac{U}{2}\left(2 - \nzieq\right)\notag\\
    \nzieq &= \left.\nzi\right|_{\delta T,\Vb = 0} = \left.\nzia\right|_{\delta T,\Vb = 0},
\end{align}
as well as on the equilibrium charge fluctuations $\delnzsqeq$,
\begin{align}
    \delnzsqeq &= \left.\Braket{(N - \nz)^2}{z}\right|_{\delta T,\Vb = 0}.
\end{align}
These are well-known from the DC, linear-response behavior of the dot~\cite{Schulenborg2017Dec}, and here directly enter properties of the driven system. This includes the work $\mathcal{W}$ corresponding to the driving cycle, which reads [\App{app_first_order_dT}]
\begin{align}
    \Weps^{(1)}
    &= -\partial_\Lambda \nz \TN\dS \approx -2\TN\frac{\INR^{(0)}}{\gamceq}\delta S + \mathcal{O}(\delta T^2),\notag \\
    \Weps^{(2)} &\approx \pi\frac{\Omega}{\gamceq}\frac{\delta\epsilon^2}{\TN}\delnzsqeq + \mathcal{O}(\dT).\label{eq_work}
\end{align}
In the limit $\dT \to 0$, both reservoirs become identical and the stationary current vanishes $I^{(0)}_{N/H\comR} \rightarrow 0$. Therefore, $\QR^{(0)} \rightarrow 0$, $\Weps^{(1)} \rightarrow 0$ and
\begin{align}
    \QR^{(1)} &\rightarrow \sin(\phi)\frac{\pi\delnzsqeq\cEeq\delta\epsilon}{2\TN}\delta\Lambda \quad, \quad \Weps^{(2)} \rightarrow \frac{\pi\Omega}{\gamceq}\frac{\delnzsqeq\delta\epsilon^2}{\TN}.\label{heat_work_lim}
\end{align}
As a consequence, unlike for a Carnot cycle, the cooling power is non-zero and the coefficient of performance $\eta$ is finite when the temperature difference vanishes:
\begin{equation}
    \eta \approx \left|\frac{\QR^{(0)}+\QR^{(1)}}{\Weps^{(1)} + \Weps^{(2)}}\right| \underset{\dT \to 0}{\longrightarrow} |\sin(\phi)|\frac{\gamceq\abs{\cEeq}\delta\Lambda}{2\Omega\delta\epsilon}\label{eta_lim}
\end{equation}
for the desired sign of $\QR$ ($\QR > 0$ for refrigeration, $\QR < 0$ for heat pump), and $\eta = 0$ otherwise. However, in the limit of infinitely slow driving $\Omega \to 0$, namely a perfectly quasistatic cycle, $\eta$ becomes infinite like the Carnot efficiency. Equations \eq{heat_work_lim} and \eq{eta_lim} are crucial analytical results in the following, more detailed discussion of the dot acting as a refrigerator or heat pump: depending only on the driving amplitudes and well known equilibrium quantities ---charge fluctuations $\delnzsqeq$, Seebeck energy $\cEeq$ and the RC time scaled by the driving frequency, $\gamceq/\Omega$--- \Eqs{heat_work_lim} and \eq{eta_lim} provide simple estimates of the device performance as a function of the system parameters.

\subsection{Refrigerator and heat pump performance}
\label{sec_performance}

\begin{figure*}
    \includegraphics{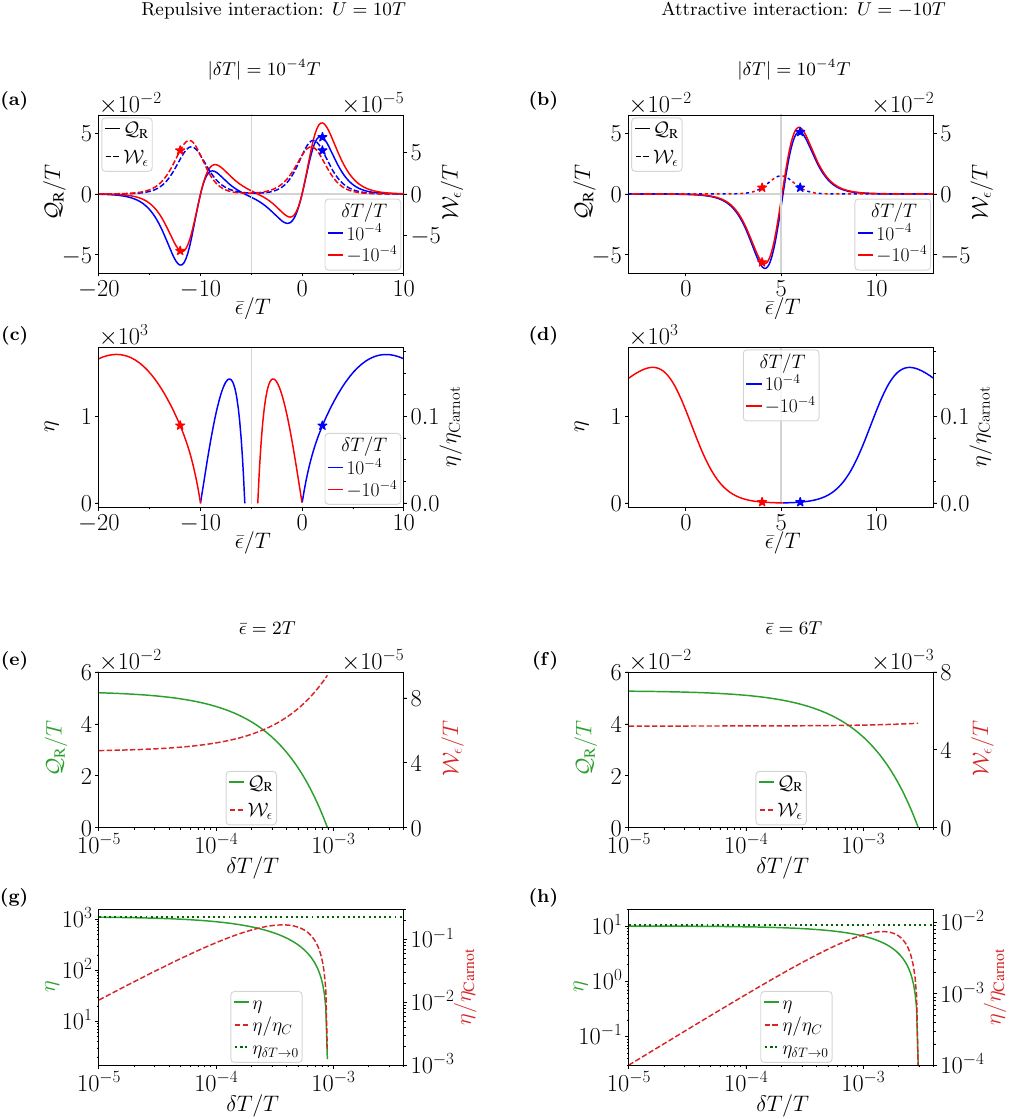}
    \caption{\label{fig_thermo_fct_epsilon}
        Thermodynamic quantities as functions of the energy $\beps$ (a-d) and temperature difference $\dT = \TL - \TN$ (e-f) for a repulsive onsite interaction (left column) and an attractive one (right column), obtained by numerically integrating the full master equation. In (a-d), we compare the refrigerator ($\dT > 0$, blue curves) and heat pump ($\dT < 0$, red curves) operating modes: (a,b) heat $\QR$ (solid lines, left axis) and work $\Weps$ (dashed lines, right axis), (c,d) coefficient of performance $\eta$, with the right axis indicating the ratio between $\eta$ and the Carnot efficiency. In (e-f), we only consider the refrigerator ($\dT > 0$), with the blue stars in (a-d) indicating the chosen $\beps$ values, and the red stars in (a-d) showing the corresponding values if one used the heat pump operating mode instead. The coefficient $\eta$ is plotted only for the values of $\beps$ for which $\QR$ has the desired sign. In (g,h), the dotted green line indicates the efficiency $\eta_{\dT\to 0}$ in the limit of small driving amplitudes for a vanishing temperature difference [Eq.~\eqref{eta_lim}].
        Parameters: $\Gamma = 10^{-2}\TN$, $\Omega =  10^{-2}\Gamma$, $\delta\epsilon = 10^{-1}\TN$, $\delta\Lambda = 1$, $\phi = \pi/2$, $|U| = 10\TN$, $\bVb = 0$, $\bLam = 0$. 
    }
\end{figure*}

To analyze the refrigeration and heat pump performance, we focus mostly on the interesting operating points for the thermal machine at $\Vb = 0$ identified in \Sec{sec_results}: mechanism A (resonance points) for a repulsive onsite interaction and mechanism C (particle-hole symmetric point) for an attractive onsite interaction. The thermodynamic quantities relevant to the device performance for these parameters are plotted in \Fig{fig_thermo_fct_epsilon} as a function of $\beps$ and $\dT$, allowing us to compare the two cases $U > 0$ and $U < 0$. 
All displayed curves were obtained by numerically solving the master equation $\partial_t\Ket{\rho} = W\Ket{\rho}$ [see Sec.~\ref{sec_master_equation}] and therefore, take into account all the orders in $\Omega/\Gamma$ (see Appendix~\ref{app_adiabatic_limit} for a comparison with the adiabatic-response limit). Note that we have always set the interaction strength to $|U|/\TN = 10$. The latter is slightly smaller than in \Sec{sec_results}, in order to avoid that corrections at higher orders in $\Omega/\Gamma$ become dominant, see below.

\begin{figure}[t!]
    \includegraphics{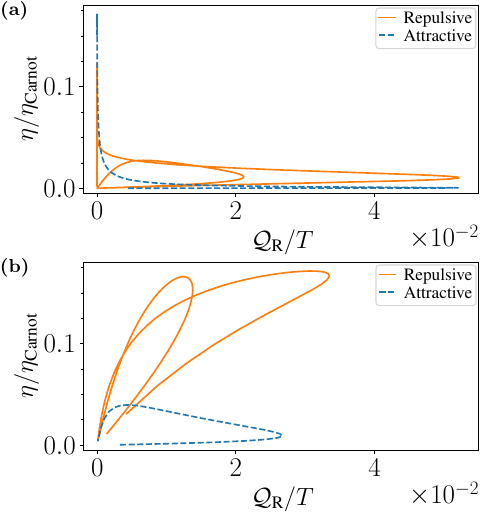}
    \caption{\label{fig_thermo_trade_off}
        Coefficient of performance $\eta$ and heat $\QR$ parametrically plotted as a function of a varied working point $\beps$, for both $U > 0$ (solid orange lines) and $U < 0$ (dashed blue lines). The curves were obtained by numerically integrating the full master equation. In (a) we fix $\dT = 10^{-5}\TN$ in the repulsive case and $\dT = 4\cdot 10^{-5}\TN$ for the attractive dot. In (b), we choose $\dT$ such that $\eta/\etaC$ is maximized in \Figs{fig_thermo_fct_epsilon}(g,h), meaning $\dT = 3.75\cdot 10^{-4} \TN$ and, respectively, $\dT = 1.5\cdot 10^{-3}\TN$ for $U > 0$, respectively, $U < 0$. The factor 4 between $\dT$ for the attractive vs. the repulsive case accounts for the increased operation range of $\dT/\TN$ observed in \Figs{fig_thermo_fct_epsilon}(f,h) compared to \Figs{fig_thermo_fct_epsilon}(e,g). The parameters not indicated are set as in \Fig{fig_thermo_fct_epsilon}.
    }
\end{figure}

Let us start by discussing the influence of the driving protocol itself, through the phase $\phi$, the amplitudes $\delta\epsilon,\delta\Lambda$, and the frequency $\Omega$. Equations \eq{heat_work_lim} and \eq{eta_lim} clearly state that $\phi$ has no influence on the $\Weps^{(2)}$ but maximizes the heat, and hence the performance $\eta$ for $\phi = \pi/2$. We therefore fix $\phi = \pi/2$ for the entirety of this analysis, including \Fig{fig_thermo_fct_epsilon} and \Fig{fig_thermo_trade_off}. The key point for the amplitude dependence is that up to the leading order in $\Omega$, the coupling asymmetry $\delta\Lambda$ only affects the heat $\QR^{(1)} \sim \delta\epsilon\delta\Lambda$, but not the work $\Weps^{(2)} \sim \delta\epsilon^2$, since it is irrelevant to which contact the energy corresponding to this work flows. As a result, the performance $\eta \sim \delta\Lambda/\delta\epsilon$ improves with larger asymmetry $\delta\Lambda$ increasing the directionality of the heat flow, but still degrades with larger $\delta\epsilon$ increasing the work done on the system to induce this heat flow. Finally, the driving frequency $\Omega$ only enters $\Weps^{(2)}$ in leading order; it provides a typical time scale for the delayed system response and the resulting work due to the driving, as further detailed below.

The dependence of $\QR$, $\Weps^{(2)}$ and $\eta$ on the working point dot level $\beps$ is shown in \Figs{fig_thermo_fct_epsilon}(a-d). For repulsive interaction $U > 0$, the proportionality $\QR,\eta \sim \cEeq$ predicted by \Eq{heat_work_lim} and \eq{eta_lim} implies the same, well-known~\cite{Staring1993Apr,Dzurak1993Sep,Dzurak1997Apr} sawtooth behavior as a function of $\beps$ as for the Seebeck coefficient $\cEeq$. Indeed, the numerical results in \Figs{fig_thermo_fct_epsilon}(a,c) confirm that apart from small $|\dT|$-corrections, the sign changes of $\cEeq$ near the resonances $\beps = 0,-U$ and near the particle-hole symmetry point $\beps = -U/2$ result in suppressed $\QR,\eta$, concomitant with switches from refrigeration (blue curve) to heat pumping (red curve) behavior. 
The slope of $\QR$ as a function of $\beps$ is, however, attenuated around $\beps = -U/2$ compared to the slopes around $\beps = 0,-U$. This is because $\QR$ is, unlike $\eta$, also proportional to the charge fluctuations $\delnzsqeq$, which are generally stronger near the single-particle resonances $\beps = 0,-U$. These fluctuations cancel out in $\eta$ because the work $\Weps^{(2)}$ as given in \Eq{heat_work_lim} is likewise proportional to $\delnzsqeq$. The latter is due to the fact that net work can only be done per $\epsilon$-driving cycle if a crossing of a single-particle resonance allows at least for temporary dot occupation changes. 

For attractive interaction $U < 0$  [\Figs{fig_thermo_fct_epsilon}(b,d)], the Seebeck energy $\cEeq$ and hence $\QR$ only change sign once close to $\beps = -U/2 = +|U|/2$, which is the pairing-related two-particle resonance~\cite{Schulenborg2020Jun}. This means that in contrast to the case $U > 0$, the system only switches once from a heat pump to a refrigerator at small $\dT/\TN \ll 1$ when sweeping through $\beps = -U/2$, with $\QR < 0 \rightarrow \QR > 0$ [\Fig{fig_thermo_fct_epsilon}(b)]. 
The second key difference to repulsive interaction visible in \Fig{fig_thermo_fct_epsilon}(d) is that $\eta$ is generally suppressed for dot levels $0 < \beps < -U$. 
This stems from the proportionality to the charge rate, $\eta \sim \gamc$, as predicted by \Eq{eta_lim}. This rate enters the coefficient of performance via the work \eq{heat_work_lim}, for which the dominant, second-order correction is, in fact, \emph{inversely} proportional to the equilibrium charge rate, $\Weps^{(2)} \sim \Omega/\gamceq$, due to the delayed system response. According to \Eq{eq_work_expansion_orders}, $\Weps^{(2)}$ depends on the first-order correction to the dot state $\rho^{(1)}$, representing the delay of the state evolution due to the external driving. 
The slower the system response, the larger the delay, as generally stated by \Eq{eq_first_correction} and as quantified by $\gamc/\Omega$ specifically for the charge-mode response. For attractive interaction and $0 < \beps < -U$, the rate $\gamc$ is exponentially suppressed. The resulting suppression of changes in the dot occupation during the $\epsilon$-drive thus increases the required work significantly, and hence suppresses the performance $\eta$. Figure~\ref{fig_thermo_fct_epsilon} indeed shows $\Weps^{(2)}$ for $U < 0$ to be approximately two orders of magnitude larger compared to the repulsive dot with $U > 0$ when comparing the points of operation indicated by the red and blue stars.
Moreover, the smaller rate $\gamc$ for $U < 0$ also requires a lower driving frequency compared to the repulsive dot in order to remain in the adiabatic-response regime, see Appendix \ref{app_adiabatic_limit}. 
The issue can be mitigated by operating at higher base temperature $\TN$, thus motivating our choice $|U| = 10\TN$ instead of $30\TN$ as in \Sec{sec_results}.

Irrespective of the interaction sign, \Figs{fig_thermo_fct_epsilon}(a-d) reveal an approximate symmetry between the refrigerator and heat pump case under a temperature inversion $\dT \rightarrow -\dT$ together with a particle-hole parameter transform $\cT$ [see \Sec{sec_symmetry}]. This is because the work $\Weps^{(2)}$ is symmetric under both $\cT$ and $\dT \rightarrow -\dT$ [\Eq{eq_work}], the pumped heat $\QR^{(1)} \sim \BHR(\bR)$ is anti-symmetric under $\cT$ but symmetric under $\dT$-inversion [\Eq{energy_pumping}], and the steady-state heat $\QR^{(0)} \sim \dT$ is instead symmetric under $\cT$ but anti-symmetric under $\dT \rightarrow -\dT$, see also \App{app_first_order_dT}. For an identical driving frequency $\Omega = 10^{-2}\Gamma$, this symmetry between the refrigerator and heat pump is less accurate for $U < 0$ than $U > 0$, which comes from higher-order contributions and confirms that the dot with an attractive onsite interaction is less in the adiabatic-response regime. This can be seen by looking at $\QR$ (solid lines) in Figs.~\ref{fig_thermo_fct_epsilon}(a,b): for $U > 0$, rotating the blue curve (refrigerator) by 180 degrees around $(-U/2, 0)$ perfectly gives the red curve (heat pump) while this is not the case for $U < 0$, in particular, the blue star is exactly on the peak whereas the red one is slightly on the side of the deep.

Nevertheless, the mapping between refrigeration and heat pumping holds well enough so that we can, from now on, focus exclusively on refrigeration to assess which temperature differences $\dT$ allow for a dot operation with reasonable performance $\eta$. First, we compare the $\dT$-dependencies of heat, work and performance coefficient for repulsive interaction [\Figs{fig_thermo_fct_epsilon}(e,g)] to the ones for attractive interaction [\Figs{fig_thermo_fct_epsilon}(f,h)]. This mainly reveals that at the individually chosen working points $\beps$ (blue stars in \Figs{fig_thermo_fct_epsilon}(a-d)), the attractive system has a sizably larger operation range for $\dT/\TN$. This is because the above mentioned suppression of the charge rate $\gamc$ for $U < 0$ also suppresses the stationary particle current $\INR^{(0)} \sim \gamceq$ [\Eq{I_N0}]. The detrimental, zeroth order steady-state heat contribution $\QR^{(0)} \sim \INR^{(0)} \sim \gamceq$ [\Eq{energy_pumping}] is therefore significantly smaller than the geometrically pumped heat $\QR^{(1)}$, even for sizable temperature differences $\dT$. Finite interaction, $|U| > 0$, has however always a detrimental effect on the cooling power due to the leakage heat current, that is the $-\kappa\dT$ term in Eq.~\eqref{energy_pumping} since the Fourier heat $\kappa$ is proportional to $U^2$ [Eq.~\eqref{eq_fourier_heat}]\footnote{Otherwise, the performances of the non-interacting case, $U = 0$, are in between the repulsive and attractive cases, e.g. for the efficiency or the maximum $\dT$ at which the refrigerator can be operated.}.
We also note that the compact analytical expression \eqref{eta_lim} (dotted green lines in \Figs{fig_thermo_fct_epsilon}(g,h)) accurately gives the efficiency in the limit of vanishing $\dT$, even for $U < 0$, though it was derived in the adiabatic-response regime and in the limit of small driving amplitudes. This therefore confirms that the device performance can be easily assessed from the system parameters.

Next, we find a $\dT$-dependent trade-off between performance $\eta$ and cooling power $\sim\QR\Omega$. This is highlighted in \Fig{fig_thermo_trade_off}, where both $\eta$ and $\QR$ are plotted parametrically as a function of the working point $\beps$. For very small $\dT/\TN$, the maximum performance $\eta$ is achieved at zero power whereas maximum power is reached at very small $\eta/\etaC$ [\Fig{fig_thermo_trade_off}(a)]. Conversely, for $\dT/\TN$ maximizing $\eta/\etaC$ at the operating point $\beps$ indicated by the blue stars in \Figs{fig_thermo_fct_epsilon}(a-d), this trade-off is much less pronounced for $U < 0$ and almost vanishes $U > 0$. Maximum power in this case almost coincides with the maximal coefficient of performance [\Fig{fig_thermo_trade_off}(b)]. Also note that sweeping $\beps$ for $U > 0$ always yields two separate $(\QR,\eta)$-curves, corresponding to the two disconnected $\beps$-ranges allowing for refrigeration as shown in \Figs{fig_thermo_fct_epsilon}(a,c); for $U < 0$, the single requirement $\beps < -U/2$ [\Figs{fig_thermo_fct_epsilon}(b,d)] analogously yields only one connected $(\QR,\eta)$-curve.

We finally emphasize that, despite the finite driving speed, we find the system to operate at finite power with more than $15\%$ of the Carnot efficiency. Furthermore, for $U > 0$ and temperature differences $\dT/\TN \sim 10^{-3}$, the device can even be operated at maximum power while keeping an $\eta$ close to the achievable maximum. This is particularly unexpected as single quantum dots are typically disregarded in favor of double-dot devices that allow for a better decoupling from the baths~\cite{Juergens2013Jun}. Quantum dots with a repulsive interaction $U \approx 5$ meV, like in Ref.~\cite{Josefsson2018Oct}, or an attractive interaction $U \approx - 0.2$ meV, like in Ref.~\cite{Prawiroatmodjo2017Aug}, would make it possible to pump heat against a tiny but finite temperature difference $\dT$ of the order of mK. Devices with larger onsite interaction strengths would allow operating at higher $\dT$.

\section{Conclusion}
We have analyzed geometric energy transport through strongly interacting, weakly-coupled quantum dots due to the slow driving of a pair of system parameters. Based on a master equation analysis combined with a fermionic duality relation, we have provided compact and intuitive analytical results linking the pumped energy to the well-known steady-state thermoelectric properties of the dot. 

We have explicitly worked out that energy pumping---in contrast to charge pumping---is the result of two different decay modes of the dot being excited due to the time-dependent driving: the charge mode, typically contributing via a "tight-coupling" term proportional to the charge current via the well-known stationary Seebeck coefficient, and the parity mode, uniquely due to the onsite Coulomb interaction. Even though the energy current is more directly susceptible to the Coulomb energy, slow, adiabatic energy pumping is typically still governed by the charge mode, even for strong local interaction. The parity mode, however, does become important if the external bias exceeds the Coulomb energy, and then ---as a result of fermionic duality--- yields a contribution symmetric in the interaction sign, i.e.,  with respect to repulsive and attractive interaction.

Using the obtained analytical results, we have identified the symmetries of the pumped energy with respect to the different (time-averaged) system parameters, and we have analyzed in detail all possible pumping schemes and pumping mechanisms. In particular, we have found a number of pumping mechanisms unique to either attractive or repulsive interaction, and pumping mechanisms that are not possible for charge pumping, but only for energy pumping. The latter emerge whenever only one or even none of the quantum dot energy levels are in resonance with the contact chemical potentials.

We have furthermore demonstrated the time-dependently driven quantum dot to be operable as a heat pump or refrigerator between biased contacts. We have derived a simple, yet insightful analytical expression for the efficiency in the limit of vanishing temperature, linking the driving characteristics to the well-known linear-response Seebeck coefficient, the equilibrium charge fluctuations, and the dot's RC time. This reveals in particular how the performance is decisively affected by the sign of the onsite interaction ---in terms of limiting driving speed, output power and efficiency.

A perspective of high interest for the field of stochastic and quantum (thermo)dynamics would be to extend this study to the statistics of energy pumping~\cite{Ren2010Apr,Riwar2021Jan}. This would yield generalized fluctuation relations for this type of interacting, fermionic, cyclic machines.

\acknowledgments
We especially thank Maarten\,R.~Wegewijs for continuous discussions. We also thank Liliana Arrachea and Valentin Bruch for helpful comments on the manuscript. We acknowledge funding from the Swedish Vetenskapsr\r{a}det and from the Knut and Alice Wallenberg foundation through the fellowship program (J. M. and J. Sp.) and from the European
Union’s H2020 research and innovation program under
grant agreement No. 862683 (J.Sp.). J. Sc. was supported by the Danish National Research Foundation and the Danish Council for Independent Research | Natural Sciences.

\appendix

\section{Time-dependent electrochemical potentials and temperatures}\label{app_macro_timedep}

Adiabatic pumping due to the driving of the reservoir potentials $\mua$ and temperatures $\Ta$ was previously studied in Refs.~\cite{Hasegawa2017Jan,Hasegawa2018Mar,Bhandari2021Jul}. These quantities are, however, macroscopic observables characterizing the (local- and time-local) equilibrium \emph{density operators} of the leads $\rho_\alpha = \rho_{\teq}(\mua,\Ta)$, and do not directly enter their Hamiltonian $H_\alpha$. Hence, including the effect of lead-potential- and temperature driving on the state of a tunnel-coupled quantum dot requires extra care. 

More specifically, the here employed master equation approach, as detailed in \Sec{sec_master_equation}, is based on the Born-Markov approximation. This implies that the only dynamics induced into the leads are due to the weakly coupled dot, causing only a negligible deviation from local equilibrium. The resulting dot dynamics after tracing out the leads then only depends on the lead potentials $\mua(t_0)$ and temperatures $\Ta(t_0)$ at some initial time $t_0$ at which dot and lead state can be assumed fully uncorrelated, $\hat{\rho}_\text{tot}(t_0) = \hat{\rho}(t_0)\cdot\prod_\alpha \hat{\rho}_\alpha(t_0)$. However, potential and temperature the choice of all other parameterspractice typically means that the lead state is also manipulated by a combination of external fields and, crucially, coupling to thermal superbaths exchanging particles and heat. A rigorous treatment would therefore consider the combined dot-lead system as an open subsystem, and trace out the external superbaths which are kept at constant temperatures and potentials. Since this seems neither analytically nor numerically feasible, we here capture the lead-superbath interaction in a simplified fashion.

On time scales much larger than the thermalization time in the leads, time-dependent $\mua(t),\Ta(t)$ can often be mimicked by instead shifting and scaling the single-particle lead spectra~\cite{Luttinger1964Sep,Eich2014Sep,Tatara2015May}:
\begin{equation}
    \hat H_\alpha \rightarrow H_\alpha(t) = c_\alpha(t)\hat H_\alpha^{0}+v_\alpha(t)\hat{N}_\alpha,
\end{equation}
with $c_\alpha(t_0) = 1$ as well as $v_\alpha(t_0) = 0$ and where $\hat H_\alpha^0$ denotes the lead Hamiltonian from \Eq{H_lead}. The corresponding Born-Markov master equation of the weakly coupled dot in the slow-driving regime then becomes equal to the master equation for constant lead parameters, except for the parameter replacement  
\begin{subequations}
    \label{eq_time_dependent_potential_temperature}
    \begin{align}
        \Ta \rightarrow \Ta(t) & =c_\alpha(t)\bar{T}_\alpha\\
        \mua \rightarrow \mua(t) & =c_\alpha(t)\bar{\mu}_\alpha+v_\alpha(t).
    \end{align}
\end{subequations}
This paper employs this replacement scheme, since it can in fact be physically motivated for the here relevant, weakly coupled Markovian leads subject to slow driving.
The key point for this is the large time-scale separation between the lead thermalization time $t_\text{therm}$ and both driving period as well as typical tunneling time: $t_\text{therm}\Omega \ll t_\text{therm}\Gamma \ll 1$. Namely, we may assume that instances of superbath-lead interactions take place at separate times $t_i$. With this, we ``slice" the full driving period into time intervals $[t_i,t_{i+1})$ for which  we assume $\Gamma t_\text{therm} \ll 1 \ll \Gamma\Delta t = \Gamma(t_{i+1} - t_i) \ll \Gamma/\Omega$. During each interval, we can approximate temperatures and potentials as constant, $\Ta(t) = \Ta(t_{i}), \mua(t) = \mua(t_{i})$, while the external superbaths effectively decouple from the leads. Due to the very fast lead rethermalization, the superbath-lead interaction should result in small yet quick changes of temperatures, $\Ta(t_i) \rightarrow \Ta(t_{i+1}) = \Ta(t_i) + \dT_{\alpha,i}$, and potentials, $\mua(t_i) \rightarrow \mua(t_{i+1}) = \mua(t_i) + \delta \mu_{\alpha,i}$, and should furthermore suppress any residual dot-lead correlation that may have built up during the interval $[t_i,t_{i+1})$. Since $\Gamma\Delta t \gg 1$ while $\Omega/\Gamma \ll 1$, this means that each time interval can be treated separately using the Born-Markov master equation as in \Sec{sec_master_equation}: dot and leads are uncorrelated at time $t_i$, the initial dot state is given either by the prepared initial state $\rho(t_i = 0) = \rho_0$ or by the solution to the master equation in the previous interval $\rho(t_i) = \rho(t_{i-1} + \Delta t)$, and the lead states at time $t_i$ are characterized by local equilibrium states with respect to $\mua(t_i)$ and $\Ta(t_i)$. Stitching all time intervals together and using $\Omega\Delta t \ll 1$, we find that on the time scale of the long driving period, the dot dynamics $\rho(t)$ obeys the single Born-Markov master equation \eq{eq_master} with the continuous parametric time-dependence  \eq{eq_time_dependent_potential_temperature}.

Interestingly, while this paper studies the driving effect on particle current $\INa$ and electronic \emph{energy} current $\IHa$, as defined below in \Eq{eq_obs_current}, the above time-sliced picture also suggests a generalization to \emph{heat currents}: For a weakly tunnel coupled dot and Markovian leads remaining approximately in local equilibrium at potential $\mua$, the stationary heat current $\IQa$ is well approximated by the \emph{excess energy current} with respect to the chemical potential: $\IQa = \IHa - \mua \INa$. The lead states within each time slice $[t_i,t_{i+1})$ can also be approximated by such local equilibrium states when setting the proper time-dependent lead parameters $\mua = \mua(t_i),\Ta = \Ta(t_i)$. Hence, on the time scale of the driving, the heat current generalizes to $\IQa(t) = \IHa(t) - \mua(t) \INa(t)$. This is the reasoning behind our claim from \Sec{sec_refrigerator} that for a time-independent $\mua = 0$, the heat current is equal to the energy current.

\section{Analytical expressions for duality quantities}\label{app_duality}
We first provide the fermionic duality relation for the kernel $W$ of the dissipative evolution due to Markovian, weakly tunnel coupled baths in the wideband limit. For the spin-degenerate dot, this duality reads~\cite{Schulenborg2016Feb,Schulenborg2018Dec}
\begin{equation}
    W + \cP\left[W^{\inv}\right]^{\sudag}\cP = -2\Gamma\mathcal{I},\label{eq_duality_relation}
\end{equation}
where $W^{\inv}$ is the kernel corresponding to the dual model, $\Gamma = \GamL + \GamR$ is the above defined, lead-summed coupling strength per spin, the bold $\sudag$ indicates Hermitian conjugation in Liouville space with respect to the Hilbert Schmidt scalar product, $\cP\bullet = \fpOp\bullet = \cP^{\sudag}\bullet$ is the Liouville-Hermitian superoperator of applying the fermion-parity operator $\fpOp$ from the left to some operator $\bullet$, and $\mathcal{I}$ is the Liouville space identity. We furthermore introduce the duality parameter transform 
\begin{equation}
    \cD:(\epsilon,U,\Vb,\Ta,\Lambda)\rightarrow (-\epsilon,-U,-\Vb,\Ta,\Lambda)\label{eq_duality_transform}
\end{equation}
to write $W^{\inv}\bullet = \cD W\cD\bullet$. The Liouville-space trace relation $\Tr W = -\Gamma\Tr\mathcal{I}$ then enables us, in analogy to how, e.g., \Ref{Prosen2012Aug} treats open-system symmetries, to formulate \Eq{eq_duality_relation} as an anti-Hermiticity relation
\begin{equation}
    \left[\cP\cD\tilde{W}\right]^{\sudag} = -\cP\cD\tilde{W}\label{eq_duality_relation_traceless}
\end{equation}
for the traceless part of the kernel $\tilde{W} = W + \Gamma\mathcal{I}$ with respect to a Hilbert-Schmidt product with modified left(dual) vectors $\Bra{x} \rightarrow \Bra{x}\cP\cD$. This links right/left eigenvectors of the inverted kernel $W^{\inv}$ to left/right eigenvectors of the original $W$. Namely, starting from the expressions in \Sec{sec_master_equation}, the right eigenvectors of $W^{\inv}$ are
\begin{gather}
    \Ket{\zin} = \frac{1}{\Gamma\gamc^\inv}\left[f_\epsilon^+f_U^+\Ket{0}+f^-_\epsilon f^+_U\Ket{1}+f^-_\epsilon f^-_U\Ket{2} \right]\label{eq_eigensystem_dual}\\
    \Ket{p^\inv} = \Ket{p} \quad,\quad \Ket{c^\inv} = \frac{1}{2}(-\hat\one)^{\hat{N}}\left[\Ket{N}-\nz\Ket{\one}\right],\notag
\end{gather}
and the left ones read
\begin{gather}
    \Bra{(\zin)'} = \Bra{\one}\quad,\quad\Bra{(p^\inv)'} = \Bra{(-\one)^{N} z}\notag\\
    \Bra{(c^\inv)'} = \Bra{N}-\nz\Bra{\one},\label{eq_eigensystem_inv_left}
\end{gather}
with the charge relaxation rate $\gamc^\inv = f_\epsilon^- + f_U^+$.

The expectation value of the quantum dot occupation and parity in the stationary state are given by
\begin{align}
    \nz &= \Braket{N}{z} = \frac{2f_\epsilon^+}{f_\epsilon^+ + f_U^-},\notag\\
    p_z &= \Braket{(-\one)^N}{z} = \frac{f_\epsilon^-f_U^- + f_\epsilon^+f_U^+ -2f_\epsilon^+f_U^- }{\Gamma(f_\epsilon^+ + f_U^-)},
\end{align}
for the original model, and
\begin{align}
    \nzi &= \Braket{N}{\zin} = \frac{2f_\epsilon^-}{f_\epsilon^- + f_U^+},\notag\\
    p_z^\inv &= \Braket{(-\one)^N}{\zin} = \frac{f_\epsilon^-f_U^- + f_\epsilon^+f_U^+ -2f_\epsilon^-f_U^+ }{\Gamma(f_\epsilon^- + f_U^+)},
\end{align}
for the dual one.

\section{Proof of the symmetries}\label{app_symmetry}

This appendix proves the particle-hole transform related symmetries of the pumping curvatures $\BHa^{\P/\C}(\bR)$ discussed in \Sec{sec_symmetry}, and the duality-related symmetry of the parity component $\BHaP(\bR)$ under inversion of the interaction $U \rightarrow -U$ identified in \Sec{sec_results_attractive}.

\subsection{Particle-hole transform}\label{app_symmetry_ph}

To prove the symmetries under the particle-hole parameter transform
\begin{equation}
    \cT:(\epsilon,U,\Vb,\Ta,\Lambda)\rightarrow (-\epsilon-U,U,-\Vb,\Ta,\Lambda),\label{eq_ph_transform}
\end{equation}
we first need expressions for its action on the rates and eigenvectors of the master equation kernel $W$ given in \Sec{sec_master_equation}.\footnote{A complete transform would also act on the initially prepared dot state. Since $\cT$ has no effect on this initial dot state, our analysis is only valid once all transient dynamics have decayed, which we, however, assume throughout the paper.} The explicit expression of the stationary state $\Ket{z}$ in \Eq{eq_eigensystem_right}, and of its dual $\Ket{z^{\inv}}$ in \Eq{eq_eigensystem_dual}, make it straightforward to verify
\begin{equation}
    \cT\Ket{z} = \pH\Ket{z} \quad,\quad \cT\Ket{\zin} = \pH\Ket{\zin}\label{eq_ph_z}
\end{equation}
with the superhermitian and unitary particle-hole transform superoperator $\pH = \pH^{\boldsymbol{\dagger}} = \pH^{-1}$ acting on the 3 spin-symmetric dot basis states $\Ket{0} = \ket{0}\bra{0}$, $\Ket{1} = \frac{1}{2}\sum_{i=\uparrow,\downarrow}\ket{i}\bra{i}$, $\Ket{2} = \ket{2}\bra{2}$ as
\begin{equation}
    \pH\Ket{0} = \Ket{2} \quad,\quad \pH\Ket{2} = \Ket{0}\quad,\quad \pH\Ket{1} = \Ket{1}.
\end{equation}
Relations analogous to \Eq{eq_ph_z} also hold for the lead$(\alpha)$-resolved states:
\begin{equation}
    \cT\Ket{\za} = \pH\Ket{\za} \quad,\quad \cT\Ket{\zia} = \pH\Ket{\zia}.\label{eq_ph_z_alpha}
\end{equation}
Furthermore, the explicit expressions for the charge and parity rate $\gamc$ and $\gamp$ given in \Sec{sec_master_equation}, as well as their lead-resolved counterparts $\gamca$ and $\gampa$, can be used to show that these rates are $\cT$-symmetric,
\begin{equation}
    \cT\gamcp = \gamcp \quad,\quad \cT\gamcpa = \gamcpa.\label{eq_ph_rates}
\end{equation}
Equation \eq{eq_ph_z} and $\Bra{N}\pH = 2\Bra{\one} - \Bra{N}$ imply that $\cT$ acts also as particle-hole transform of the dot occupation number, $\cT \nz = \Bra{N}\cT\Ket{z} = \Bra{N}\pH\Ket{z} = 2 - \nz$. Summarized with analogous relations for the lead-resolved charges and all duals, we find  
\begin{align}
    \cT \nz = 2 - \nz \quad&,\quad \cT \nza = 2 - \nza\notag\\
    \cT \nzi = 2 - \nzi \quad&,\quad \cT \nzia = 2 - \nzia.\label{eq_ph_occupation}
\end{align}
Equations \eq{eq_ph_z} and \eq{eq_ph_occupation} together with the eigenvector expressions from \Sec{sec_master_equation} and the $\cT$-invariance of parameter-independent operators yield the $\cT$-action on all left and right eigenvectors of the kernels $W$ and $\Wa$:
\begin{align}
    \cT\Ket{z} = \pH\Ket{z} \quad&,\quad \cT\Bra{\one} = \Bra{\one} = \Bra{\one}\pH\notag\\
    \cT\Ket{c} = -\pH\Ket{c} \quad&,\quad \cT\Bra{c'} = -\Bra{c'}\pH\notag\\
    \cT\Ket{p} = \Ket{p} = \pH\Ket{p} \quad&,\quad \cT\Bra{p'} = \Bra{p'}\pH,\label{eq_ph_eigenvectors}  
\end{align}
and, equivalently, for the lead$(\alpha)$-resolved versions, where we have used that the unit operator $\one$ and the fermion parity $\fpOp$ are invariant under the particle-hole transform $\pH$. The rate symmetries \eq{eq_ph_rates} and the eigenvector relations \eq{eq_ph_eigenvectors} combine to the key symmetry relations
\begin{align}
    \cT W = \pH W \pH  \quad&,\quad \cT \Wa = \pH \Wa \pH\notag\\
    \cT \frac{1}{\tW} = \pH \frac{1}{\tW} \pH  \quad&,\quad \cT \frac{1}{\tWa} = \pH \frac{1}{\tWa} \pH.\label{eq_ph_kernels}
\end{align}
Together with \Eq{eq_ph_z}, this rigorously proves our statement from \Sec{sec_symmetry} that $\cT$ effectively implements a particle-hole transform of the entire system evolution and its steady state. Apart from these kernel relations \eq{eq_ph_kernels}, the only remaining ingredient required for our proof is the commutation relation of $\cT$ and the parameter gradient $\nr = (\partial_\epsilon,\partial_U,\partial_{\Vb},\partial \Ta, \partial_\Lambda)^T$. By the very definition of $\cT$ in \Eq{eq_ph_transform}, we have
\begin{equation}
    \cT\nr = \tilde{\sigma}\nr\cT \quad,\quad \tilde{\sigma} = \begin{pmatrix}-1 & 0 & 0 & 0_{1\times 2} \\ -1 & 1 & 0 & 0_{1\times 2} \\ 0 & 0 & -1 & 0_{1\times 2} \\ 0_{2\times 1} & 0_{2\times 1} & 0_{2\times 1} & 1_{2\times 2} \end{pmatrix}.\label{eq_ph_gradient}
\end{equation}
The diagonal elements of $\tilde{\sigma}$ correspond to the signs introduced by $\cT$, whereas the single off-diagonal element in the second row stems from the additional $U$ shift of $\epsilon$. Note that with respect to the main text, we here introduced a five-dimensional gradient $\nr$. Whenever considering a concrete driving scheme or calculating curvatures from cross-products, the relevant subspace needs to be considered, both for $\nr$ and $\tilde{\sigma}$.

The most general form of the particle-hole symmetry relations discussed in \Sec{sec_symmetry} now follows straightforwardly from the definition of the lead-resolved geometric connections in \Eq{eq_obs_connection} and the corresponding pumping curvatures \eq{B_O,alpha}. Namely
\begin{align}
    \cT\AOa(\bR) &= \tilde{\sigma} \Bra{\cT\cO}[\cT \Wa]\left[\cT\frac{1}{\tW}\right]\nr\cT\Ket{z}\notag\\
    &= \tilde{\sigma}\Bra{\cT\cO}\pH \Wa \pH\pH \frac{1}{\tW} \pH\pH \nr\Ket{z}\notag\\
    &= \tilde{\sigma}\mathbf{A}_{\left[\pH\cT\cO\right],\alpha}(\bR).\label{eq_ph_geometric}
\end{align}
For the two observables $\hat{\cO}$ discussed in this paper---energy $\hat{H}$ and charge $\hat{N}$---we have
\begin{equation}
    \pH\cT \hat{H} = \hat{H} - (2\epsilon + U)\one \quad,\quad \pH\cT \hat{N} = 2\cdot\one - \hat{N},
\end{equation}
which, when using $\Bra{\one}\Wa = 0$ in \Eq{eq_obs_connection}, gives
\begin{align}
    \cT\AHa(\bR) &= +\tilde{\sigma}\AHa(\bR) \notag\\
    \cT\ANa(\bR) &= -\tilde{\sigma}\ANa(\bR).\label{eq_ph_geometric_charge_energy}
\end{align}
The only difference between energy and charge under action of $\cT$ is the relative minus sign. This formally explains why all particle-hole symmetry relations for the energy pumping curvatures are the same as for charge pumping except for an overall sign. 

Finally, the concrete form of the symmetry relations for the pumping curvatures readily follows from the definition \eq{B_O,alpha}:
\begin{align}
    \cT\BOa(\bR) &= \left[\tilde{\sigma}\nr\right] \times \cT\AOa(\bR)\notag\\
    &= \left[\tilde{\sigma}\nr\right] \times \left[\tilde{\sigma}\mathbf{A}_{\left[\pH\cT\cO\right],\alpha}(\bR)\right].\label{eq_ph_curvature}
\end{align}
The (anti-)symmetries for $\hat{\cO} = \hat{H}$ stated in \Sec{sec_symmetry} are simply the respective, non-zero components of \Eq{eq_ph_curvature} for the individual two-parameter driving schemes. The diagonal elements of $\tilde{\sigma}$ provide the parameter-dependent symmetry signs $\sigma(R_i)$; the special form \eq{B_eps_B_U} for $U$-driving arises from the off-diagonal element of $\tilde{\sigma}$ and from the properties of the cross-product. To see that equations \eq{eq_ph_geometric} and \eq{eq_ph_curvature} also hold individually for the charge and parity component, we can simply repeat the steps in \Eq{eq_ph_geometric} with the respective projectors $\Ket{c}\Bra{c'}$ and $\Ket{p}\Bra{p'}$ inserted precisely to the left of the gradient $\nr$. Since \Eq{eq_ph_eigenvectors} separately ensures $\cT\left[\Ket{c}\Bra{c'}\right] = \pH\Ket{c}\Bra{c'}\pH$ and $\cT\left[\Ket{p}\Bra{p'}\right] = \pH\Ket{p}\Bra{p'}\pH$, the computation proceeds analogously:
\begin{align}
    \cT\AOa^{\C/\P}(\bR) &= \tilde{\sigma}\mathbf{A}^{\C/\P}_{\left[\pH\cT\cO\right],\alpha}(\bR)\notag\\
    \cT\AHa^{\C/\P}(\bR) &= +\tilde{\sigma}\AHa^{\C/\P}(\bR) \notag\\
    \cT\ANa^{\C/\P}(\bR) &= -\tilde{\sigma}\ANa^{\C/\P}(\bR) \notag\\
    \cT\BOa^{\C/\P}(\bR) &= \left[\tilde{\sigma}\nr\right] \times \left[\tilde{\sigma}\mathbf{A}^{\C/\P}_{\left[\pH\cT\cO\right],\alpha}(\bR)\right].\label{eq_ph_component}
\end{align}
This completes our proof of the symmetries related to the particle-hole parameter transform $\cT$.

\subsection{Interaction-sign symmetric parity component}\label{app_symmetry_dual}

We can prove the $U$-symmetric behavior of the parity-like energy pumping curvature $\BHaP(\bR)$ highlighted in \Sec{sec_results_attractive} by studying the action of the duality parameter transform \eq{eq_duality_transform} on the ingredients of $\BHaP(\bR)$. First, it is clear by definition that
\begin{equation}
    \cD\Ket{z} = \Ket{\zin} \quad,\quad \cD\Ket{\zin} = \Ket{z}.\label{eq_duality_z}
\end{equation}
It is furthermore evident from \Eq{eq_duality_transform} that
\begin{equation}
    \cD\gamp = \gamp \quad,\quad \cD\fpOp = \fpOp \quad,\quad \cD U = -U,\label{eq_duality_misc}
\end{equation}
and that the commutation relation between $\cD$ and the gradient $\nr$ reads
\begin{equation}
    \cD\nr = \tilde{\sigma}_{\cD}\nr\cD \quad,\quad \tilde{\sigma}_{\cD} = \text{diag}(-1,-1,-1,1,1).\label{eq_duality_gradient}
\end{equation}
Equations \eq{eq_duality_z}-\eq{eq_duality_gradient} together show that the action of $\cD$ on the parity-like excitation $\bxP$ as introduced in the first line of \Eq{x_p} is given by
\begin{align}
    \cD\bxP &= -\frac{1}{\cD\gamp}\Bra{\cD \zin\fpOp}\cD\nr\Ket{z}\notag\\
    &= -\tilde{\sigma}_{\cD}\frac{1}{\gamp}\Bra{z\fpOp}\nr\Ket{\zin} = -\tilde{\sigma}_{\cD}\bxP.\label{eq_duality_excitation}
\end{align}
The last line in \Eq{eq_duality_excitation} follows from \Eq{x_p_anti}, with the key difference being the matrix prefactor $\tilde{\sigma}_{\cD}$. This factor signifies that when applying the dual transform \emph{after} taking the gradient $\nr$, $\bxP$ is not anymore anti-symmetric in all components, but attains an additional sign according to \Eq{eq_duality_gradient}. This likewise translates to the parity component $\AHaP(\bR)$ of the geometric connection given in \Eq{eq_A_p_overlap},
\begin{equation}
    \cD\AHaP(\bR) = U\gampa\cD\bxP = +\tilde{\sigma}_{\cD}\AHaP (\bR)\label{eq_duality_geometric},
\end{equation}
and to the corresponding pumping curvature,
\begin{align}
    \cD\BHaP(\bR) &= \tilde{\sigma}_{\cD}\nr\times \cD\AHaP(\bR)\notag\\
    &= \left[\tilde{\sigma}_{\cD}\nr\right]\times \left[\tilde{\sigma}_{\cD}\AHaP(\bR)\right].\label{eq_duality_curvature}
\end{align}

To show the symmetry of $\BHaP(\bR)$ with respect to the sign of a \emph{time-independent} interaction $U$, we now apply the particle hole parameter transform $\cT$ defined in \Eq{eq_ph_transform} to \Eq{eq_duality_curvature} from the left and shift it through to the right. Using \Eq{eq_ph_gradient} and \Eq{eq_ph_component}, this gives
\begin{align}
    \left.\left.\BHaP(\bR)\right|_{\beps \rightarrow \beps - \bU}\right|_{\bU\rightarrow -\bU} &= \cT\cD\BHaP(\bR) \label{eq_symmetry_curvature}\\
    &= \left[\tilde{\sigma}_{\cD}\tilde{\sigma}\nr\right]\times\left[\tilde{\sigma}_{\cD}\tilde{\sigma}\AHaP(\bR)\right]\notag
\end{align}
with
\begin{equation}
    \tilde{\sigma}_{\cD}\tilde{\sigma} = \tilde{\sigma}\tilde{\sigma}_{\cD} = \begin{pmatrix}1 & 0 & 0 & 0_{1\times 2} \\ 1 & -1 & 0 & 0_{1\times 2} \\ 0 & 0 & 1 & 0_{1\times 2} \\ 0_{2\times 1} & 0_{2\times 1} & 0_{2\times 1} & 1_{2\times 2} \end{pmatrix}.\label{eq_symmetry_matrix}
\end{equation}
As long as $U$ is not driven, the second row and column of $\tilde{\sigma}_{\cD}\tilde{\sigma}$ are excluded, the matrix $\tilde{\sigma}_{\cD}\tilde{\sigma}$ becomes a $4\times4$ identity, and the interaction-inversion symmetry 
\begin{equation}
    \BHaP(\bR) = \left.\left.\BHaP(\bR)\right|_{\beps \rightarrow \beps - \bU}\right|_{\bU\rightarrow -\bU} \text{ if } \partial_t U = 0\label{eq_symmetry_bP}
\end{equation}
follows. In other words, the parity-like curvature $\BHaP$ for \emph{constant} attractive interaction $\bU < 0$ is equal to $\BHaP$ evaluated for repulsive interaction $-\bU > 0$ and a working point $\beps$ shifted by $\bU = -|\bU|$. A comparison of the small panels in \Fig{fig_pumping_curvature_repulsive_R} to the corresponding panels in \Fig{fig_pumping curvature attractive R} indeed confirms this. 
Beyond this, \Eq{eq_symmetry_curvature} also reveals the behavior with $U$ as one driving parameter:
\begin{align}
    \BHaP(\{U,R_2\}) &= -\left.\left.\BHaP(\{U,R_2\})\right|_{\beps \rightarrow \beps - \bU}\right|_{\bU\rightarrow -\bU} \notag\\
    &\phantom{=} +\BHaP(\{\epsilon,R_2\}).\label{eq_symmetry_advanced}
\end{align}
For the special case $R_1 = \epsilon, R_2 = U$, we find
\begin{equation}
    \BHaP(\{\epsilon,U\}) = -\left.\left.\BHaP(\{\epsilon,U\})\right|_{\beps \rightarrow \beps - \bU}\right|_{\bU\rightarrow -\bU},\label{eq_symmetry_bP_U}
\end{equation}
as can be seen by comparing \Fig{fig_pumping_curvature_repulsive_R}(i) to \Fig{fig_pumping curvature attractive R}(i). Altogether, equations \eq{eq_symmetry_bP} and \eq{eq_symmetry_bP_U} are explicit, analytical manifestations of the strong similarities between energy pumping for attractive and repulsive interaction observed at large bias $|\Vb| \geq |U|$.

\section{Charge pumping}\label{app_charge}

\begin{figure*}[h!]
    \includegraphics{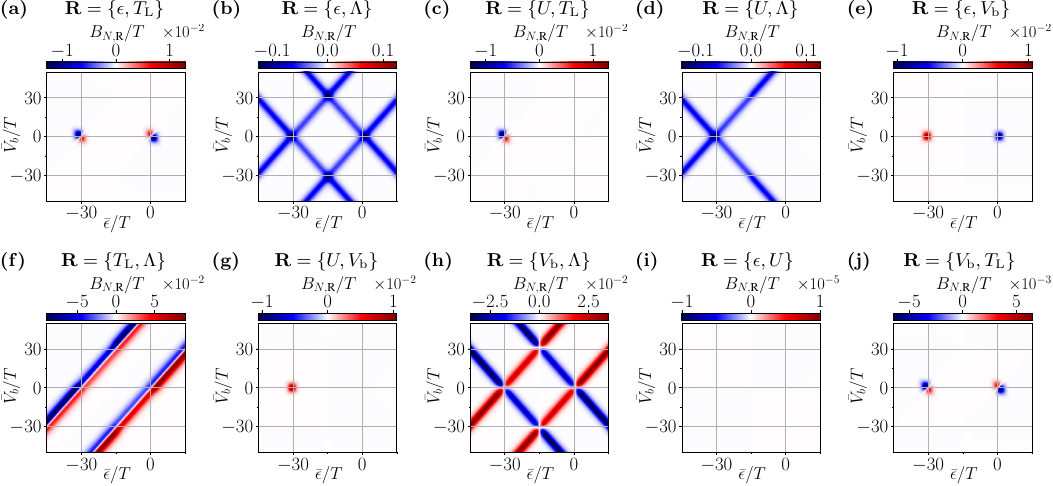}
    \captionof{figure}{
        \label{fig_charge_pumping_curvature_repulsive_R}
        Pumping curvatures $\BNR$ for a quantum dot with repulsive onsite interaction, $U>0$, in the $(\beps, \bVb)$-space for all possible two-parameter driving schemes $\bR = \{R_1, R_2\}$. We have plotted the component of $\BNR$ perpendicular to the driving plane defined by $(R_1, R_2)$. The way the driving is done  and the choice of all other parameters are the same as in \Fig{fig_pumping_curvature_repulsive_R}.
    }
\end{figure*}

\begin{figure*}[h!]
    \includegraphics{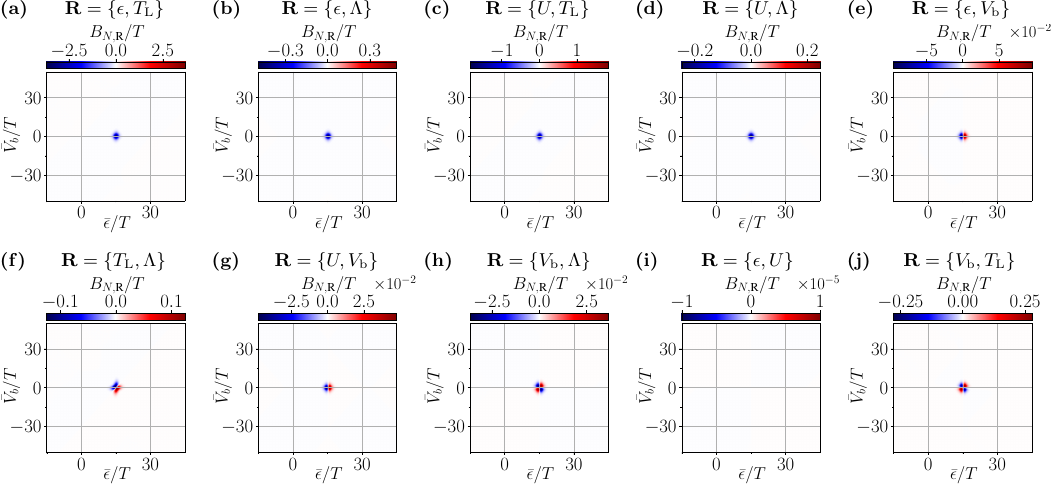}    
    \captionof{figure}{
        \label{fig_charge_pumping_curvature_attractive_R}
        Pumping curvatures $\BNR$ for a quantum dot with attractive onsite interaction, $U<0$, in the $(\beps, \bVb)$-space for all possible two-parameter driving schemes $\bR = \{R_1, R_2\}$. We have plotted the component of $\BNR$ perpendicular to the driving plane defined by $(R_1, R_2)$. The way the driving is done and the choice of all other parameters are the same as in \Fig{fig_pumping_curvature_repulsive_R}.
    }
\end{figure*}

Geometrical charge pumping for both a repulsive and attractive interaction has already been studied in Ref.~\cite{Reckermann2010Jun,Calvo2012Dec,Placke2018Aug}. Nevertheless, to make the comparison with our results for energy pumping in \Sec{sec_results} easier, we also plot the pumping curvatures $\BNR(\bR)$ in \Fig{fig_charge_pumping_curvature_repulsive_R} for $U > 0$ and in \Fig{fig_charge_pumping_curvature_attractive_R} for $U < 0$. There are no surfaces of finite pumping curvature due to off-resonant pumping (absence of mechanisms G and H). For $U > 0$, the most present mechanism is A, namely at a double resonance at zero bias voltage, while lines at single resonances can be obtained when $\Lambda$ is a driving parameter, see Sec.~\ref{sec_features_repulsive}. For $U < 0$, the only mechanism is C, see Sec.~\ref{sec_attractive_C} and Appendix \ref{app_two_particle_resonance_charge}.

\section{Quasi-stationary exchange with environment}\label{app_work}
We argue at several points in \Sec{sec_features_repulsive} that the charge pumping curvature $\bB_N=\sum_\alpha\BNa$ for the total, lead-summed system must vanish due to total charge conservation.\footnote{In the fully left-right symmetric case, the lead-resolved curvature $\BNa$ is proportional to $\bB_N$ and hence also vanishes.} Furthermore, we argue that this exact vanishing is not generally true for the energy pumping curvature $\bB_H$, as dot parameter driving does work on particles occupying the dot. This appendix derives both statements.

First of all, for any observable $\hat\cO$, the total \emph{stationary} current entering the dot is zero by definition: $I_\cO^{(0)} =\IOL^{(0)} + \IOR^{(0)} = \melt{\cO}{W}{z} = 0$. However, for the total, lead-summed first non-adiabatic correction, $I_\cO^{(1)} = \bA_{\cO} \cdot\partial_t \bR$, \Eq{eq_obs_connection} and $\sum_\alpha \Wa/\tW = 1 - \Ket{z}\Bra{\one}$ dictate
\begin{align}
    \bA_\cO =\Bra{\cO}\nr\Ket{z}.\label{eq_disp_one}
\end{align}
This means
\begin{align}
    \bB_\cO =\,& \left[\nr\Bra{\cO}\right]\times\left[\nr\Ket{z}\right].\label{eq_disp_two}
\end{align}
Since the occupation number $\hat\cO = \hat{N}$ is parameter independent, $\nr \hat{N} = 0$, the curvature vanishes, $\bB_N = 0$, expressing that in the stationary limit, total charge conservation prohibits any net charge build-up in the dot averaged over a driving cycle. However, for $\hat\cO = \hat H$, any derivative with respect to $\epsilon$ or $U$ is generally finite.
More explicitly, inserting a unit formed by the spin-degenerate basis operators $\Ket{\one}/2,\Ket{N - \one}/\sqrt{2},\Ket{\fpOp}/2$ into \Eqs{eq_disp_one}, \eq{eq_disp_two}, and using that $\Braket{H}{N - \one}/2 = \epsilon + U/2$ as well as $\Braket{H}{\fpOp}/4 = U/4$, we find
\begin{align}
    \bA_H &= \left[\epsilon + \frac{U}{2}\right]\nr \nz + \frac{U}{4} \nr p_z,\notag\\
    \bB_H &= \nr\left[\epsilon + \frac{U}{2}\right]\times \nr \nz + \frac{1}{4}\nr U \times \nr p_z.\label{eq_disp_three}
\end{align}
Hence, if $\epsilon$ or $U$ are among the driving parameters, there is generally a finite net work done on the dot that results in a net cycle-averaged energy flow into the leads, $\Delta H^{(1)} \neq 0$. This is, however, not the case for a completely unbiased environment, with fixed $\dT = \Vb = 0$: here, $\Ket{\za} = \Ket{z}$ means that a driven coupling asymmetry $\Lambda$ has no effect on $\Ket{z}$, and that $f^\pm_{\epsilon} \rightarrow f^\pm_\alpha(\epsilon)$ and $f^\pm_{U} \rightarrow f^\pm_\alpha(\epsilon + U)$ in \Eq{eq_eigensystem_right} implies for \Eq{eq_disp_three} that
\begin{equation}
    \bB_H \overset{\dT,\Vb\rightarrow 0}{\rightarrow} 0.
\end{equation}
This is confirmed by \Figs{fig_pumping curvature attractive R}(b,d,i) and shown approximately in \App{app_two_particle_resonance_energy} for $U < 0$, but it holds exactly and for any interaction sign. It is also consistent with the usual understanding that equilibrium heating is an effect of at least second order in the driving frequency~\cite{Juergens2013Jun}.

\section{Charge and energy pumping through a non-interacting dot}
\label{app_non_interacting}
\begin{figure}[htb!]
    \includegraphics{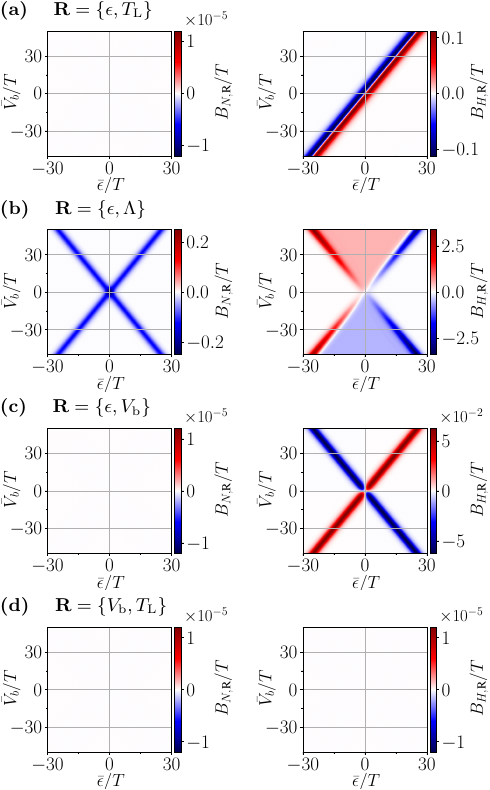}
    \caption{\label{fig_pumping_curvature_non_interacting_R}
        Pumping curvatures $\BNR$ (left column) and  $\BHR$ (right column) for a non-interacting quantum dot, namely $U = 0$, in the $(\beps, \bVb)$-space. The way the driving is done and the choice of all other parameters are the same as in \Fig{fig_pumping_curvature_repulsive_R}.
    }
\end{figure}

The charge and energy pumping curvatures, $\BNR$ and $\BHR$, for a non-interacting quantum dot, namely with $U = 0$, are plotted in \Fig{fig_pumping_curvature_non_interacting_R} for a few driving schemes. This figure evidences the existence of interaction induced pumping terms. For instance, the contribution of mechanism A to charge pumping for $\bR= \{\epsilon, \TL\}$ and $U > 0$ [\Fig{fig_charge_pumping_curvature_repulsive_R}(a)] is absent in \Fig{fig_pumping_curvature_non_interacting_R}(a). For energy pumping, mechanism B, which requires a double resonance [\Fig{fig_pumping mechanisms}], also vanishes in the absence of interaction (\Fig{fig_pumping_curvature_repulsive_R} compared to the bottom row of \Fig{fig_pumping_curvature_non_interacting_R}). Similarly, mechanism C for $U < 0$ [\Figs{fig_charge_pumping_curvature_attractive_R} and \ref{fig_pumping curvature attractive R}] is absent from  \Fig{fig_pumping_curvature_non_interacting_R}. However, when $\Lambda$ is one of the driving parameters, both charge and energy pumping are always possible, with and without interaction, like for instance in \Fig{fig_pumping_curvature_non_interacting_R}(b). Additionally, \Fig{fig_pumping_curvature_non_interacting_R}(a) and (c) are simple examples of energy pumping in the absence of charge pumping. 

\renewcommand{\arraystretch}{1}

\section{Analytical results for single-resonance pumping}\label{app_non_resonant}

In this appendix, we derive analytical pumping curvature expressions for repulsive interaction $U > 0$ and constant coupling asymmetry $\Lambda$ in the single-resonance case, that is, with one lead $\alpha'$ being far off-resonant. We start by noting that the kernel $\Wap$ for this non-resonant lead and all its derived quantities become independent of any driving parameter $R_i \neq \Lambda$, namely
\begin{equation}
    \pr[i] \Wap/\Gamma \sim \pr[i]\gamcap/\Gamma \sim \pr[i] \nzap \approx 0.\label{eq_non_resonant_1}
\end{equation}
Note that we cannot generally approximate the dual occupation $\pr[i] \nziap \approx 0$, since the latter changes at the resonances of the dual attractive system, different from the resonances in the repulsive system of interest. Using these results, we aim at expressing the pumping curvature in term of lead resolved quantities to make them easier to interpret and, in particular, show that $\BNR(\bR)$ vanishes.
Using that the total occupation number can be decomposed into the~\cite{Schulenborg2017Dec},
\begin{equation}
    \nz = \sum_{\alpha''}\frac{\gamcapp}{\gamc}\nzapp\label{eq_non_resonant_2},
\end{equation}
and that $\sum_{\alpha''}\gamcapp = \gamc$ means $\gamca/\gamc = 1 - \gamcap/\gamc$ for two leads, we can use \Eq{eq_non_resonant_1} to rewrite
\begin{align}
    \pr[i] \nz &\rightarrow \pr[i]\left(\frac{\gamca}{\gamc}\right)\left(\nza - \nzap\right) + \frac{\gamca}{\gamc}\pr[i] \nza.\label{eq_non_resonant_3}
\end{align}
Near a resonance of the repulsive dot, i.e., for parameters deviating from the particle-hole symmetric point, the lead-resolved charge rate $\gamca$ can be expressed~\cite{Schulenborg2018Dec} in terms of the lead-resolved coupling $\Gama$, stationary occupation $\nza$ and dual occupation $\nzia$:
\begin{equation}
    \gamca = \gampa\frac{\nzia - 1}{\nzia - \nza}.\label{eq_non_resonant_4}
\end{equation}
For $\pr[i]\gamcap \rightarrow 0$, the derivative of the charge rate ratio in \Eq{eq_non_resonant_3} thereby can be decomposed into
\begin{align}
    \pr[i]\left(\frac{\gamca}{\gamc}\right) &\rightarrow -\frac{\gamcap}{\gamc^2}\pr[i]\gamca\notag\\
    &= -\frac{\gamcap\gamca\gamcia}{\gampa\gamc^2}\frac{1}{\nzia - 1}\pr[i] \nzia\notag\\
    &\phantom{=} -\frac{\gamcap\gamca}{\gamc^2}\frac{1}{\nzia - \nza}\pr[i] \nza, \label{eq_non_resonant_5}
\end{align}
with $\gamcia = \gampa - \gamca$. For the charge pumping curvature of the resonant lead $\alpha$, using \Eqs{eq_non_resonant_3} and \eq{eq_non_resonant_5} gives
\begin{align}
    \BNa(\bR) &\rightarrow \nr\left(\frac{\gamca}{\gamc}\right)\times \left[\left(\nza - \nzap\right)\left(\nr\frac{\gamca}{\gamc}\right)\right.\notag\\
    &\phantom{\rightarrow \nr\left(\frac{\gamca}{\gamc}\right)\times \left[\right.} +\left.\frac{\gamca}{\gamc}\nr\nza\right]\notag\\
    &= \nr\left(\frac{\gamca}{\gamc}\right)\times \frac{\gamca}{\gamc}\nr\nza\label{eq_non_resonant_5_a}\\
    &= -\frac{\gamcap\gamca^2\gamcia}{\gampa\gamc^3}\frac{1}{\nzia - 1}\nr\nzia \times \nr\nza.\notag
\end{align}
This quantity vanishes in the single-resonance case because the finite gradient $\nr \nza$ near resonance is overcompensated by the nearly vanishing gradient of the \emph{inverted} dual charge, $\nr \nzia \approx 0$, whose inverted interaction causes its transition energies to be far away from resonance with $\mua$ for $U \gg \Ta > 0$.  

When driving temperature and bias voltage for constant $U,\Gama$, the above derivation of the vanishing curvature from the properties of the dual charges $\nzia,\nzi$ can also be extended to the energy current. To do so, we are going to collect all terms in $\nr \nz$ separately from terms in $\nr p_z$ in the geometric connection $\AHR(\bR)$, unlike in \Eq{decomposition_A}.
The key ingredient for this is that according to \Ref{Schulenborg2018Dec}, \Eq{eq_non_resonant_4} generalizes in a remarkably simple fashion to the full charge rate
\begin{equation}
    \gamc = \sum_\alpha\gamca = \gamp\frac{\nzi - 1}{\nzi - \nz}.
\end{equation}
This together with $\gampa/\gamp = \Gama/\Gamma$ enables us to rewrite the finite-bias correction to the coefficient $\aCNHa$ given in \Eq{aH_CN NTC} as
\begin{align}
    \aCNTCHa 
    &= \frac{U}{2}\frac{\gampa}{\gamc}[(\nzia - 1) - (\nzi - 1)]\notag\\
    &= \frac{U}{2}\frac{\gampa}{\gamc}\left[\frac{\gamca}{\gampa}(\nzia - \nza) - \frac{\gamc}{\gamp}(\nzi - \nz)\right]\notag\\
    &=  \frac{\gamca}{\gamc}\frac{U}{2}(\nzia - \nza) + \frac{U\Gama}{2\Gamma}(\nz - \nzi).\label{eq_non_resonant_6}
\end{align}
One can now straightforwardly sum \Eq{eq_non_resonant_6} with $\aPNHa$ given in \Eq{aH_PN}, yielding 
\begin{align}
    \taCNHa &= \aCNHa + \aPNHa\label{eq_geometric_alt2}\\
    &= \left[\epsilon + \frac{U}{2}(2 - \nza)\right]\frac{\gamca}{\gamc} + \frac{U}{2}\frac{\gampa}{\gamp}(\nz - 1).\notag
\end{align}
We can thus rewrite the geometric connection
\begin{equation}
    \AHa(\bR) = \taCNHa\nr \nz + \aPpHa\nr p_z.\label{eq_geometric_alt1}
\end{equation}
The key difference between $\taCNHa$ and $\aCNHa$ [\Eq{aH_CN}] is that, unlike for the tight-coupling energy \eq{eq_tight_coupling_energy}, the prefactor in front of the charge rate ratio $\gamca/\gamc$ in \Eq{eq_geometric_alt2} depends on the actual, lead-resolved occupation $\nza$, and not on the dual $\nzia$, and the finite-bias correction $\sim \gampa/\gamp$ likewise depends on $\nz$ instead of $\nzi$. This difference simplifies the analytical expression for the energy pumping curvature $\BHa(\bR)$ whenever $U$ and the couplings $\Gama$ are constant during the driving cycle. Namely, with $\nr \aPpHa = \nr\frac{U}{4}\frac{\gampa}{\gamp} \rightarrow 0$ and $\nr \nz \times \nr \nz = 0$, the only terms $\sim\nr \nz$ which survive when inserting \Eq{eq_geometric_alt1} into \Eq{decomposition_B} are
\begin{align}
    \BHa(\bR) &\overset{U,\Gama \text{ const.}}{\rightarrow}\nr \taCNHa \times \nr \nz\label{eq_geometric_alt3}\\
    &\overset{U,\Gama \text{ const.}}{\rightarrow} \left[\epsilon + \frac{U}{2}(2 - \nza)\right]\BNa(\bR)\notag\\
    &\phantom{\overset{U,\Gama \text{ con}}{\rightarrow}} + \left[\nr\epsilon - \frac{U}{2}\nr\nza\right]\frac{\gamca}{\gamc} \times \nr \nz.\notag
\end{align}

The form \eq{eq_geometric_alt3} provides useful estimates of the pumped energy for constant tunnel couplings $\Gama$ in the \emph{single-resonance} case\footnote{However,  at first glance, \Eq{eq_geometric_alt3} suggests $\epsilon + (U/2)(2 - \nza)$ to be the energy effectively transported with the pumped charge current which is misleading. For example, for $U > 0$, this energy would approach $\epsilon + U$, close to a resonance at which $\nza \approx 0$ and at which we would instead expect a pumped energy $\sim \epsilon$ as correctly predicted by the tight-coupling term \eq{eq_tight_coupling_energy}. Moreover, the second line of \Eq{eq_geometric_alt3} typically contributes significantly whenever $\BNa(\bR) \neq 0$.}. Most importantly, if we set $\Vb$ as well as any temperature $\Tapp$ of the two leads as driving parameters, and assume the non-resonant lead $\alpha'$ to fulfill the condition \eq{eq_non_resonant_1}, we can use \Eq{eq_non_resonant_5_a} in \Eq{eq_geometric_alt3} to simplify $\BHa(\{\Vb,\Tapp\})$ to
\begin{align}
    \BHa(\{\Vb,\Tapp\}) &\rightarrow \left[\epsilon +\frac{U}{2}(2 - \nza)\right]\BNa(\{\Vb,\Tapp\})\notag\\
    &\phantom{\rightarrow} -\frac{U}{2}\frac{\gamca}{\gamc}\nr\nza \times \nr \nz\notag\\
    &= \left[\epsilon +\frac{U}{2}(2 - \nza)\right]\BNa(\{\Vb,\Tapp\}) \notag\\
    &\phantom{=} + \frac{U}{2}\left(\nza - \nzap\right)\left(\nr\frac{\gamca}{\gamc}\right)\notag\\
    &\phantom{= + \frac{U}{2}}\times \frac{\gamca}{\gamc}\nr\nza\\
    &\rightarrow\left[\epsilon +\frac{U}{2}(2 - \nzap)\right]\BNa(\{\Vb,\Ta\}).\notag
\end{align}
A vanishing charge pumping curvature thereby dictates a vanishing energy pumping curvature.

\section{Analytical results for the two-particle resonance}\label{app_two_particle_resonance}

This appendix analyzes more in detail charge and energy pumping for an attractive interaction $U < 0$ close to the two-particle resonance $\epsilon/\TN \approx -U/(2\TN)$, by studying the analytical expressions of the pumping curvatures for $|\Vb| \le |U|$ (mechanism C).

\subsection{Charge pumping}\label{app_two_particle_resonance_charge}

The identities $\nz = \sum_\alpha (\gamca/\gamc)\nza$ [\Eq{eq_non_resonant_2}] and $\sum_\alpha(\gamca/\gamc) = 1$ for arbitrary interaction energy $U$ show that charge pumping is suppressed at finite bias voltage [mechanisms C$^\pm]$. Indeed, at $|U| > |\Vb| \geq \TN$ and $\epsilon/\TN \approx -U/(2\TN)$, the dot is actually off-resonant with each individual lead due to the bias. This off-resonance condition and the strong pairing effect exponentially suppress any parameter-dependence of the individual $\nza$. The total charge gradient $\nr \nz$ is therefore proportional to the gradient of the charge rate ratio $\gamca/\gamc$, 
\begin{equation}
    \nr \nz \approx (\nza - \nzap) \nr\left(\frac{\gamca}{\gamc}\right).
\end{equation} 
This together with the expression \eq{aN} for $\aCNNa$ dictates $\nr \aCNNa$ and $\nr\nz$ to be parallel, and hence a vanishing charge pumping curvature \eq{BNR}. 

Around $\Vb = 0$, all driving schemes yield a sizable pumped charge, except for $\bR = \{\epsilon, U\}$, as shown by \Fig{fig_charge_pumping_curvature_attractive_R}. The latter exception is, just as for $U > 0$, due to both leads being effectively identical when $|\Vb| = \TL - \TN = 0$: the contacts thereby combine to a single equilibrium environment with vanishing cycle-averaged particle transfer to the dot due to total charge conservation [\App{app_work}].

However, if only one of the dot energies $\epsilon$ or $U$ is driven together with $\TL$ or $\Lambda$, the latter two parameters affect the left-right lead-dot coupling balance. The $\epsilon$- or $U$-driving at mechanism C with $|\Vb|/\TN \approx 0$ then makes the dot more likely to accept particles when coupled more strongly to one lead compared to when coupled more strongly to the other. This yields a net pumped charge current [\Fig{fig_charge_pumping_curvature_attractive_R}(a-d)] whose sign is only determined by the driving sequence, but not by the precise working point level $\beps$ and bias voltage $\bVb$ around $\beps = -\bU/2$, $\bVb = 0$. 

The finite pumped charge for the driving schemes $\bR =\{R_1, \Vb\}$, with $R_1 = \epsilon, U$, arises because the dot energy driving changes the transmission strength for times with positive bias, relatively to the transmission strength for times with negative bias. Depending on the driving phase $\phi$ in \Eq{Ri}, a working point $\beps < -\bU/2$ could then mean that the dot approaches the two-particle resonance while $\Vb(t) < 0$, but moves further away from this resonance while $\Vb(t) > 0$, which would favor charge transport from right to left. The situation is reversed for $\beps > -\bU/2$ and the same driving phase $\phi$, thereby introducing an $\beps$-dependent sign change of $\BNa(\bR)$ [\Figs{fig_charge_pumping_curvature_attractive_R}(e,g)].

If the dot energies remain constant, and $\Vb$ is instead driven together with $\TL$ or $\Lambda$ affecting the left-right coupling balance, the sign of $\BNa(\bR)$ changes both as a function of $\beps$ and $\bVb$ around $\beps = -\bU/2, \bVb = 0$ [\Figs{fig_charge_pumping_curvature_attractive_R}(h,j)]. The positive sign for $\beps > -\bU/2,\bVb > 0$ and for $\beps < -\bU/2,\bVb < 0$  means that the left/right lead potential $\bmua$ is, on average, closer to a two-particle resonance with the dot during cycle times in which the left/right lead is also more strongly coupled; the complementary working-point regimes with negative charge pumping curvature sign correspond to the reversed relation between resonance and coupling.

Finally, for $\bR = \{\TL, \Lambda\}$, the pumped charge switches signs when crossing the line $\bmuL =  \beps + \bU/2$ [\Fig{fig_charge_pumping_curvature_attractive_R}(f)]. Indeed, in this case, using \Eq{eq_non_resonant_2}, we get
\begin{align}
    \partial_\Lambda \nz &= (\nza - \nzap) \partial_\Lambda\left(\frac{\gamca}{\gamc}\right),\nonumber\\
    \partial_{\TL} \nz &= (\nza - \nzap) \partial_{\TL}\left(\frac{\gamca}{\gamc}\right) + \frac{\gamcL}{\gamc}\partial_{\TL}\nzL,
\end{align} 
and, therefore,
\begin{align}
    B_{N\comal}(\{\TL, \Lambda\}) &= -\TN\partial_\Lambda\left(\frac{\gamca}{\gamc}\right)\frac{\gamcL}{\gamc}\partial_{\TL}\nzL.
\end{align}
The derivative $\partial_\Lambda\left({\gamca}/{\gamc}\right)$ has a constant sign while the sign of $\partial_{\TL}\nzL$ depends on the working point level position $\beps$ relatively to the two-particle resonance with the left lead: $\nzL$ increases with $\TL$ for $\beps + \bU/2 < \bmuL$ due to the increase in the number of available electrons in the lead at energy $\beps$ and, on the contrary, $\nzL$ decreases with $\TL$ for $\beps + \bU/2 > \bmuL$ due to the increased number of available holes in the lead at energy $\beps + \bU$.

\subsection{Energy pumping}\label{app_two_particle_resonance_energy}

The energy pumping curvature $|\Vb| < |U|$ is purely charge-like, $\BHa(\bR) \approx \BHaC(\bR)$, just as for repulsive interaction. However, close to zero bias, $|\Vb|/\TN \approx 0$, $\BHaC(\bR)$ is \emph{not} dominated by the tightly-coupled contribution,
\begin{equation}
    \BHa^{\C,\TC}(\bR) = \cEa\BNa(\bR) + \frac{\gamca}{\gamc}\nr \cEa \times \nr \nz, \label{eq_tpr_1}
\end{equation}
coming from \Eq{aH_CN TC}. On the contrary, the main contribution comes from the second term in \eq{aH_CN TC}, $\BHa^{\C,\NTC}(\bR) = \nr \aCNTCHa \times \nr \nz$, where $\aCNTCHa$ is defined by \Eq{aH_CN NTC}. To see this, we first note that, like for $U > 0$, the tightly-coupled contribution $\BHa^{\C,\TC}(\bR)$
is small around $\epsilon/\TN \approx -U/(2\TN)$, $|\Vb|/\TN \approx 0$, since the Seebeck energy $\cEa$ vanishes at this point, and its derivative $\pr[i]\cEa$ is non-vanishing only for $R_i = \epsilon, U$ since $\nzia$ is approximately constant there. Knowing the parameter dependence of $\nz$ for $U < 0$ from the one of $\nzi$ for $U > 0$, it is clear from \Eq{eq_tpr_1} that only driving $\epsilon$ or $U$ in combination with $\TL$ or $\Lambda$ gives sizable tight-coupling contributions. However, for $U > 0$ the non-tightly coupled term $\BHa^{\C,\NTC}(\bR)$ vanishes at $|\Vb| < U/2$ while it is typically the main contribution to the energy pumping curvature for $U < 0$.

Quantifying the non-tightly coupled contribution with the expression \eq{aH_CN NTC} for $\aCNTCHa$ is difficult for $U < 0$: both $(\nzia - \nzi)$ and $\gamc$ are individually suppressed for mechanism C, but their ratio is actually finite. Indeed, $\nzi \approx \nzia \approx 1$ at the point of interest $\epsilon/\TN \approx -U/(2\TN)$, which lies in the Coulomb blockade regime for the inverted, \emph{repulsive} model. We therefore use the rewriting from \Eq{eq_non_resonant_6} as a starting point. Then, using \Eq{eq_non_resonant_2} and labeling the other lead by $\alpha' \neq \alpha$, we obtain
\begin{align}
    \aCNTCHa  &\approx  \frac{U}{2}\left(\frac{\Gamap\gamca}{\Gamma\gamc}(1 - \nza) -\frac{\Gama\gamcap}{\Gamma\gamc}(1 - \nzap)\right).\label{eq_tpr_2}
\end{align}
For $\Lambda = 0$, $\TL = T$ and  $|U| \gg \TN$, we use the explicit expressions of the Fermi functions and neglect the exponential terms in $U/(2\TN)$, since we are interested in the parameters such that $|U|/(2\TN) \gg |2\epsilon + U|/(2\TN)$ and $|U|/\TN \gg |\Vb|/\TN$, to obtain
\begin{equation}
    \aCNTCHa  \approx \frac{U}{4}\tanh(\frac{\Vb}{2\TN}).\label{eq_tpr_3}
\end{equation}
For $|Vb| < |U|$, $\nz$ depends only on the sign of $\epsilon + U/2$, that is whether the dot is above or below the particle-hole symmetric point and, as a consequence, $\partial_{\Vb}\nz/\TN \approx 0$. Therefore, when driving the bias voltage, using \Eq{eq_tpr_3}, we get
\begin{equation}
    \BCNTCHa(\{R_1,\Vb\})=  -\frac{\bU}{8}\left(1 - \tanh[2](\frac{\bVb}{2\TN})\right)\pr[1]\nz.\label{eq_tpr_4}
\end{equation}
This pumping curvature has the sign of $\pr[1]\nz$ since $U$ is negative.
If $R_1 = \epsilon, U$, then $\pr[1]\nz$ is negative at C since the dot goes from double occupation to empty when crossing the particle-hole symmetric point at $\epsilon = -U/2$ (and $\partial_{\epsilon}\nz \approx 2\partial_{U}\nz$). The curvature $\BCNTCHa(\{R_1,\Vb\})$ is thus always negative, and suppressed for $|\bVb| \gg \TN$, as visible in \Figs{fig_pumping curvature attractive R}(e,g). On the other hand, if $R_1 = \Lambda, \TL$, $\pr[1]\nz$ changes sign with $\bVb$ and so does the energy pumping curvature, see  \Figs{fig_pumping curvature attractive R}(h,j).

For $(\epsilon, U)$-driving, \Eq{eq_tpr_3} gives
\begin{equation}
    \BCNTCHa(\{\epsilon, U\})= - \frac{\TN^2}{4} \tanh(\frac{\bVb}{2\TN})\partial_{\epsilon}\nz.\label{eq_tpr_5}
\end{equation}
The energy pumping curvature thereby has the sign of the bias, vanishes at $\bVb = 0$ and takes finite values for $\TN \leq |\bVb| < |U|$ around the particle-hole symmetric point, see \Fig{fig_pumping curvature attractive R}(i). In other words, as soon as the system deviates from the perfectly balanced case $\bVb/\TN = \bTL/\TN - 1 = 0$ shown in \App{app_work}, the energy pumped by $(\epsilon,U)$-driving is finite, with the sign of $\bVb$ determining the sign of the cycle-averaged energy flow.

We now look at $(\TL, \Lambda)$-driving and compute the derivatives of $\aCNTCHa$ from \Eq{eq_tpr_2}. At $\bLam=0$, $\bTL=\TN$, for $|U|\gg \TN$ $|U|/(2\TN) \gg |2\epsilon + U|/(2\TN)$ and $|U|/\TN \gg |\Vb|/\TN$,
\begin{align}
    \partial_{\Lambda}\aCNTCHa\!&\!= \frac{U}{4\gamc^2}(\gamca\! - \!\gamcap)(\gamca(1\! - \!\nza)\! -\! \gamcap(1\! -\! \nzap))\nonumber\\
    &\approx \frac{U}{4}\tanh[2](\frac{Vb}{2\TN})\tanh(\frac{2\epsilon  + U}{2\TN})\label{eq_tpr_6}
\end{align}
and
\begin{equation}
    \partial_{\TL}\aCNTCHa\approx U\frac{2\beps + U - \Vb + U\tanh(\frac{2\epsilon + U}{2\TN})}{16\TN\cosh[2](\frac{Vb}{2\TN})}.\label{eq_tpr_7}
\end{equation}
The derivative $\partial_{\Lambda}\aCNTCHa$ [\Eq{eq_tpr_6}] expectedly vanishes at $\bVb = 0$, since both leads are identical at the working point $\bTL = \TN$, and thus together behave as one unique equilibrium environment. The derivative precisely at $\epsilon = -U/2$ also goes to zero for $\Vb \neq 0$, as $f^+_{\alpha/\alpha'}(\epsilon) \overset{\epsilon = -U/2}{=} f^-_{\alpha'/\alpha}(\epsilon + U)$ for $\bTL = \bTR = \TN$ means $\gamca \overset{\epsilon = -U/2}{=} \gamcap$. Flipping the sign of $\Vb$ amounts to exchanging the leads, again because $\bTL = \TN$, and it is clear from \Eq{eq_tpr_6} that $\partial_{\Lambda}\aCNTCHa$ is unchanged under $\alpha \leftrightarrow \alpha'$. Consequently, the sign of the $\Lambda$-derivative is independent of $\Vb$, but changes with $\epsilon + U/2$. Turning to $\partial_{\TL}\aCNTCHa$, \Eq{eq_tpr_7} indicates a strong suppression for $|\Vb| \gg \TN$. At low bias, $|\Vb|/\TN \ll 1$, the derivative $\partial_{\TL}\aCNTCHa$ changes sign as a function of $\epsilon$ close to the particle-hole symmetric point, since $|2\epsilon + U  - \Vb| \rightarrow |\Vb| \ll |U|$, while the sign of $\Vb$ has no influence. 
As discussed above, both $\partial_{\TL}\nz$ and $\partial_{\Lambda}\nz$ have the same sign as $\bVb$ but do not change sign with $\beps + \bU/2$. Since  $\partial_{\Lambda}\aCNTCHa$ and  $\partial_{\TL}\aCNTCHa$ have opposite signs as a function of $\beps + \bU/2$, this results in a pumping curvature with four quadrants with alternating signs around the particle-hole symmetric point, see \Fig{fig_pumping curvature attractive R}(f).

The energy pumping curvature for $(\epsilon, \Lambda)$-driving is given by $\BCNTCHa(\{\epsilon, \Lambda\})= - \partial_{\Lambda}\aCNTCHa\partial_\epsilon \nz$. Since $\partial_\epsilon\nz$ has a constant negative value for $\epsilon/\TN \approx -U/(2\TN)$,  $\BCNTCHa(\{\epsilon, \Lambda\})$ is mainly determined by $\partial_{\Lambda}\aCNTCHa$ [\Eq{eq_tpr_6}]. This derivative has been shown above to vanish both at $\Vb = 0$ and for $\epsilon = -U/2$, but to only change sign with $\epsilon + U/2$, and this is confirmed by \Fig{fig_pumping curvature attractive R}(b). For $(U, \Lambda)$-driving, \Eqs{eq_tpr_2} and \eq{eq_tpr_6} give, for $|U|\gg \TN$ $|U|/(2\TN) \gg |2\epsilon + U|/(2\TN)$ and $|U|/\TN \gg |\Vb|/\TN$,
\begin{equation}
    \BCNTCHa(\{U, \Lambda\}) \approx \frac{\tanh[2](\frac{\bVb}{2\TN})\left(\TN + \bU\tanh(\frac{2\beps + \bU}{2\TN})\right)}{4\cosh[2](\frac{2\beps + \bU}{2\TN})},
\end{equation}
which predicts a similar behavior as for $(\epsilon, \Lambda)$-driving given that $|U| \gg \TN$, see \Fig{fig_pumping curvature attractive R}(d).

The non-tightly coupled contribution to the curvature $\BCNTCHa(\{\epsilon, \TL\})= - \TN^2\partial_{\TL}\aCNTCHa\partial_\epsilon \nz$ for $(\epsilon, \TL)$-driving is suppressed at finite bias voltage [\Eq{eq_tpr_7}], $|\Vb| \gg \TN$. The pumped energy is therefore dominated by the tightly-coupled contribution [\Eq{eq_tpr_1}], $B_{H\comal}^\text{C,TC}(\{\epsilon, \TL\}) \approx [(1-\bLam)/2] \partial_{\TL}\nz$, which has the same sign as $\bVb$, see \Fig{fig_pumping curvature attractive R}(a). Conversely, at zero bias, $|\Vb|/\TN \approx 0$, the non-tightly coupled term $\BCNTCHa(\{\epsilon, \TL\})$ dominates and changes sign with $\beps + \bU/2$ but not with $\bVb$ [\Eq{eq_tpr_7}]. For $(U, \TL)$-driving, the fact that $\partial_{U}\aCNTCHa$ does not vanish at $\Vb > \TN$ and is of the same order of magnitude as $B_{H\comal}^\text{C,TC}(\{U, \TL\})$ makes it difficult to understand the origin of the features in \Fig{fig_pumping curvature attractive R}(c). Focusing on $\Vb = 0$ only, $\partial_{U}\aCNTCHa \sim \Vb/8$ [\Eq{eq_tpr_3}], $|\TN \partial_{\TL}\nz| \ll 1$ and $\partial_{U}\nz \approx \partial_\epsilon\nz/2$ imply $\BCNTCHa(\{U, \TL\})\approx \BCNTCHa(\{\epsilon, \TL\})/2$. This in particular predicts the same sign change of the energy pumping curvature as a function of $\beps + \bU/2$ as for $(\epsilon, \TL)$-driving.

\allowdisplaybreaks
\section{Analytical expressions at the first order in the temperature difference} \label{app_first_order_dT}

This appendix calculates the transport quantities of \Eq{energy_pumping} and \Eq{eq_work} from \Sec{sec_refrigerator}, i.e., the heat contributions $Q^{(0,1)}_\R$ and work terms $\Weps^{(1,2)}$ up to linear order in a small left-right temperature difference $\dT$, such that $\bTL= \TN + \dT$ and $\abs{\dT} \ll \TN$. We are considering $\bLam = 0$ and $\bVb = 0$.

The first step is therefore to expand all relevant relaxation rates and stationary averages up to first order in $\dT$ around the full-equilibrium values. Since we are looking at the linear response of the system, these expansion will make the equilibrium fluctuations $\delnzsqeq$ , $\delnzisqeq$ appear. We denote
\begin{gather}\label{eq_eqquantities}
    \nzeq = \left.\nz\right|_{\teq} = \left.\nza\right|_{\teq}\;\,,\;\, \nzieq = \left.\nzi\right|_{\teq} = \left.\nzia\right|_{\teq}\\
    \gamceq = \left.\gamc\right|_{\teq} = \frac{\Gamma}{\Gama}\left.\gamca\right|_{\teq}\notag\\
    \cEeq = \left.\cEa\right|_{\teq} = \epsilon + \frac{U}{2}\left(2 - \nzieq\right)\notag\\
    \ciEeq = \left.\cEeq\right|_{H,\Vb\rightarrow -H,-\Vb} = -\epsilon - \frac{U}{2}\left(2 - \nzeq\right)\notag\\
    \delnzsqeq = \left.\Braket{(N-\nz)^2}{z}\right|_{\teq} = \left.\Braket{(N - \nza)^2}{\za}\right|_{\teq} \notag\\
    \delnzisqeq = \left.\Braket{(N-\nzi)^2}{\zin}\right|_{\teq} = \left.\Braket{(N-\nzia)^2}{\zia}\right|_{\teq}\notag\\
    \lamnzcbeq = \left.\Braket{(N-\nz)^3}{z}\right|_{\teq} = \left.\Braket{(N - \nza)^3}{\za}\right|_{\teq} \notag\\
    \lamnzicbeq = \left.\Braket{(N-\nzi)^3}{\zin}\right|_{\teq} = \left.\Braket{(N-\nzia)^3}{\zia}\right|_{\teq},\notag
\end{gather}
where $|_{\teq}$ corresponds to the limit $\dT,\Vb \rightarrow 0$, and $\Ket{\za},\Ket{\zia}$ are the stationary state and its dual for the situation in which only the lead $\alpha$ couples to the dot. The skewnesses $\lamnzcbeq$, $\lamnzicbeq$ will appear in the derivatives of the equilibrium charge fluctuations.
The expansion of the stationary averages up to linear order in $\dT$ around $\dT,\Vb \rightarrow 0$ is achieved with the help of the state-linearization derived in Eqs.~(A31)-(A33) of \Ref{Schulenborg2017Dec} for the spin-degenerate single-level quantum dot with energy-independent couplings: 
\begin{align}
    \Ket{z_{\R}} &= \Ket{z}|_{\teq} = \Ket{\zeq} \notag\\
    \left[\partial_{\dT}\Ket{z_\L}\right]|_{\teq} &= \frac{1}{\TN^2}\left[H - \Braket{H}{\zeq}\one\right]\Ket{\zeq}\label{eq_transport_zero}
\end{align}
and
\begin{align}
    \left[\partial_{\dT}\Ket{z}\right]|_{\teq} &= \sum_\alpha \frac{\Gama}{\Gamma}\left[\partial_{\dT}\Ket{\za}\right]|_{\teq}\notag\\
    &= \frac{1 + \Lambda}{2\TN^2}\left[H - \Braket{H}{\zeq}\one\right]\Ket{\zeq}\notag\\
    \Rightarrow \Ket{z} &\approx \Ket{\zeq} + \frac{1 + \Lambda}{2}\frac{\dT}{\TN^2}\left[H - \Braket{H}{\zeq}\one\right]\Ket{\zeq}\notag\\
    &\phantom{=} + \mathcal{O}(\dT^2),\label{eq_transport_one}
\end{align}
having used $\Vb = 0$ and that for a constant $\TR = \TN$, $\partial_{\dT}$ only acts on terms with $\alpha = \L$ of the lead sums in \Eq{eq_transport_one}. 

Together with Eqs.~(A47) and (A48) from \Ref{Schulenborg2017Dec}, equations \eq{eq_transport_zero} and \eq{eq_transport_one} imply
\begin{align}
    \nzR &= \nzeq \quad,\quad \nzL \approx \nzeq + \delnzsqeq\frac{\cEeq\dT}{\TN^2} + \mathcal{O}(\dT^2) \notag\\
    \nz &\approx \nzeq + \frac{1 + \Lambda}{2}\delnzsqeq\frac{\cEeq\dT}{\TN^2} + \mathcal{O}(\dT^2)\label{eq_transport_two}.
\end{align}
Since energy inversion commutes with $\partial_{\dT}$, we obtain the corresponding dual quantities from \Eq{eq_transport_two} by simply applying the dual parameter transform:
\begin{align}
    \nziR &= \nzieq \quad, \quad \nziL \approx \nzieq + \delnzisqeq\frac{\ciEeq\dT}{\TN^2} + \mathcal{O}(\dT^2)\notag\\
    \nzi &\approx \nzieq + \frac{1 + \Lambda}{2}\delnzisqeq\frac{\ciEeq\dT}{\TN^2} + \mathcal{O}(\dT^2).\label{eq_transport_three}
\end{align}

To calculate the parity derivative $\partial_{\dT}p_z = \Braket{\fpOp}{\partial_{\dT}z}$, we first insert the Liouville space unit operator expansion with respect to the eigenmodes of right-lead kernel $\WR$, and then use \Eq{eq_transport_two} as well as Eqs.~(A55) and (A61) from \Ref{Schulenborg2017Dec}:
\begin{align}
    \partial_{\dT}p_z|_{\teq} &= \left.\Braket{\fpOp}{c_{\R}}\right|_{\teq}\partial_{\dT}\nz|_{\teq}\label{eq_transport_four}\\
    &\phantom{=}+ 4\left[\partial_{\dT}\Braket{z^{\inv}_\R\fpOp}{z}\right]|_{\teq}\notag\\
    &= \frac{1 + \Lambda}{\TN^2}\left[\cEeq(1 - \nzieq) + \frac{U}{2}\delnzisqeq\right]\delnzsqeq,\notag
\end{align}
where we have used
\begin{equation}
    \left[\partial_{\dT}\Braket{z^{\inv}_\R\fpOp}{z}\right]|_{\teq} = \frac{1 + \Lambda}{2}\frac{U}{4\TN^2}\delnzsqeq\delnzisqeq.\label{eq_transport_four_one}
\end{equation}
This means
\begin{align}
    \pzR &= \pzeq\label{eq_transport_five}\\
    p_z &\approx \pzeq + \frac{1 + \Lambda}{\TN^2}\left[\cEeq(1 - \nzieq) +\frac{U}{2}\delnzisqeq\right] \delnzsqeq\dT\notag\\
    &+ \mathcal{O}(\dT^2)\notag.
\end{align}

Next, we need to expand the charge rate ratio $\gamcR/\gamc$. We use \Eqs{eq_non_resonant_4}, \eq{eq_transport_one}, \eq{eq_transport_three}, $\gamc = \gamcR + \gamcL$, $\gamcR = (\GamR/\Gamma)\gamceq$ and $\Gamma_{\text{L/R}} = \Gamma(1 \pm \Lambda)/2$ to derive
\begin{align}
    \frac{\gamcR}{\gamc} &= \left[1 + \frac{1 + \Lambda}{1 - \Lambda}\frac{\gamp}{\gamceq}\frac{\gamcL}{\gampL}\right]^{-1},\\
    \frac{\gamcL}{\gampL} &\approx \frac{\gamceq}{\gamp} + \frac{\gamceq^2}{\gamp^2}\frac{\delnzsqeq}{\nzieq - 1}\frac{\cEeq\dT}{\TN^2} + \mathcal{O}(\dT^2) + \text{inv.},\notag
\end{align}
which means
\begin{align}
    \frac{\gamcR}{\gamc} &\approx \frac{1 - \Lambda}{2} - \frac{(1 - \Lambda^2)\gamceq\delnzsqeq}{4(\nzieq - 1)\gamp}\frac{\cEeq\dT}{\TN^2} + \mathcal{O}(\dT^2) + \text{inv.}\label{eq_transport_six}
\end{align}
The terms ``inv.'' are contributions $\sim\delnzisqeq$ that further below will only appear in products $\sim\delnzsqeq\delnzisqeq$ which are strongly suppressed by the very nature of the dual fluctuations $\delnzisqeq$ being sizable at different points in parameter space than the fluctuations $\delnzsqeq$.

Having determined all relevant quantities up to linear order in $\dT$, we still require the derivatives of these quantities with respect to $\epsilon$ and $\Lambda$ in order to compute the quantities entering \Eq{energy_pumping} and \Eq{eq_work}. Any quantity derived from the equilibrium state $\Ket{\zeq} = \Ket{z}|_{\teq}$ is independent of the tunnel couplings, hence giving a vanishing $\Lambda$-derivative. For the $\epsilon$-derivative, equation A42 from \Ref{Schulenborg2017Dec} implies
\begin{align}
    \partial_{\epsilon}\Ket{\zeq} &= -\frac{N - \nzeq\hat\one}{\TN}\Ket{\zeq} \,\,,\,\, \partial_{\epsilon}\Ket{z^{\inv}_{\teq}} = \frac{N - \nzieq\hat\one}{\TN}\Ket{z^{\inv}_{\teq}}\label{eq_transport_seven}
\end{align}
and thus
\begin{gather}
    \partial_\epsilon \nzeq = -\frac{1}{\TN}\delnzsqeq \quad,\quad \partial_\epsilon \nzieq = +\frac{1}{\TN}\delnzisqeq\notag\\
    \partial_\epsilon \cEeq = 1 - \frac{U}{2\TN}\delnzisqeq
    \label{eq_transport_eight}.
\end{gather}
For the equilibrium parity $\pzeq = \Braket{\fpOp}{\zeq}$, we insert the unit expansion of the equilibrium eigenmodes and use that $\Bra{\zieq\fpOp}\partial_{\epsilon}\Ket{\zeq} = -\Braket{\zieq\fpOp N}{\zeq}/\TN = 0$ according to the eigenmode orthogonality $\Braket{\zieq\fpOp}{\zeq} = 0$ and Eq.~A46 of \Ref{Schulenborg2017Dec}:
\begin{align}
    \partial_{\epsilon}\pzeq &= \Braket{\fpOp}{c_{\teq}}\partial_{\epsilon}\nzeq = -\frac{2(1 - \nzieq)}{\TN}\delnzsqeq.\label{eq_transport_eight_two}
\end{align}
Moreover, using the definition \eq{eq_eqquantities} of the fluctuations $\delnzsqeq,\delnzisqeq$, the derivative $\partial_\epsilon(N - x\one)^n = -n(\partial_\epsilon x)(N - x\one)^{n-1}$, and \Eq{eq_transport_seven}, we obtain the $\epsilon$-derivative of the fluctuations in terms of the skewnesses $\lamnzcbeq$, $\lamnzicbeq$ 
at full equilibrium:
\begin{gather}
    \partial_\epsilon\delnzsqeq = -\frac{\lamnzcbeq}{\TN} \quad, \quad \partial_\epsilon\delnzisqeq = +\frac{\lamnzicbeq}{\TN}
    \label{eq_transport_nine}
\end{gather}
The derivative of $\gamceq$ with respect to $\epsilon$ is given by
\begin{align}
    \partial_{\epsilon}\gamceq &\approx \frac{-\delnzsqeq\gamceq^2}{\TN(\nzieq - 1)\gamp} + \text{inv.},
\end{align}
where "inv." indicates, as before, terms $\sim\delnzisqeq$ which will later drop out.
With this, we are finally in the position to compute the transport quantities in \Eqs{energy_pumping} and \eq{eq_work} up to linear order in $\dT$ around full equilibrium $\dT = \Vb = 0$. The stationary charge and energy current between dot and right lead $\alpha = \R$ read
\begin{align}
    \INR^{(0)} &= \Bra{N}\WR\Ket{z} = -\gamcR(\nz - \nzR)\notag\\
    &\approx -\frac{1-\Lambda^2}{4}\gamceq\delnzsqeq\frac{\cEeq\dT}{\TN^2} + \mathcal{O}(\dT^2)\notag\\
    \IHR^{(0)} &= \Bra{H}\WR\Ket{z} = \cEeq \INR^{(0)} - U\gampR\Braket{z^{\text{\inv}}_\R\fpOp}{z}\notag\\
    &\approx \cEeq \INR^{(0)} - \kappa \dT + \mathcal{O}(\dT^2),\label{eq_transport_ten}
\end{align}
having recalled \Eq{eq_transport_four_one} and identified the Fourier heat
\begin{equation}
    \kappa = \frac{1 - \Lambda^2}{4}\gamp\delnzsqeq\delnzisqeq\frac{U^2}{4\TN^2}\label{eq_fourier_app}
\end{equation}
as defined in, e.g., \Ref{Schulenborg2017Dec}. The corresponding derivatives of the currents and $\kappa$ with respect to $\epsilon$ and $\Lambda$ around $\Lambda = 0$ up to linear order in $\dT$ are
\begin{gather}
    \partial_{\Lambda}\kappa|_{\Lambda} = \partial_{\epsilon}\partial_{\Lambda}\kappa|_{\Lambda = 0} = 0 \notag\\
    \partial_\Lambda \INR^{(0)}|_{\Lambda = 0} \sim \partial_\Lambda \IHR^{(0)}|_{\Lambda = 0} \sim \partial_{\epsilon}\partial_{\Lambda}\INR^{(0)}|_{\Lambda = 0} \approx 0 + \mathcal{O}(\dT^2)\notag\\
    \partial_{\epsilon}\partial_{\Lambda}\IHR^{(0)}|_{\Lambda = 0} \approx \partial_{\epsilon}\cEeq\partial_{\Lambda}\INR^{(0)}|_{\Lambda = 0} \approx 0 + \mathcal{O}(\dT^2) \notag\\
    \left.\frac{\partial_\epsilon \INR^{(0)}}{\INR^{(0)}/\TN}\right|_{\Lambda=0} \approx \frac{\TN}{\cEeq} - \frac{\partial_\epsilon\gamceq}{\gamceq/\TN} - \frac{\lamnzcbeq}{\delnzsqeq} + \mathcal{O}(\dT^2)\notag\\
    \left.\frac{\partial_\epsilon \kappa}{\kappa/\TN}\right|_{\Lambda=0} = \frac{\lamnzicbeq}{\delnzisqeq} - \frac{\lamnzcbeq}{\delnzsqeq}
\end{gather}
as well as
\begin{align}
    \left.\frac{\partial^2_\epsilon \INR^{(0)}}{\INR^{(0)}/\TN^2}\right|_{\Lambda = 0} \!\!&\approx \left[\frac{\TN}{\cEeq} - \frac{\partial_\epsilon\gamceq}{\gamceq/\TN} - \frac{\lamnzcbeq}{\delnzsqeq}\right]^2\\
    &+\TN\partial_\epsilon\left[\frac{\TN}{\cEeq} - \frac{\partial_\epsilon\gamceq}{\gamceq/\TN} - \frac{\lamnzcbeq}{\delnzsqeq}\right] + \mathcal{O}(\dT^2)\notag\\
    \left.\frac{\partial^2_\epsilon \kappa}{\kappa/\TN^2}\right|_{\Lambda = 0} &= \left[\frac{\lamnzicbeq}{\delnzisqeq} - \frac{\lamnzcbeq}{\delnzsqeq}\right]^2\\
    &+\TN\partial_\epsilon\left[\frac{\lamnzicbeq}{\delnzisqeq} - \frac{\lamnzcbeq}{\delnzsqeq}\right]\notag\\
    \left.\frac{\partial^2_\Lambda\kappa}{\kappa}\right|_{\Lambda = 0} &= -2 \;,\; \left.\frac{\partial^2_\Lambda\INR^{(0)}}{\INR^{(0)}}\right|_{\Lambda = 0} \approx -2 + \mathcal{O}(\dT^2)
\end{align}
for the charge current $\INR^{(0)}$ and Fourier heat $\kappa$, and
\begin{align}
    \left.\partial_\epsilon \IHR^{(0)}\right|_{\Lambda = 0} &\approx \left.\INR^{(0)}\right|_{\Lambda = 0} + \left.\cEeq\partial_\epsilon\INR^{(0)}\right|_{\Lambda = 0}-\left.(\partial_\epsilon\kappa)\right|_{\Lambda = 0}\dT\notag\\
    &+ \mathcal{O}(\dT^2)\label{eq_transport_ten_two_step}\\
    \left.\partial^2_\epsilon \IHR^{(0)}\right|_{\Lambda = 0} &\approx \left.\partial_\epsilon \INR^{(0)}\right|_{\Lambda = 0} + \cEeq\left.\partial_\epsilon^2\INR^{(0)}\right|_{\Lambda = 0} \notag\\
    & - \left.(\partial^2_\epsilon\kappa)\right|_{\Lambda = 0}\dT + \mathcal{O}(\dT^2)\\
    \left.\partial^2_\Lambda \IHR^{(0)}\right|_{\Lambda = 0} &\approx -2\left.\IHR^{(0)}\right|_{\Lambda = 0} + \mathcal{O}(\dT^2)\label{eq_transport_ten_two}
\end{align}
for the steady-state energy current $\IHR^{(0)}$. Note that we have neglected terms $\sim\delnzsqeq\delnzisqeq$ and $\sim\delnzsqeq\lamnzicbeq$ that are, when not scaled by the large prefactor $(U/T)^2$ as in the Fourier heat $\kappa$, strongly suppressed compared to the other terms, since the quantities of the dual, inverted system have finite support at very different levels $\epsilon$ compared to the actual dot system,  see \Ref{Schulenborg2017Dec}. Inserting the driving protocol \eq{eq_driving_protocol} into the integral \Eq{eq_heat_order} and Taylor expanding the integrand $\INR^{(0)}(t)$ in the small driving amplitudes $\delta\epsilon,\delta\Lambda$ around the working point $\beps$ and $\bLam = 0$ using \Eqs{eq_transport_ten}-\eq{eq_transport_ten_two}, we find  the $\ell = 0$ heat contribution
\begin{align}
    \QR^{(0)} &\approx\frac{2\pi}{\Omega}\left[\left(\cEeq \INR^{(0)} -\kappa\dT\right)\left(1 - \frac{(\delta\Lambda)^2}{2}\right)\right.\notag\\
    &\phantom{\approx} \left. + \frac{(\delta\epsilon)^2}{4}\left(\cEeq\partial^2_\epsilon \INR^{(0)} + \partial_\epsilon \INR^{(0)} - (\partial^2_\epsilon\kappa)\dT\right)\right]_{\Lambda = 0}\notag\\
    &\phantom{\approx}+ \mathcal{O}(\dT^2).
\end{align}
The $\ell = 1$ heat component depends on the energy pumping curvature \eq{decomposition_B}, i.e., on the $(\epsilon,\Lambda)$-derivatives of $\nz$, $p_z$ and of the coefficients \eq{aH} in the geometric connection. The derivatives of $\nz$ follow from \Eqs{eq_transport_two} and \eq{eq_transport_eight} as well as \Eq{eq_transport_nine}. Again dropping any terms $\sim\delnzsqeq\delnzisqeq$ which are not scaled by a large prefactor $\sim(U/T)^2$, we find
\begin{align}
    \partial_\Lambda \nz|_{\Lambda = 0} &\approx \frac{1}{2}\delnzsqeq\cEeq\frac{\dT}{\TN^2} + \mathcal{O}(\dT^2)\label{eq_transport_11}\\
    \partial_\epsilon \nz|_{\Lambda = 0} &\approx -\frac{\delnzsqeq}{\TN} + \left[\delnzsqeq - \frac{\cEeq\lamnzcbeq}{\TN}\right]\frac{\dT}{2\TN^2} + \mathcal{O}(\dT^2),\notag
\end{align}
Differentiating the parity $p_z$ is performed analogously using \Eq{eq_transport_five} together with \Eqs{eq_transport_eight_two} and \eq{eq_transport_nine}:
\begin{align}
    \partial_\Lambda p_z|_{\Lambda = 0} &\approx \frac{\cEeq}{\TN}(1 - \nzieq)\delnzsqeq\frac{\dT}{\TN} + \mathcal{O}(\dT^2)\label{eq_transport_12}\\
    \partial_\epsilon p_z|_{\Lambda = 0} &\approx \frac{2}{\TN}(\nzieq - 1)\left[\delnzsqeq\left(1 -  \frac{\dT}{2\TN}\right) + \frac{\cEeq}{\TN}\lamnzcbeq\frac{\dT}{2\TN}\right]\notag\\
    &+ \mathcal{O}(\dT^2),\notag
\end{align}
For the charge mode coefficients in the mode-decomposed geometric connection \eq{aH}, \eq{aH_CN}, \Eqs{eq_transport_three},\eq{eq_transport_six} yield
\begin{align}
    \partial_\Lambda \aCNHR|_{\Lambda = 0} &\approx -\frac{\cEeq}{2} + \mathcal{O}(\dT^2)\notag\\
    \partial_\epsilon \aCNHR|_{\Lambda = 0} &\approx \frac{1}{2} - \frac{U}{4\TN}\delnzisqeq + \mathcal{O}(\dT).\label{eq_transport_13}
\end{align}
The derivatives of the parity-mode terms \eq{aH_Pp},\eq{aH_PN} are
\begin{align}
    \partial_\Lambda \aPNHR|_{\Lambda = 0} &\approx -\frac{U}{4}(\nzieq - 1) + \mathcal{O}(\dT^2)\notag\\
    \partial_\epsilon \aPNHR|_{\Lambda = 0} &\approx \frac{U}{4\TN}\delnzisqeq + \mathcal{O}(\dT)
\end{align}
and
\begin{align}
    \partial_\Lambda \aPpHR|_{\Lambda = 0}  &= -\frac{U}{8}\quad,\quad \partial_\epsilon \aPpHR|_{\Lambda = 0} = 0.\label{eq_transport_14}
\end{align}
\begin{widetext}
    Combining equations \eq{eq_transport_11}-\eq{eq_transport_14}, we obtain:
    \begin{align}
        \frac{Q^{(1)}_\R}{\delta S} &=  B_{H\comR}(\{\epsilon,\Lambda\})|_{\Lambda = 0}\notag\\
        &= \TN\left[(\partial_\epsilon \aCNHR + \partial_\epsilon \aPNHR)\partial_\Lambda \nz
        -(\partial_\Lambda \aCNHR + \partial_\Lambda \aPNHR)\partial_\epsilon \nz\right]|_{\Lambda = 0}
        + \TN\left[\partial_\epsilon \aPpHR\partial_\Lambda p_z - \partial_\Lambda \aPpHR\partial_\epsilon p_z\right]|_{\Lambda = 0}\notag\\
        &\approx -\frac{\delnzsqeq\cEeq}{2}\left(1 - \frac{\dT}{\TN}\right) - \frac{\lamnzcbeq\cEeq}{4}\frac{\cEeq}{\TN}\frac{\dT}{\TN} + \mathcal{O}(\dT^2),
    \end{align}
    with the surface element $\dS = \iint_\cS \dd \Lambda (\dd \epsilon/\TN) = -\pi \sin(\phi) (\delta\epsilon/\TN)\delta\Lambda$ normalized by $\TN$ as in the main text. The final step is to compute the work, as defined in \Eq{eq_work}, with the help of \Eq{eq_transport_11}. We arrive at
    \begin{align}
        \Weps^{(1)} &= -\partial_\Lambda \nz \TN\dS|_{\Lambda = 0} \approx -2\TN\frac{\INR^{(0)}}{\gamceq}\dS + \mathcal{O}(\dT^2)\\
        \Weps^{(2)} &=  -\pi\Omega \delta\epsilon^2 \frac{\partial_\epsilon \nz}{\gamc} -\pi \Omega\cos(\phi)\delta\epsilon\delta\Lambda \frac{\partial_\Lambda \nz}{\gamc}\\
        &\approx \pi\frac{\Omega}{\gamceq}\frac{\delta\epsilon^2}{\TN}\delnzsqeq\left(1 - \frac{\dT}{2\TN}\right)
        + \pi\frac{\Omega}{\gamceq}\frac{\delta\epsilon^2\cEeq\dT}{2\TN^3}\left(\lamnzcbeq -\frac{\gamceq}{\gamp}\frac{[\delnzsqeq]^2}{\nzieq - 1}\right)
        + 2\pi\frac{\Omega}{\gamceq}\cos(\phi)\frac{\INR^{(0)}}{\gamceq}\delta\epsilon\delta\Lambda + \mathcal{O}(\dT^2).\notag
    \end{align}
    This proves all relations necessary to derive the efficiency given in \Eq{eta_lim} of the main text.
\end{widetext}

\section{Validity of the adiabatic approximation} \label{app_adiabatic_limit}

\renewcommand{\arraystretch}{1}
\begin{figure}[htb!]
    \includegraphics{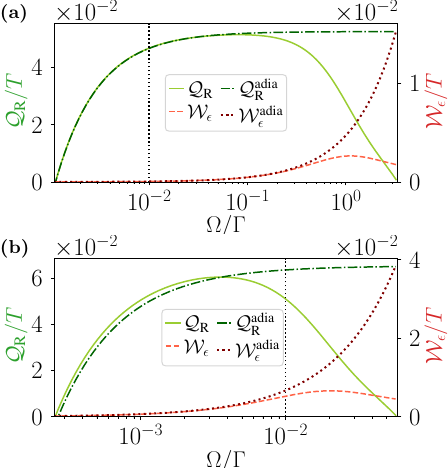}
    \caption{\label{fig_adiabatic_validity}
        Comparison of the heat $\QR$ (left axis, in green) and work (right axis, in red) computed with and without the adiabatic approximation (see legend). The interaction is repulsive in (a) and  attractive in (b). All the parameters, except for $\Omega$, are the ones corresponding to the blue stars in Fig.~\ref{fig_thermo_fct_epsilon} and the driving frequency used in the figures of Sec.~\ref{sec_refrigerator} is indicated by the vertical dotted black line.
    }
\end{figure}

In Sec.~\ref{sec_refrigerator}, we study energy pumping in the context of cyclically driven refrigerators, using the master equation in all orders in frequency. 
This appendix studies to which extent the work and heat  from this full equation deviates from the adiabatic approximation, that is truncating at the first order the expansion of $\Ket{\rho}$ in $\Omega/\Gamma$, when using it to compute  thermodynamic quantities. Fig.~\ref{fig_adiabatic_validity} shows the heat and work as functions of the driving frequency computed with and without the adiabatic approximation. More precisely, we have plotted $\QR^\text{adia} = \QR^{(0)} + \QR^{(1)}$ [Eq.~\eqref{eq_heat_order}] and $\Weps^\text{adia} = \Weps^{(1)} + \Weps^{(2)}$ [Eq.~\eqref{eq_work_expansion_orders}] and the $\QR$ and $\Weps$ obtained by numerically solving the master equation $\partial_t\Ket{\rho} = W\Ket{\rho}$ like in the figures of Sec.~\ref{sec_refrigerator}. The adiabatic approximation gives good results for $\delta R_i \Omega \ll \Gamma$ but, as explained in Sec.~\ref{sec_performance}, requires a lower driving frequency when the interaction is attractive. For the value of $\Omega$ used in the figures of Sec.~\ref{sec_refrigerator} (dotted black line), the repulsive dot is perfectly in the adiabatic-response regime while there is a significant deviation for the attractive dot. We nevertheless chose not to take a smaller $\Omega$ because the detrimental stationary contribution $\QR^{(0)}$ scales like $\Gamma/\Omega$.

\bibliography{main.bib}

\end{document}